\documentclass[a4paper, 11 pt]{article}
\pdfoutput = 1 

\usepackage{jheppub, bm}
\usepackage[T1]{fontenc} 
\usepackage[table]{xcolor}

\usepackage{graphicx}
\usepackage{subcaption} 
\usepackage{caption}
\usepackage{hhline}
\usepackage{multirow}
\usepackage{makecell}
\usepackage[export]{adjustbox}

\usepackage{mathtools}
\usepackage{MnSymbol}
\usepackage{mathrsfs}
\usepackage{slashed} 
\usepackage{physics} 
\usepackage{amsmath}

\usepackage[capitalise]{cleveref} 

\usepackage{diagbox}

\newcommand{\nnu}{\nonumber\\}
\allowdisplaybreaks
\usepackage{atbegshi,picture}

\title{Polarized fragmenting jet functions in Inclusive and Exclusive Jet Production}

\author[a,b,c]{Zhong-Bo Kang}
\author[d,e,f]{, Hongxi Xing}
\author[g]{, Fanyi Zhao}
\author[d,e,a]{ and Yiyu Zhou}

\affiliation[a]{Department of Physics and Astronomy, University of California, Los Angeles, California 90095, USA}
\affiliation[b]{Mani L. Bhaumik Institute for Theoretical Physics, University of California, Los Angeles, California 90095, USA}
\affiliation[c]{Center for Frontiers in Nuclear Science, Stony Brook University, Stony Brook, New York 11794, USA}
\affiliation[d]{Key Laboratory of Atomic and Subatomic Structure and Quantum Control (MOE), Guangdong Basic Research Center of Excellence for Structure and Fundamental Interactions of Matter, Institute of Quantum Matter, South China Normal University, Guangzhou 510006, China}
\affiliation[e]{Guangdong-Hong Kong Joint Laboratory of Quantum Matter, Guangdong Provincial Key Laboratory of Nuclear Science, Southern Nuclear Science Computing Center, South China Normal University, Guangzhou 510006, China}
\affiliation[f]{Southern Center for Nuclear-Science Theory (SCNT), Institute of Modern Physics, Chinese Academy of Sciences, Huizhou 516000, China}
\affiliation[g]{Center for Theoretical Physics, Massachusetts Institute of Technology, Cambridge, MA 02139, USA}

\emailAdd{zkang@ucla.edu}
\emailAdd{hxing@m.scnu.edu.cn}
\emailAdd{fanyi@mit.edu}
\emailAdd{zyiyu@m.scnu.edu.cn}

\abstract{
In this work, we present a complete theoretical framework for analyzing the distribution of polarized hadrons within jets, with and without measuring the transverse momentum relative to the standard jet axis.
Using soft-collinear effective theory (SCET), we derive the factorization and provide the theoretical calculation of both semi-inclusive and exclusive fragmenting jet functions (FJFs) under longitudinal and transverse polarization.
With the polarized FJFs, one gains access to a variety of new observables that can be used for extracting both collinear and transverse momentum dependent parton distribution functions (PDFs) and fragmentation functions (FFs).
As examples, we provide numerical results for the spin asymmetry $A_{TU,T}^{\cos(\phi _S - \hat{\phi}_{S_h})}$ from polarized semi-inclusive hadron-in-jet production in polarized $pp$ collisions at RHIC kinematics, where a transversely polarized quark would lead to the transverse spin of the final-state hadron inside the jet and is thus sensitive to the transversity fragmentation functions. Similarly, another spin asymmetry, $A_{TU, L}^{\cos(\phi _q - \phi _{S})}$ from polarized exclusive hadron-in-jet production in polarized $ep$ collisions at EIC kinematics would allow us to access the helicity fragmentation functions. These observables demonstrate promising potential in investigating transverse momentum dependent PDFs and FFs and are worthwhile for further measurements.
}

\AtBeginShipoutNext{\AtBeginShipoutUpperLeft{%
\put(\dimexpr\paperwidth-0.5cm\relax,-4.5cm){\makebox[0pt][r]{MIT-CTP/5633}}%
}}
\begin{document} 
\maketitle

\section{Introduction}
\label{s.introduction}

Jet substructure, originally developed to exploit the flow of energy within jets of particles at the Large Hadron Collider (LHC) to enhance new physics searches \cite{Butterworth:2008iy}, has since emerged as a primary tool for studying Quantum Chromodynamics (QCD). Not only has jet substructure been vastly studied in ongoing experiments at the Relativistic Heavy Ion Collider (RHIC) and the LHC~\cite{Newman:2013ada,Kogler:2018hem,Larkoski:2017jix,Connors:2017ptx,Cunqueiro:2021wls,Belmont:2023fau}, but also in future facilities such as the Electron-Ion Collider (EIC)~\cite{AbdulKhalek:2021gbh, AbdulKhalek:2022hcn}. For example, at the LHC, the jet substructure has been successfully used to tag the origin of jets in precision measurements and searches for new physics. At both the RHIC and the LHC, the jet substructure has been an important tool for studying the properties of the hot and dense quark-gluon plasma. On the other hand, at the EIC, as a good proxy of parton-level dynamics, jet allows for a clean factorization between the target and current-fragmentation regions, which is generally difficult in hadron productions~\cite{Aschenauer:2019kzf, Wang:2019bvb, Whitehill:2022mpq} in deep inelastic scatterings (DIS).

Most recently, one particular set of jet substructure observables, the distribution of hadrons inside the jet, has received a lot of attention~\cite{Boussarie:2023izj,LHCb:2022rky,ATLAS:2017pgl,ATLAS:2019dsv,ATLAS:2011myc,CMS:2014jjt,ALICE:2023jgm}. Studying hadron distribution inside the jet enables us to pinpoint fragmentation functions of various kinds and to explore the elusive hadronization process. They are complementary to the more standard processes, such as semi-inclusive hadron production in deep inelastic scattering and hadron production in $e^+e^-$ collisions, and allow us to study QCD factorization and to test the universality properties of the associated fragmentation functions in different processes~\cite{Collins:2011zzd}. For example, one can measure the distribution of hadrons inside the jet as a function of the hadron momentum fraction $z_h$, where $z_h$ is the ratio of the transverse momentum of the hadron ($p_{hT}$) to that of the jet ($p_{JT}$), both relative to the beam axis. Such a $z_h$ dependence can be used for extracting the $z_h$-dependence of the collinear fragmentation functions (FFs)~\cite{Kang:2016ehg,Anderle:2017cgl}. On the other hand, if one measures both $z_h$ and the transverse momentum $\boldsymbol{j}_{\bot}$ distribution of the hadrons with respect to the jet axis, one would be able to study the transverse momentum-dependent fragmentation functions (TMD FFs)~\cite{Kang:2017glf}.

The theoretical object which describes the momentum distribution of hadrons inside a fully reconstructed jet is called fragmenting jet functions (FJFs).
For the $z_h$-dependence while integrating over the hadron's transverse momentum with respect to the jet axis, we would have collinear FJFs, which can be matched onto the standard collinear FFs.
These collinear FJFs have been studied in the so-called semi-inclusive jet production, as well as the exclusive jet production.
For semi-inclusive jet production, one measures the signal jet while completely inclusive on other particles in the final state, see the studies in \textit{e.g.}, \cite{Arleo:2013tya, Kaufmann:2015hma, Kang:2016ehg, Dai:2016hzf, Kang:2019ahe, Zhang:2022bhq, Liu:2023fsq}. On the other hand, in the context being discussed, ``exclusive'' jet production refers to situations where a fixed number of jets are produced in the final state but one vetoes additional jets. For example, in the back-to-back electron-jet production in electron-proton collisions, our factorization would depend on the exclusive FJFs, where we have an electron and a single signal jet in the final state. See the studies along this line~\cite{Procura:2009vm, Jain:2011xz, Jain:2011iu, Chien:2015ctp, Kang:2020xyq}. On the other hand, measuring both $z_h$ and $\boldsymbol{j}_{\bot}$ distribution of the hadrons with respect to the standard jet axis would give us the transverse momentum dependent fragmenting jet functions (TMD FJFs). There are close relationships~\cite{Bain:2016rrv, Kang:2017glf, Makris:2017arq, Neill:2018wtk, Kang:2019ahe} between these TMD FJFs and the standard TMD FFs~\cite{Bacchetta:2000jk, Mulders:2000sh, Metz:2016swz}. More recently, studies have been advanced to the polarized sector, including the polarization of both the parton initiating the jet and the hadron produced within the jet~\cite{Kang:2020xyq,Kang:2021ffh,Kang:2021kpt,Zhao:2023lav}. For a recent study exploring dihadron fragmentation functions (or two hadrons) inside the jet, see~\cite{Bacchetta:2023njc,Chien:2021yol}. For the TMD study of the hadron with respect to the Winner-Take-All jet axis, see~\cite{Neill:2016vbi, Neill:2018wtk, Gutierrez-Reyes:2019vbx, Makris:2021drz}. As for the TMD study inside the groomed jet, see \cite{Makris:2017arq, Makris:2018npl, Gutierrez-Reyes:2019msa}. More recently, exploring the connections between TMD physics and the energy-energy correlators has seen a lot of progress~\cite{Gao:2019ojf,Li:2020bub,Kang:2023big}.

For the polarized collinear FJFs and TMD FJFs, we have performed studies in previous publications, mainly in~\cite{Kang:2020xyq,Kang:2021ffh} for semi-inclusive and exclusive jet productions. Even though we provided theoretical formalism and illustrated some phenomenological studies there, for the computations of FJFs, only partial results were available in those previous publications, but not the complete full results. This is the goal of our current paper. We will perform all the relevant computations and give all associated matching coefficients that allow us to connect these FJFs to the standard FFs. In doing so, we also update our formalism using the more recent approach outlined in the TMD Handbook~\cite{Boussarie:2023izj}. We provide additional phenomenological studies at the RHIC and the future EIC. 

The rest of the paper is organized as follows:
In \cref{s.collinear_FJFs}, we establish the theoretical foundation for collinear FJFs, elucidate a number of correlations between the polarization of the hadrons and fragmenting partons, and illuminate their physical significance and connections to standard fragmentation functions.
In \cref{s.TMD_FJFs}, we introduce the semi-inclusive and exclusive TMD FJFs. We discuss their factorization and evolution properties, as well as their relations to the standard TMD FFs. 
In \cref{s.phenomenology}, we employ this framework to predict the spin asymmetry $A_{TU,T}^{\cos(\phi _S - \hat{\phi}_{S_h})}$ for the semi-inclusive production of transversely polarized $\Lambda$ baryons inside jet in transversely polarized $pp$ collisions at RHIC kinematics, and $A_{TU, L}^{\cos(\phi _q - \phi _{S})}$ for the exclusive production of longitudinally polarized $\Lambda/\overline{\Lambda}$ inside jet in polarized $ep$ collisions at EIC kinematics. Finally, we provide a summary of our findings in \cref{s.conclusion}. We collect a few additional details in the Appendix. 

\section{The collinear fragmenting jet functions (FJFs)}
\label{s.collinear_FJFs}
In this section, we introduce the definition of the semi-inclusive and exclusive fragmenting jet functions (FJFs) in soft-collinear effective theory (SCET)~\cite{Bauer:2000ew, Bauer:2000yr, Bauer:2001ct, Bauer:2001yt, Bauer:2002nz} with both unpolarized and polarized fragmenting hadron.
Such FJFs are used to describe the longitudinal momentum fraction distribution of hadrons within jets.
We first provide the operator definitions of FJFs, then perform the calculation to NLO, and lastly derive and solve the RG evolution equations.

\subsection{Collinear FJFs in semi-inclusive jet productions}
\label{ss.collinear_FJF}

\begin{figure}[htb]
\centering
\includegraphics[width = 0.56 \textwidth]{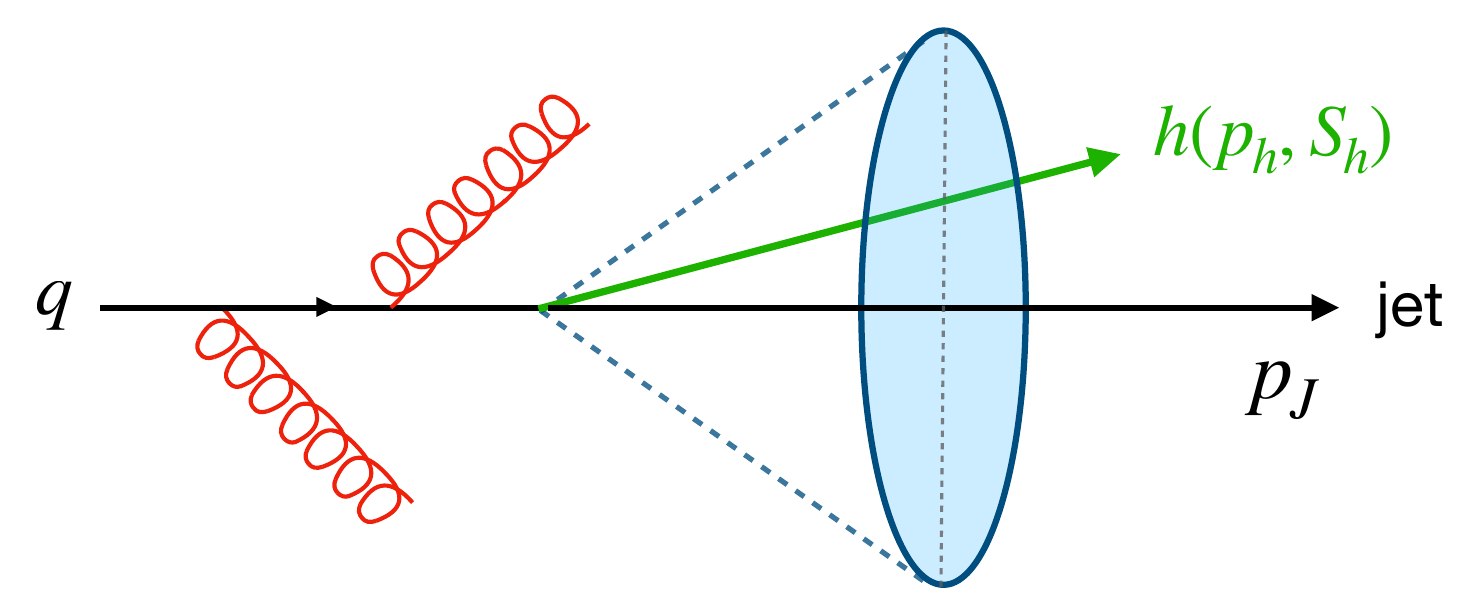}
\caption{Illustration for the distribution of hadrons $h$ inside a jet that is initiated by a parton $q$. The hadron has a momentum $p_h$ and spin $S_h$, while the jet has a momentum $p_J$.}
\label{f.jet_hadron}
\end{figure}

We can construct the semi-inclusive fragmenting quark and gluon jet functions using the corresponding gauge invariant quark and gluon fields in SCET which are given by:
\begin{equation}
\label{e.SCET_fields}
\chi _n = W_n^{\dagger} \xi _n
\, ,
\quad
\mathcal{B}_{n \perp}^{\mu} = \frac{1}{g} \bqty{W_n^{\dagger} i D_{n \perp}^{\mu} W_n}
\, ,
\end{equation}
where the subscript $n$ indicates that the field is along the $n$-collinear direction, and $n^{\mu} = \pqty{1, 0, 0, 1}$ is the light-cone vector whose spatial component is aligned with the jet axis. In \cref{e.SCET_fields}, the covariant derivative is $i D_{n \perp}^{\mu} = \mathcal{P}_{n \perp}^{\mu} + g A_{n \perp}^{\mu}$, where $\mathcal{P}^{\mu}$ is the label momentum operator.
Moreover, $W_{n}$ is the Wilson line of collinear gluons:
\begin{equation}
W_n \pqty{x}
=
\sum_{\textrm{perms}} \exp(-g \frac{1}{\overline{n} \cdot \mathcal{P}} \overline{n} \cdot A_n \pqty{x})
\, .
\end{equation}
It is also useful to define the conjugate light-cone vector $\overline{n}^{\mu} = (1,0,0,-1)$, such that $n^2 = \overline{n}^2 = 0$ and $n \cdot \overline{n} = 2$.
Thus, any four-vector $p^{\mu}$ can be decomposed as $p^{\mu} = \bqty{p^+, p^-, \boldsymbol{p}_{\perp}}$
\footnote{We use square bracket for vectors written in light cone coordinates.}
, where $p^+ = n \cdot p$ and $p^- = \overline{n} \cdot p$, namely:
\begin{equation}
\label{e.light_cone_definition}
p^{\mu} = p^- \frac{n^{\mu}}{2} + p^+ \frac{\overline{n}^{\mu}}{2} + p_{\perp}^{\mu}
\, .
\end{equation}
For a hadron in the reference frame where it has no transverse momentum and moves along the $+z$-direction, it will have a large $p_h^-$ component and a small $p_h^+$ component, \textit{i.e.}, $p_h^+ \ll p_h^-$.
Thus the momentum $p_h$ and spin $S_h$ of the hadron inside the jet can be parameterized as:
\begin{align}
p_h = \bqty{\frac{M_h^2}{p_h^-}, p_h^-, \boldsymbol{0}}
\, ,
\quad
S_h = \bqty{- \lambda_h \frac{M_h}{p_h^-}, \lambda_h \frac{p_h^-}{M_h}, \boldsymbol{S}_{h\perp}}
\, .
\end{align}
Here $\lambda_h$ is the helicity of the hadron with mass $M_h$, and $\boldsymbol{S}_{h\perp}$ is the transverse polarization of the hadron inside the jet.

Having these collinear quark and gluon fields readily available, we can formulate the correlator definitions for the quark and gluon semi-inclusive FJFs, each with different polarizations \cite{Procura:2009vm, Jain:2011iu, Kang:2016mcy, Kang:2016ehg}:
\begingroup
\allowdisplaybreaks
\begin{align}
\mathcal{G}_q^h \pqty{z, z_h, p_TR, \mu}
& =
\frac{z}{2 N_c} \delta \pqty{z_h - \frac{\omega _h}{\omega _J}}\Tr \bigg[\frac{\slashed{\overline{n}}}{2}
\mel**{0}{\delta \pqty{\omega - \overline{n} \cdot \mathcal{P}} \chi _n \pqty{0}}{\pqty{Jh} X}
\nonumber \\
& \qquad \qquad \qquad \qquad \qquad \times
\mel**{\pqty{Jh} X}{\overline{\chi}_n \pqty{0}}{0} \bigg]
\, , \label{e.collinear_FJFs_q} \\
\Delta \mathcal{G}_q^h \pqty{z, z_h, p_TR, \mu}
& =
\frac{z}{2N_c} \delta \pqty{z_h - \frac{\omega _h}{\omega _J}} \Tr \bigg[\frac{\slashed{\overline{n}}}{2} \gamma _5\mel**{0}{\delta \pqty{\omega - \overline{n} \cdot \mathcal{P}} \chi _n \pqty{0}}{\pqty{Jh} X}
\nonumber \\
& \qquad \qquad \qquad \qquad \qquad \times
\mel**{\pqty{Jh} X}{\overline{\chi}_n \pqty{0}}{0} \bigg]
\, , \label{e.collinear_Delta_FJFs_q} \\
\boldsymbol{S}_{h\perp}^i \Delta _T \mathcal{G}_q^h \pqty{z, z_h, p_TR, \mu}
& =
\frac{z}{2 N_c} \delta \pqty{z_h - \frac{\omega _h}{\omega _J}} \Tr \bigg[\frac{\slashed{\overline{n}}}{2} \gamma _{\perp}^i\gamma _5
\mel**{0}{\delta \pqty{\omega - \overline{n} \cdot \mathcal{P}} \chi _n \pqty{0}}{\pqty{Jh} X}
\nonumber \\
& \qquad \qquad \qquad \qquad \qquad \times
\mel**{\pqty{Jh} X}{\overline{\chi}_n \pqty{0}}{0}\bigg]
\, , \label{e.collinear_Delta_T_FJFs_q} \\
\mathcal{G}_g^h \pqty{z, z_h, p_TR, \mu}
& =
- \frac{z \omega}{\pqty{d - 2} \pqty{N_c^2 - 1}} \delta \pqty{z_h - \frac{\omega _h}{\omega _J}}
\nonumber \\
& \quad \times
\mel**{0}{\delta \pqty{\omega - \overline{n} \cdot \mathcal{P}} \mathcal{B}_{n \perp , \, \mu} \pqty{0}}{\pqty{Jh} X}
\mel**{\pqty{Jh} X}{\mathcal{B}_{n \perp}^{\mu} \pqty{0}}{0}
\, , \label{e.collinear_FJFs_g}\\
\Delta \mathcal{G}_g^h \pqty{z, z_h, p_TR, \mu}
& =
\frac{\epsilon^{\mu\nu\alpha\beta}\overline{n}_\alpha n_\beta}{2}\frac{z \omega}{\pqty{d - 2}\pqty{N_c^2 - 1}} \delta \pqty{z_h - \frac{\omega _h}{\omega _J}}
\nonumber \\
& \quad \times
\mel**{0}{\delta \pqty{\omega - \overline{n} \cdot \mathcal{P}} \mathcal{B}_{n \perp , \, \mu} \pqty{0}}{\pqty{Jh} X}
\mel**{\pqty{Jh} X}{\mathcal{B}_{n \perp , \, \nu}\pqty{0}}{0}
\, , \label{e.collinear_Delta_FJFs_g}
\end{align}
\endgroup
where $N_c$ is the number of colors for quarks and $(d-2)$ is the number of polarizations for gluons in $d$ dimensional space-time. Note that we only consider massless quark flavors.
For studies about heavy meson, see \textit{e.g.}, Refs. \cite{vonKuk:2023jfd, Makris:2018npl, Dai:2018ywt, Anderle:2017cgl}.
The state $\ket{\pqty{Jh} X}$ denotes the final state, encompassing both unobserved particles labeled as $X$, and the observed jet $J$ containing an identified hadron $h$, which is collectively referred to as $\pqty{Jh}$. 
Additionally, $\omega$ represents the large light-cone component for the momentum of the quark or gluon initiating the jet, whereas $\omega_J=p_J^-$ and $\omega_h=p_h^-$ correspond respectively to the large light-cone component for the momentum of the jet itself and the identified hadron within the jet, as shown in \cref{f.jet_hadron}. In our convention and the perturbative calculations performed below, we choose a frame where the jet has no transverse momentum and $\omega_J  \simeq 2E_J$ with $E_J$ the jet energy. However, in the actual experimental measurements at the collider experiments such as proton-proton collisions at the LHC, the jet has a transverse momentum $p_T$ with a jet radius $R$~\cite{Cacciari:2008gp} in the lab frame. Since we will perform phenomenological studies for the relevant collider experiments, we include $p_TR$ as an argument in our collinear FJFs in \cref{e.collinear_FJFs_q,e.collinear_Delta_FJFs_q,e.collinear_Delta_T_FJFs_q,e.collinear_FJFs_g,e.collinear_Delta_FJFs_g}.

At the same time, the energy fractions $z$ and $z_h$ are defined as:
\begin{equation}
\label{e.z_zh}
z = \frac{\omega _J}{\omega}\,,
\quad
z_h = \frac{\omega _h}{\omega _J}\, .
\end{equation}
As given in \cref{e.collinear_FJFs_q,e.collinear_Delta_FJFs_q,e.collinear_Delta_T_FJFs_q}, to obtain the helicity and transversity distributions of the hadron in quark FJFs, namely $\Delta \mathcal{G}_q^h$ and $\Delta _T \mathcal{G}_q^h$, we replace the $\slashed{\overline{n}}$ in \cref{e.collinear_FJFs_q} by $\slashed{\overline{n}} \gamma _5$ and $\slashed{\overline{n}} \gamma _{\perp}^i \gamma _5$. The unpolarized collinear FJF $\mathcal{G}_{q,g}^h$ represents the situation where an unpolarized quark or gluon initiates a jet that carries a momentum fraction $z$ of the parent parton, and we further observe an unpolarized hadron $h$ inside the jet which carries the momentum fraction $z_h$ of the jet. Similarly, $\Delta \mathcal{G}_{q,g}^h$ stands for the case where a longitudinally polarized parton initiates a jet in which a longitudinally polarized hadron $h$ is observed. Finally, $\Delta _T \mathcal{G}_q^h$ is the case for a transversely polarized quark going into a jet in which a transversely polarized hadron is observed. Note that we do not have the corresponding $\Delta _T \mathcal{G}_g^h$ for gluons when hadron $h$ is a spin-1/2 particle, due to the helicity-conservation constraint. This is the same reason why the gluon transversity does not exist for the spin-1/2 nucleon~\cite{Barone:2003fy}.
\begin{figure}[htb]
\centering
\includegraphics[width = 0.73 \textwidth]{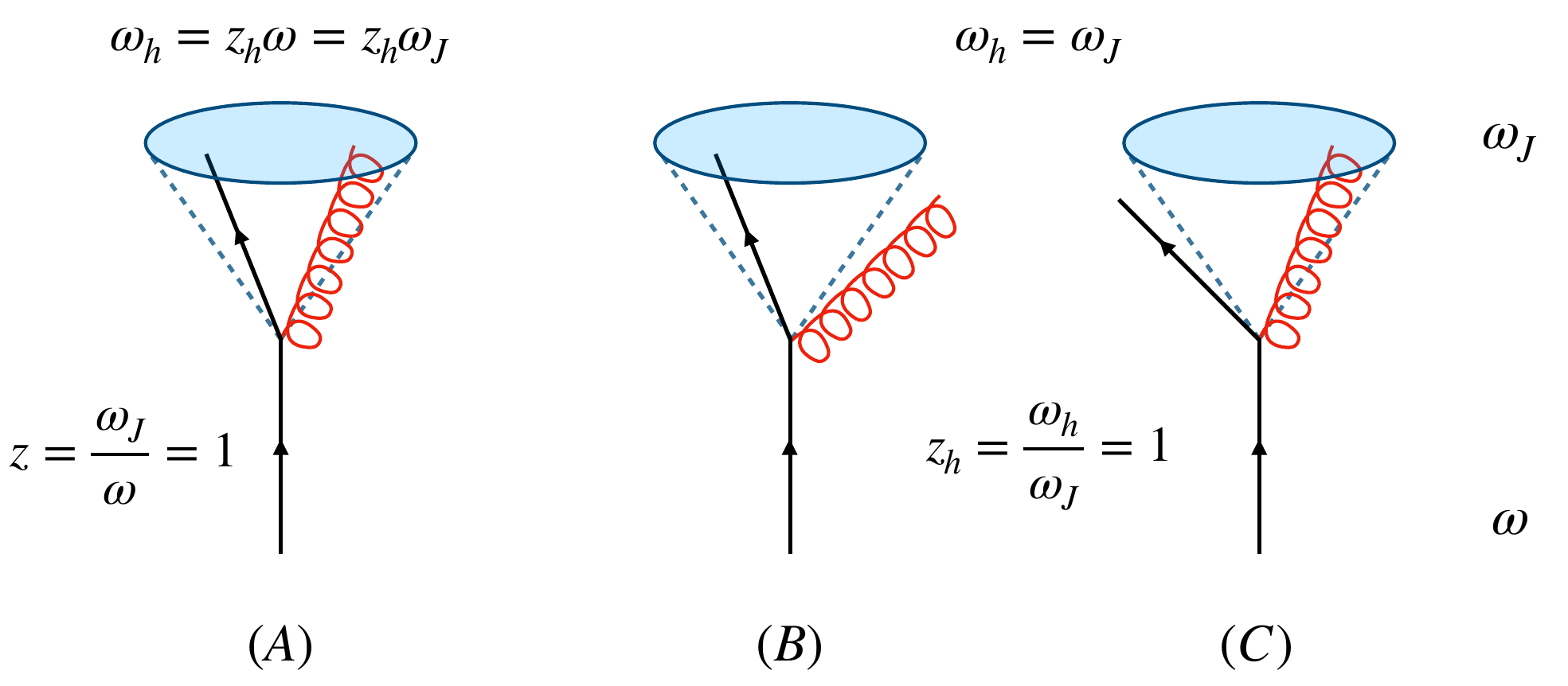}
\caption{
Contributions required for studying semi-inclusive FJFs.
In (A), both the quark and the gluon are inside the jet, while in (B) and (C) only the quark or gluon is inside the jet.
}
\label{f.jet_NLO_Feynman}
\end{figure}

\subsubsection{NLO calculation}
\label{sss.collinear_siFJFs_calculation}
Although the semi-inclusive FJFs $\mathcal{G}_i^h \pqty{z, z_h, p_TR, \mu}$ are not perturbatively calculable, they can be matched onto the standard collinear fragmentation functions as long as $p_TR\gg \Lambda_{\mathrm{QCD}}$~\cite{Kang:2016ehg}. The matching coefficients can be computed by replacing the hadron $h$ by a parton $j$ and then computing $\mathcal{G}_i^j \pqty{z, z_h, p_TR, \mu}$ perturbatively. Prior to describing the process of computing the semi-inclusive FJFs, it is important to emphasize that the outcomes are influenced by the choice of jet algorithms.
To streamline our explanation, however, we solely concentrate on the anti-$k_T$ algorithm~\cite{Cacciari:2008gp} in this study.

We will now compute the bare partonic semi-inclusive FJFs $\mathcal{G}_i^j \pqty{z, z_h, p_TR}$ in order to determine the UV counter terms. At the leading order (LO), one has a single parton. It is this parton $i$ that forms the jet as well as presenting the parton $j$ inside the jet. Thus we have $z=1$ and $z_h=1$ at LO and bare polarized semi-inclusive FJFs are given by:
\begin{equation}
\Delta _{(T)} \mathcal{G} _i^{j, (0)} (z, z_h, p_TR) = \delta_{ij}\delta (1 - z) \delta (1 - z_h)\, .
\end{equation}
Here and throughout the rest of the paper, we use the symbol $\Delta _{(T)}$ to collectively represent both the longitudinally ($\Delta$) and transversely ($T$) polarized cases.

At the next-to-leading order (NLO), we have to consider the $1\to 2$ splitting process in the real emission diagrams as shown in~\cref{f.jet_NLO_Feynman}, as well as virtual diagrams. In dimensional regularization, since virtual diagrams become scaleless integrals and vanish, we thus only consider the real diagrams. Following the discussion in~\cite{Kang:2016ehg}, we consider the cases of both partons in the jet and only one parton in the jet as shown in~\cref{f.jet_NLO_Feynman}.
They are given by the same integrals over the transverse momentum $q_{\bot}$ of the parton $j$, but are constrained by different Heaviside functions.
Working in pure dimensional regularization with $d = 4 - 2 \epsilon$ dimensions, we have:
\begin{align}
\label{e.bare_collinear_FJF_1}
\Delta _{(T)} \mathcal{G}_i^{j, (1)} (z, z_h, \omega_J)
& =
\frac{\alpha _s}{2 \pi} \frac{(e^{\gamma _E} \mu^2)^{\epsilon}}{\Gamma (1-\epsilon)}
\bigg(
\delta (1-z) \Delta _{(T)} \hat{P}_{ji} (z_h, \epsilon)
\int\frac{\dd{q_{\perp}^2}}{(q_{\perp}^2)^{1 + \epsilon}} \Theta ^{\text{anti-}k_T}_{\mathrm{both}}
\nonumber \\
& \qquad \qquad \qquad \quad +
\delta (1-z_h) \Delta _{(T)} \hat{P}_{ji}(z, \epsilon)
\int \frac{\dd{q_{\perp}^2}}{(q_{\perp}^2)^{1 + \epsilon}} \Theta ^{\text{anti-}k_T}_{j}
\bigg)
\, ,
\end{align}
where the $\Theta ^{\text{anti-}k_T}_{\mathrm{both}}$ and $\Theta ^{\text{anti-}k_T}_{j}$ are the jet algorithm constraints with both partons in jet (\cref{f.jet_NLO_Feynman}A) and only one parton in jet (\cref{f.jet_NLO_Feynman}B, \labelcref{f.jet_NLO_Feynman}C), respectively, and the superscript ``anti-$k_T$'' indicates that anti-$k_T$ algorithm is used.
The form of $\Theta ^{\text{anti-}k_T}_{\mathrm{both}}$ and $\Theta ^{\text{anti-}k_T}_{j}$ are given by the Heaviside functions \cite{Ellis:2010rwa,Kang:2016ehg,Kang:2017mda}:
\begin{align}
\Theta ^{\text{anti-}k_T}_{\textrm{both}}
& =
\theta \pqty{z_h (1 - z_h) \omega _J \tan{\frac{\mathcal{R}}{2}} - q_{\perp}}
\, , \\
\Theta ^{\text{anti-}k_T}_{j}
& =
\theta \pqty{q_{\perp} - (1 - z) \omega _J \tan\frac{\mathcal{R}}{2}}
\, , 
\end{align}
where $\mathcal{R}$ is the angular separation between three vector momenta inside the jet~\cite{Ellis:2010rwa}. This is to be compared with the jet radius $R$ in the hadron collider that specifies the separation in pseudo-rapidity and azimuthal angle as in $\sqrt{(\Delta\eta)^2+(\Delta \phi)^2}$. For jets with a small jet radius $R\ll 1$, we have
\begin{align}
\mathcal{R} = \frac{R}{\cosh(\eta)}\,,
\end{align}
where $\eta$ is the jet pseudo-rapidity. Note that
\begin{align}
    \omega_J \tan\frac{\mathcal{R}}{2} \approx  2 E_J \,\frac{\mathcal{R}}{2}
    = p_T\cosh(\eta) \frac{R}{\cosh(\eta)} = p_T R\,,
\end{align}
and we will thus include the argument $p_TR$ in the FJFs.

The longitudinally polarized functions $\Delta  \hat{P}_{ji} (z , \epsilon)$ in \cref{e.bare_collinear_FJF_1} are given in~\cite{Vogelsang:1996im}:
\begin{align}
\Delta \hat{P}_{qq} (z, \epsilon)
& =
C_F \bqty{\frac{1 + z^2}{1 - z} - \epsilon (1 - z)}
\, , \label{e.Delta_P_qq} \\
\Delta \hat{P}_{gq} (z, \epsilon)
& =
C_F \bqty{2 - z + 2 \epsilon (1 - z)}
\, , \label{e.Delta_P_gq} \\
\Delta \hat{P}_{qg} (z, \epsilon)
& =
T_F \bqty{2 z - 1 - 2 \epsilon (1 - z)}
\, , \label{e.Delta_P_qg} \\
\Delta \hat{P}_{gg} (z, \epsilon)
& =
2 C_A \bqty{\frac{1}{1 - z} - 2 z + 1 + 2 \epsilon (1 - z)}
\, . \label{e.Delta_P_gg}
\end{align}
The transversely polarized splitting functions $\Delta _{T} \hat{P}_{ji} (z , \epsilon)$ only exist for $\Delta _{T} \hat{P}_{qq}$ because there is no gluon transversity fragmentation function for spin-1/2 hadron as we mentioned above. We have the following expression for $\Delta _{T} \hat{P}_{qq}(z , \epsilon)$~\cite{Vogelsang:1997ak}
\begin{equation}
\label{e.Delta_T_P_qq}
\Delta _T \hat{P}_{qq} (z, \epsilon)
=
C_F \left[\frac{2 z}{1 - z}\right]
\, .
\end{equation}
By inserting the $\Theta$ functions for anti-$k_T$ algorithm and carrying out the integration in \cref{e.bare_collinear_FJF_1}, we obtain the bare results
\footnote{More details on the $q_{\perp}$ integral and relevant expansions as $\epsilon \to 0$ can be found in \cref{appendix.siFJFs}.}
for longitudinally polarized semi-inclusive FJFs $\Delta \mathcal{G}^{j}_{i, \mathrm{bare}} (z, z_h, p_TR)$ with $i, j \in \Bqty{q, g}$:
\begin{align}
\Delta \mathcal{G}^q_{q, \mathrm{bare}} & (z, z_h, p_TR)
=
\delta (1-z) \delta (1-z_h)
\nonumber \\
& +
\frac{\alpha _s}{2 \pi} \bqty{\pqty{\frac{1}{\epsilon} + L} \Delta P_{qq} (z) \delta (1-z_h) - \pqty{\frac{1}{\epsilon} + L} \Delta P_{qq} (z_h) \delta (1-z)}
\nonumber \\
& +
\delta (1-z) \frac{\alpha _s}{2 \pi} \bqty{2 C_F (1 + z_h^2) \pqty{\frac{\ln(1-z_h)}{1-z_h}}_+ + C_F (1-z_h) + 2 \Delta P_{qq} (z_h) \ln(z_h)}
\nonumber \\
& - \delta (1-z_h) \frac{\alpha _s}{2 \pi} \bqty{2 C_F (1+z^2) \pqty{\frac{\ln(1-z)}{1-z}}_+ + C_F (1-z)}
\, , \label{e.bare_collinear_G_qq} \\
\Delta \mathcal{G}^g_{q, \mathrm{bare}} & (z, z_h, p_TR) = \frac{\alpha _s}{2 \pi} \bqty{\pqty{\frac{1}{\epsilon} + L} \Delta P_{gq} (z) \delta (1-z_h) - \pqty{\frac{1}{\epsilon} + L} \Delta P_{gq} (z_h) \delta(1-z)}
\nonumber \\
& +
\delta (1-z) \frac{\alpha _s}{2 \pi} \bigg[2 \Delta P_{gq}(z_h) \ln(z_h (1-z_h)) - 2 C_F (1-z_h) \bigg]
\nonumber \\
& -
\delta (1-z_h) \frac{\alpha _s}{2 \pi} \bigg[2 \Delta P_{gq} (z) \ln(1-z) - 2 C_F (1-z) \bigg]
\, , \label{e.bare_collinear_G_qg} \\
\Delta \mathcal{G}^q_{g, \mathrm{bare}} & (z, z_h, p_TR)
=
\frac{\alpha _s}{2 \pi} \bqty{\pqty{\frac{1}{\epsilon} + L} \Delta P_{qg} (z) \delta (1-z_h) - \pqty{\frac{1}{\epsilon} + L} \Delta P_{qg} (z_h) \delta (1-z)}
\nonumber \\
& +
\delta (1-z) \frac{\alpha _s}{2 \pi} \bigg[2 \Delta P_{qg}(z_h) \ln(z_h (1-z_h)) + 2 T_F (1 - z_h) \bigg]
\nonumber \\
& -
\delta (1-z_h) \frac{\alpha _s}{2 \pi} \bigg[2 \Delta P_{qg} (z) \ln(1-z) + 2 T_F (1-z) \bigg]
\, , \label{e.bare_collinear_G_gq} \\
\Delta \mathcal{G}^g_{g, \mathrm{bare}} & (z, z_h, p_TR)
=
\delta (1-z) \delta (1-z_h)
\nonumber \\
& +
\frac{\alpha _s}{2 \pi} \bqty{\pqty{\frac{1}{\epsilon} + L} \Delta P_{gg} (z) \delta (1-z_h) - \pqty{\frac{1}{\epsilon} + L} \Delta P_{gg} (z_h) \delta (1-z)}
\nonumber \\
& +
\delta (1-z) \frac{\alpha _s}{2 \pi} \bqty{4 C_A \pqty{2 (1-z_h)^2 + z_h} \pqty{\frac{\ln (1-z_h)}{1-z_h}}_+}
\nonumber \\
& +
\delta (1-z) \frac{\alpha _s}{2 \pi} \bigg[2 \Delta P_{gg}(z_h) \ln(z_h) - 4 C_A (1-z_h) \bigg]
\nonumber \\
& -
\delta (1-z_h) \frac{\alpha _s}{2 \pi} \bqty{4 C_A \pqty{2 (1-z)^2 + z} \pqty{\frac{\ln(1-z)}{1-z}}_+ - 4 C_A (1-z)}
\, , \label{e.bare_collinear_G_gg}
\end{align}
where the logarithm 
\begin{align}
    L \equiv \ln(\frac{\mu ^2}{\left(p_TR\right)^2})\,.
\end{align}
This indicates that the natural scale for the collinear FJFs is given by $\mu_{\mathscr{G}}\sim p_TR$. As for the transversely polarized semi-inclusive FJFs, we obtain:
\begin{align}
\Delta _T \mathcal{G}^q_{q, \mathrm{bare}} & (z, z_h, p_TR)
= 
\delta (1-z) \delta(1-z_h)
\nonumber \\
& +
\frac{\alpha _s}{2 \pi} \bqty{\pqty{\frac{1}{\epsilon} + L} \Delta _T P_{qq} (z) \delta (1-z_h) - \pqty{\frac{1}{\epsilon} + L} \Delta _T P_{qq} (z_h) \delta (1-z)}
\nonumber \\
& +
\delta (1-z) \frac{\alpha _s}{2 \pi} \bqty{4 C_F z_h \pqty{\frac{\ln(1-z_h)}{1-z_h}}_+ + 2 \Delta _T P_{qq} (z_h) \ln(z_h)}
\nonumber \\
& -
\delta (1-z_h) \frac{\alpha _s}{2 \pi} \bqty{4 C_F z \pqty{\frac{\ln(1-z)}{1-z}}_+}
\, . \label{e.bare_collinear_G_qq_T}
\end{align}
Here the functions $\Delta_{(T)} P_{ji}(z)$ are the longitudinally (transversely) polarized Altarelli-Parisi splitting kernels:
\begin{align}
\Delta P_{qq}(z)
& =
C_F \bqty{\frac{1+z^2}{(1-z)_+} + \frac{3}{2} \delta (1-z)}
\, , \label{e.AP_longitudinal_qq} \\
\Delta P_{gq}(z)
& =
C_F \bqty{2 - z}
\, , \label{e.AP_longitudinal_gq} \\
\Delta P_{qg}(z)
& =
T_F \bqty{2z-1}
\, , \label{e.AP_longitudinal_qg} \\
\Delta P_{gg}(z)
& =
2 C_A \bqty{\frac{1}{(1-z)_+} - 2z + 1}
+
\frac{\beta _0}{2} \delta (1-z)
\, , \label{e.AP_longitudinal_gg} \\
\Delta _T P_{qq}(z)
& =
C_F \bqty{\frac{2z}{(1-z)_+} + \frac{3}{2} \delta (1-z)}
\, , \label{e.AP_transverse_qq}
\end{align}
where $\beta _0 \equiv \frac{11}{3} C_A - \frac{4}{3} T_F n_f$ and $n_f$ is number of flavors.
The ``plus'' distributions are defined as usual by:
\begin{equation}
\int _0^1 \dd{z} f \pqty{z} \bqty{g \pqty{z}}_+
=
\int _0^1 \dd{z} \Big(f \pqty{z} - f \pqty{1} \Big) g \pqty{z}
\, .
\end{equation}

\subsubsection{Renormalization and matching onto collinear FFs}
\label{sss.collinear_siFJFs_renormalization}
In this section, we will renormalize the bare semi-inclusive FJFs obtained in the previous section, and then match the results to the standard collinear FFs. In doing so, it is important to point out that the $1 / \epsilon$ poles with a factor of $\Delta _{\pqty{T}} P_{ij} \pqty{z_h} \delta \pqty{1 - z}$ in~\cref{e.bare_collinear_G_qq,e.bare_collinear_G_qg,e.bare_collinear_G_gq,e.bare_collinear_G_gg,e.bare_collinear_G_qq_T} are the infrared (IR) poles, that would be matched onto the standard longitudinally (transversely) polarized collinear FFs. On the other hand, the poles with a factor of $\Delta _{\pqty{T}} P_{ij} \pqty{z} \delta \pqty{1 - z_h}$ are the ultraviolet (UV) poles which will be taken care of by renormalization. Since the UV poles do not involve the variable $z_h$, we should expect that $z_h$ is merely a parameter in renormalization. This is exactly the same as the unpolarized situation studied in~\cite{Kang:2016ehg}. With this in mind, the relationship between the bare and renormalized semi-inclusive FJFs is given by
\begin{equation}
\Delta _{\pqty{T}} \mathcal{G}_{i, \mathrm{bare}}^j \pqty{z, z_h, p_T R}
=
\sum _k \int _z^1 \frac{\dd{z'}}{z'}
\Delta _{\pqty{T}} Z_{ik} \pqty{\frac{z}{z'}, \mu}
\Delta _{\pqty{T}} \mathcal{G} _k^j \pqty{z', z_h, p_T R, \mu}
\, ,
\end{equation}
where $\Delta _{\pqty{T}} Z_{ik} \pqty{\frac{z}{z'}, \mu}$ is the renormalization matrix and $\Delta _{\pqty{T}} \mathcal{G} _k^j \pqty{z', z_h, p_T R, \mu}$ are the renormalized semi-inclusive FJFs. For the aforementioned reason, the convolution only involves the variable $z$.
The renormalized semi-inclusive FJFs satisfy the following RG evolution equations:
\begin{equation}
\mu \frac{\dd{}}{\dd{\mu}}
\Delta _{\pqty{T}} \mathcal{G} _i^j \pqty{z, z_h, p_TR, \mu}
=
\sum _k \int _z^1 \frac{\dd{z'}}{z'}
\Delta _{\pqty{T}}\gamma _{ik}^{ \mathcal{G}} \pqty{\frac{z}{z'}, \mu}
\Delta _{\pqty{T}} \mathcal{G} _k^j \pqty{z', z_h, p_T R, \mu}
\, ,
\end{equation}
with the anomalous dimension matrix given by:
\begin{equation}
\Delta _{\pqty{T}}\gamma _{ik}^{ \mathcal{G}} \pqty{\frac{z}{z'}, \mu}
=
- \sum _k \int _z^1 \frac{\dd{z'}}{z'}
\pqty{\Delta _{\pqty{T}} Z}_{il}^{-1} \pqty{\frac{z}{z'}, \mu}
\mu \frac{\dd{}}{\dd{\mu}} \Delta _{\pqty{T}} Z_{lk} \pqty{z', \mu}
\, ,
\end{equation}
and $\pqty{\Delta _{\pqty{T}} Z}_{il}^{-1}$ is the inverse of the renormalization matrix that is defined such that:
\begin{equation}
\sum _k \int _z^1 \frac{\dd{z'}}{z'}
\pqty{\Delta _{\pqty{T}} Z}_{il}^{-1} \pqty{\frac{z}{z'}, \mu}
\Delta _{\pqty{T}} Z_{lj} \pqty{z', \mu}
=
\delta _{ij} \delta \pqty{1 - z}
\, .
\end{equation}
Up to $\order{\alpha _s}$, the renormalization matrix is
\begin{equation}
\Delta _{\pqty{T}} Z_{ij} \pqty{z, \mu}
=
\delta _{ij} \delta \pqty{1 - z}
+
\frac{\alpha _s \pqty{\mu}}{2 \pi} \frac{1}{\epsilon}
\Delta _{\pqty{T}} P_{ji} \pqty{z}
\, ,
\end{equation}
and therefore the anomalous dimension matrix is given by:
\begin{equation}
\Delta _{\pqty{T}}\gamma _{ij}^{ \mathcal{G}} \pqty{z, \mu}
=
\frac{\alpha _s \pqty{\mu}}{\pi}
\Delta _{\pqty{T}} P_{ji} \pqty{z}
\, .
\end{equation}
We therefore observe that the evolution of the renormalized polarized semi-inclusive FJFs conforms to the time-like DGLAP equation for collinear polarized FFs \cite{Altarelli:1977zs}:
\begin{equation}
\mu \frac{\dd{}}{\dd{\mu}}
\Delta _{\pqty{T}} \mathcal{G} _i^h \pqty{z, z_h, p_TR,  \mu}
=
\frac{\alpha _s \pqty{\mu}}{\pi}
\sum _k \int _z^1 \frac{\dd{z'}}{z'}
\Delta _{\pqty{T}} P_{ji} \pqty{\frac{z}{z'}}
\Delta _{\pqty{T}} \mathcal{G} _j^h \pqty{z', z_h, p_TR,  \mu}
\, ,
\end{equation}
where the leading order splitting kernels are given in \cref{e.AP_longitudinal_qq,e.AP_longitudinal_gq,e.AP_longitudinal_qg,e.AP_longitudinal_gg,e.AP_transverse_qq}.
Notice that the hadronic FJFs have been reinstated since the hadronic and partonic FJFs would contain the same UV counter terms and thus follow the same RG evolution equations.
As we mentioned in the previous section that the natural scale for the collinear FJFs is given by $\mu_{\mathscr{G}} = p_T R$, thus one would use the above renormalization group equations to evolve these FJFs from their natural scale $p_TR$ to the hard scale $p_T$ for the jet production.
The effect of this would be to resum the logarithms of the jet radius $R$ for narrow jets with $R \ll 1$.
Analogous behaviors were observed in the context of the semi-inclusive unpolarized FJFs in \cite{Kang:2016ehg}.

With the UV poles eliminated by renormalization, we can proceed to dealing with the IR poles, which can be addressed by matching onto the collinear polarized FFs.
Such matching can be done at a scale $\mu \gg \Lambda _{\mathrm{QCD}}$ as follows:
\begin{equation}
\Delta _{\pqty{T}} \mathcal{G}_i^h \pqty{z, z_h, p_TR,  \mu}
=
\sum _j \int _{z_h}^1 \frac{\dd{z_h'}}{z_h'}
\Delta_{\pqty{T}} \mathcal{J}_{ij} \pqty{z, z_h', p_TR,  \mu}
\Delta _{\pqty{T}} D_{h/j} \pqty{\frac{z_h}{z_h'}, \mu}
\, ,
\end{equation}
where again, the hadronic semi-inclusive FJFs and collinear FFs are reinstated.
Here, $\Delta_{(T)} D_{h/j} (z_h, \mu)$ is the collinear fragmentation function characterizing a parton $j$ fragmenting into a hadron $h$.
Specifically, $\Delta D_{h/j}$ ($\Delta_{T} D_{h/j}$) is the helicity (transversity) fragmentation function describing a longitudinally (transversely) polarized parton fragmenting into a longitudinally (transversely) polarized hadron.
And the matching relation above holds, with a power correction of order $\order{\Lambda _{\mathrm{QCD}}^2 / \pqty{p_TR)^2}}$~\cite{Jain:2011xz, Chien:2015ctp, Kang:2016ehg}.
In this case, however, different from the renormalization procedure, $z_h$ is being convoluted and $z$ is merely a parameter.
This process is similar to the approach used for unpolarized FJFs described in \cite{Kang:2016ehg}, except that instead of the unpolarized FFs, collinear polarized FFs are used in this case.
The perturbative results at the parton level for the polarized collinear FFs are given by:
\begin{equation}
\Delta _{\pqty{T}} D_{j/i} \pqty{z_h, \mu}
=
\delta _{ij} \delta \pqty{1-z_h}
+
\frac{\alpha _s}{2 \pi}
\Delta _{\pqty{T}} P_{ji} \pqty{z_h} 
\pqty{-\frac{1}{\epsilon}}
\, .
\end{equation}
Finally, the matching coefficients $\Delta _{\pqty{T}} \mathcal{J}_{ij}$ for anti-$k_T$ jet algorithm are as follows
\begin{align}
\Delta \mathcal{J}_{qq} (z, z_h, p_TR,  \mu)
& =
\delta (1-z) \delta (1-z_h) + \frac{\alpha _s}{2 \pi} \bigg\{L \bigg[\Delta P_{qq} (z) \delta (1-z_h) - \Delta P_{qq} (z_h) \delta (1-z) \bigg]
\nonumber \\
& \quad +
\delta (1-z) \bqty{2 C_F (1+z_h^2) \pqty{\frac{\ln(1-z_h)}{1-z_h}}_+ + C_F (1-z_h) + \Delta \mathcal{I}_{qq}^{\text{anti-}k_T}(z_h)} \nonumber \\
& \quad -
\delta (1-z_h) \bqty{2 C_F (1+z^2) \pqty{\frac{\ln(1-z)}{1-z}}_+ + C_F (1-z)} \bigg\}
\, , \\
\Delta \mathcal{J}_{qg} (z, z_h, p_TR,  \mu)
& =
\frac{\alpha _s}{2 \pi} \bigg\{L \bigg[\Delta P_{gq}(z) \delta (1-z_h) - \Delta P_{gq}(z_h) \delta (1-z) \bigg]
\nonumber \\
& \quad +
\delta (1-z) \bigg[2 \Delta P_{gq}(z_h) \ln(1-z_h) - 2 C_F (1-z_h) + \Delta \mathcal{I}_{gq}^{\text{anti-}k_T}(z_h) \bigg]
\nonumber \\
& \quad
- \delta (1-z_h) \bigg[2 \Delta P_{gq}(z) \ln(1-z) - 2 C_F (1-z) \bigg] \bigg\}
\, , \\
\Delta \mathcal{J}_{gq} (z, z_h, p_TR,  \mu)
& =
\frac{\alpha _s}{2 \pi} \bigg\{L \bigg[\Delta P_{qg}(z) \delta (1-z_h) - \Delta P_{qg}(z_h) \delta (1-z) \bigg]
\nonumber \\
& \quad
+ \delta (1-z) \bigg[2 \Delta P_{qg}(z_h) \ln(1-z_h) + 2 T_F (1-z_h) + \Delta \mathcal{I}_{qg}^{\text{anti-}k_T}(z_h) \bigg]
\nonumber \\
& \quad
- \delta (1-z_h) \bigg[2 \Delta P_{qg}(z) \ln(1-z) + 2 T_F (1-z) \bigg] \bigg\}
\, , \\
\Delta \mathcal{J}_{gg} (z, z_h, p_TR,  \mu)
& =
\delta (1-z) \delta (1-z_h) + \frac{\alpha _s}{2 \pi} \bigg\{L \bigg[\Delta P_{gg}(z) \delta (1-z_h) - \Delta P_{gg}(z_h) \delta (1-z) \bigg]
\nonumber \\
& \quad
+ \delta (1-z) \bigg[4 C_A \pqty{2 (1-z_h)^2 + z_h} \pqty{\frac{\ln(1-z_h)}{1-z_h}}_+
\nonumber \\
& \qquad \qquad \qquad -
4 C_A (1-z_h) + \Delta \mathcal{I}_{gg}^{\text{anti-}k_T}(z_h) \bigg]
\nonumber \\
& \quad -
\delta (1-z_h) \bqty{4 C_A \pqty{2(1-z)^2 + z} \pqty{\frac{\ln(1-z)}{1-z}}_+ - 4 C_A (1-z)} \bigg\}
\, , \\
\Delta _T \mathcal{J}_{qq} (z, z_h, p_TR,  \mu)
& =
\delta (1-z) \delta (1-z_h) + \frac{\alpha _s}{2 \pi} \bigg\{L \bigg[\Delta _T P_{qq}(z) \delta (1-z_h) - \Delta _T P_{qq}(z_h) \delta (1-z) \bigg]
\nonumber \\
& \quad +
\delta (1-z) \bqty{4 C_F z_h \pqty{\frac{\ln(1-z_h)}{1-z_h}}_+ + \Delta _T \mathcal{I}_{qq}^{\text{anti-}k_T}(z_h)}
\nonumber \\
& \quad -
\delta (1-z_h) \bqty{4 C_F z \pqty{\frac{\ln(1-z)}{1-z}}_+} \bigg\}
\, ,
\end{align}
where we have defined $\Delta_{(T)} \mathcal{I}_{ij}^{\text{anti-}k_T}$ as:
\begin{equation}
\begin{split}
\Delta \mathcal{I}_{ij}^{\text{anti-}k_T} \pqty{z_h}
& =
2 \Delta P_{ji} \pqty{z_h} \ln(z_h)
\, , \\
\Delta _T \mathcal{I}_{ij}^{\text{anti-}k_T} \pqty{z_h}
& =
2 \Delta _T P_{ji} \pqty{z_h} \ln(z_h)
\, . \\
\end{split}
\end{equation}
The matching coefficient functions $\Delta\mathcal{J}_{ij}$ for the longitudinally polarized case have been given in the previous publication~\cite{Kang:2020xyq} with a different notation $\mathcal{J}^L_{ij}$. Here, we provide the detailed steps of how such matching coefficients are derived. At the same time, the matching coefficient for the transversely polarized case $\Delta_T\mathcal{J}_{qq}$ is presented here for the first time.

\subsection{Collinear FJFs in exclusive jet productions}
\label{ss.collinear_exclusive_FJFs}

Exclusive jet production, such as back-to-back dijet production in proton-proton collisions~\cite{Kang:2020xez,Gao:2023ulg} or back-to-back electron-jet production in electron-proton collisions~\cite{Liu:2018trl}, can provide valuable insight into the understanding of the fundamental dynamics of hadron structure and interactions. However, in such situations, typically the radiation outside the jet is restricted. For example, in back-to-back dijet production, one usually restricts the transverse momentum imbalance $q_T$ to be much smaller than the average transverse momentum of the jets. In this case, the hard/collinear gluon emission outside the jet would move the imbalance $q_T$ away from the back-to-back region. Because of that, the collinear radiation can only happen inside the jet. Consequently, for jet function calculations in exclusive jet production, we only have to consider~\cref{f.jet_NLO_Feynman}A while B and C would not contribute. We refer to the corresponding FJFs as exclusive FJFs below, which are key ingredients in the factorization formalism for these processes. They describe the probability distribution for a parton in the jet to fragment into a hadron with a given collinear momentum fraction in exclusive jet production. 

The perturbative computations for exclusive FJFs $\Delta_{(T)}\mathscr{G}_i^j(z_h, p_TR, \mu)$ are identical to those for the semi-inclusive FJFs, except that we keep only the contributions from~\cref{f.jet_NLO_Feynman}A. Since both partons are always inside the jet, we always have $z=1$, \textit{i.e.}, the jet carries all the energy of the parton that initiates the jet in this case. Because of this, we no longer keep the variable $z$ in the FJFs and thus the exclusive FJFs only depend on $z_h$. The RG equations for exclusive FJFs can also be derived in the same manner as above, and we find that they satisfy the following equation 
\begin{equation}
\mu \frac{\dd{}}{\dd{\mu}} \Delta _{(T)} \mathscr{G}_i^h(z_h, p_TR, \mu)
=
\gamma _{\mathscr{G}}^i(\mu) \Delta _{(T)} \mathscr{G}_i^h (z_h, p_TR, \mu)
\, ,
\end{equation}
where the index $i$ is not summed over, and the anomalous dimensions $\gamma _{\mathscr{G}}^i(\mu)$ with $i=q$, $g$ at this order are given by:
\begin{align}
\gamma_{\mathscr{G}}^q (\mu)
& =
\frac{\alpha_s (\mu)}{\pi}
\pqty{C_F L + \frac{3}{2}C_F}
\, , \\
\gamma_{\mathscr{G}}^g (\mu)
& =
\frac{\alpha_s (\mu)}{\pi}
\pqty{C_A L + \frac{\beta_0}{2}}
\, .
\end{align}
A few comments are in order.
First of all, it is important to emphasize that the RG equation for exclusive FJFs is a multiplicative renormalization, instead of a convolution over $z$ as in the semi-inclusive FJFs given in the previous section.
Secondly, the anomalous dimension $\gamma _{\mathscr{G}}^i(\mu)$ is the same as that of the exclusive jet functions in \cite{Ellis:2010rwa} and exclusive unpolarized FJFs~\cite{Kang:2019ahe,Chien:2015ctp,Waalewijn:2012sv}.
Since in the factorization formalism, one simply replaces the exclusive jet function with the exclusive FJFs, thus they should have the same anomalous dimensions simply because of the RG consistency.
At the same time, we find that the UV divergent terms are all proportional to $\delta(1-z_h)$ where the radiated gluon becomes soft, and thus the result should be independent of the polarization of the partons.
This explains the anomalous dimensions for both longitudinally and transversely polarized FJFs are the same as the unpolarized FJFs. 
The solution to the RG equation for the exclusive FJFs is then:
\begin{equation}
\Delta_{(T)}\mathscr{G}_i^h (z_h, p_TR, \mu)
=
\Delta_{(T)}\mathscr{G}_i^h \pqty{z_h, p_TR, \mu_{\mathscr{G}}} \exp \bqty{\int_{\mu_{\mathscr{G}}}^\mu \frac{\dd{\mu^{\prime}}}{\mu^{\prime}} \gamma_{\mathscr{G}}^i\pqty{\mu^{\prime}}}
\, , \label{eq:exclu-rg}
\end{equation}
where the scale $\mu_{\mathscr{G}}$ is the characteristic scale that eliminates the large logarithms in the fixed-order perturbative calculations. The logarithm $L$ indicates that the natural scale for the exclusive FJFs is also given by $\mu_{\mathscr{G}} = p_T R$, the same as for the semi-inclusive FJFs. One would thus evolve the FJFs from the natural scale $\mu_{\mathscr{G}}\sim p_T R$ to the hard scale of the process $\mu\sim p_T$ and effectively resum the logarithm of jet radius, $\ln R$. 

Finishing the discussion on the RG equation for the exclusive FJFs, let us now turn to their matching to the standard collinear FFs. Similar to the semi-inclusive FJFs, the exclusive fragmenting jet functions $\Delta_{(T)} \mathscr{G}_i^h(z_h, p_TR, \mu)$ are also closely related to the fragmentation functions $\Delta_{(T)} D_{h/j}$ and in this work we label the corresponding matching coefficients as $\Delta_{(T)} \mathscr{J}_{i j}$.
Following~\cite{Jain:2011xz,Waalewijn:2012sv,Chien:2015ctp,Kang:2019ahe}, one has the following expansion for exclusive FJFs:
\begin{align}
\Delta _{(T)} \mathscr{G}_i^h (z_h, p_TR, \mu)
 =
\sum _j \int _{z_h}^1 \frac{\dd{z'_h}}{z'_h} \Delta_{(T)} \mathscr{J}_{ij} (z'_h, p_TR, \mu) \Delta _{(T)} D_{h/j} \pqty{\frac{z_h}{z'_h}, \mu}
\, . \label{e.exclusive_collinear_FJF_factorization}
\end{align}
Again this equation is valid up to power corrections of order $\order{\Lambda _{\mathrm{QCD}}^2 / \pqty{p_TR)^2}}$ for light hadrons. For heavy meson fragmenting jet functions, $\Lambda_{\mathrm{QCD}}$ should be replaced by the heavy quark mass $m_Q$ in the above equation \cite{Chien:2015ctp, Baumgart:2014upa}. We provide the NLO results of matching coefficients $\Delta_{(T)}\mathscr{J}_{i j}$, which depend on the jet algorithm.
The results for unpolarized FJFs with cone algorithm were given in \cite{Procura:2011aq}, while those for anti-$k_T$ jets were first written down in~\cite{Waalewijn:2012sv}. While the results for the unpolarized FJFs are available in the literature, the coefficients for the polarized FJFs are given for the first time in our current paper. We provide the detailed derivations of exclusive FJFs with polarizations for anti-$k_T$ jets in \cref{s.exclusive_FJF_calculation}. Here we only list the final results:
\begin{align}
\Delta \mathscr{J}_{qq} (z_h, p_TR, \mu)
& =
\delta (1-z_h) + \frac{\alpha _s C_F}{2 \pi} \bigg[\delta(1-z_h) \pqty{\frac{L^2}{2} - \frac{\pi^2}{12}} - \Delta P_{qq}(z_h) L
\nonumber \\
& \qquad \qquad \qquad \qquad \quad +
1 - z_h + \Delta \hat{\mathscr{I}}_{qq}^{\text{anti-}k_T}(z_h) \bigg]
\, , \\
\Delta \mathscr{J}_{q g} (z_h, p_TR, \mu)
& =
\frac{\alpha _s C_F}{2 \pi} \bigg[-\Delta P_{gq}(z_h) L - 2 (1-z_h) + \Delta \hat{\mathscr{I}}_{qg}^{\text{anti-}k_T}(z_h) \bigg]
\, , \\
\Delta \mathscr{J}_{g q} (z_h, p_TR, \mu)
& =
\frac{\alpha _s T_F}{2 \pi} \bigg[-\Delta P_{qg}(z_h) L + 2 (1-z_h) + \Delta \hat{\mathscr{I}}_{gq}^{\text{anti-}k_T}(z_h) \bigg]
\, , \\
\Delta \mathscr{J}_{gg} (z_h, p_TR, \mu)
& =
\delta (1-z_h) + \frac{\alpha _s C_A}{2 \pi} \bigg[\delta (1-z_h) \pqty{\frac{L^2}{2} - \frac{\pi ^2}{12}} - \Delta P_{gg}(z_h) L
\nonumber \\
& \qquad \qquad \qquad \qquad \quad - 4 (1-z_h) + \Delta \hat{\mathscr{I}}_{gg}^{\text{anti-}k_T}(z_h) \bigg]
\, , \\
\Delta _T \mathscr{J}_{qq} (z_h, p_TR, \mu)
& =
\delta (1-z_h) + \frac{\alpha _s C_F}{2 \pi} \bigg[\delta (1-z_h) \pqty{\frac{L^2}{2} - \frac{\pi ^2}{12}}
\nonumber \\
& \qquad \qquad \qquad \qquad \quad -
\Delta _T P_{qq}(z_h) L + \Delta _T \hat{\mathscr{I}}_q^{\text{anti-}k_T}(z_h) \bigg]
\, ,
\end{align}
where for anti-$k_T$ jets, the jet-algorithm dependent pieces, $\Delta_{(T)}\hat{\mathscr{I}}_{i j}^{\text{anti-}k_T}(z_h)$, are given by
\begin{align}
\Delta \hat{\mathscr{I}}_{qq}^{\text{anti-}k_T} \pqty{z_h}
& =
2 \Delta P_{qq}(z_h) \ln(z_h) + 2 \pqty{1+z_h^2} \pqty{\frac{\ln (1-z_h)}{1-z_h}}_+
\, , \\
\Delta \hat{\mathscr{I}}_{qg}^{\text{anti-}k_T} \pqty{z_h}
& =
2 \Delta P_{gq}(z_h) \ln{\Big(z_h (1-z_h)\Big)}
\, , \\
\Delta \hat{\mathscr{I}}_{gq}^{\text{anti-}k_T} \pqty{z_h}
& =
2 \Delta P_{qg}(z_h) \ln{\Big(z_h (1-z_h)\Big)}
\, , \\
\Delta \hat{\mathscr{I}}_{gg}^{\text{anti-}k_T} \pqty{z_h}
& =
2 \Delta P_{gg}(z_h) \ln(z_h) + 4 \Big(2 (1-z_h)^2 + z_h \Big) \pqty{\frac{\ln(1-z_h)}{1-z_h}}_+
\, , \\
\Delta _T \hat{\mathscr{I}}_{qq}^{\text{anti-}k_T} \pqty{z_h}
& =
2 \Delta _T P_{qq}(z_h) \ln(z_h) + 4 z_h \pqty{\frac{\ln(1-z_h)}{1-z_h}}_+
\, ,
\end{align}
where the splitting kernels $\Delta_{(T)}{P}_{ji}(z_h)$ \cite{Vogelsang:1996im,Vogelsang:1997ak} have been introduced in~\cref{e.AP_longitudinal_qq,e.AP_longitudinal_gq,e.AP_longitudinal_qg,e.AP_longitudinal_gg,e.AP_transverse_qq}. 

\section{The transverse momentum dependent FJFs (TMD FJFs)}
\label{s.TMD_FJFs}

In this section, we study TMD FJFs in both the semi-inclusive jet production and the exclusive jet production, including polarizations of the initiating parton and the final observed hadron inside the jet. They are termed as polarized TMD FJFs~\cite{Kang:2020xyq}. Some partial results are available in the previous publications~\cite{Kang:2020xyq,Kang:2021ffh}, but never the complete results. For example, the hard matching functions for both longitudinally and transversely polarized cases are presented for the first time. Below, we provide the complete results for all the relevant polarized semi-inclusive and exclusive TMD FJFs. The operator definitions of these functions in SCET are presented, followed by their factorization formalism, which involves hard functions, soft functions, and TMD FFs. 

\begin{figure}[htb]
\centering
\includegraphics[width = 0.6 \textwidth]{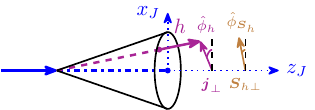}
\caption{Illustration for the distribution of hadrons inside a jet with transverse momentum $\boldsymbol{j}_{\perp}$ and azimuthal angle $\hat{\phi}_h$, and transverse spin $\boldsymbol{S}_{h \bot}$ and azimuthal angle $\hat{\phi}_{S_h}$.
Both azimuthal angles are measured with respect to the jet coordinate $x_J$ and within the jet transverse plane.}
\label{f.jet_cone_with_hadron}
\end{figure}

The kinematics is set up as in \cref{f.jet_cone_with_hadron}, a hadron $h$ is observed inside the jet, carrying a fraction $z_h$ of the jet longitudinal momentum (not labeled in the figure), a transverse momentum $\boldsymbol{j}_{\perp}$ with azimuthal angle $\hat{\phi}_h$, and a transverse spin $\boldsymbol{S}_{h \bot}$ with azimuthal angle $\hat{\phi}_{S_h}$.
Both $\boldsymbol{j}_{\perp}$ and $\boldsymbol{S}_{h \bot}$ are measured with respect to the jet axis $z_J$, while both azimuthal angles $\hat{\phi}_h$ and $\hat{\phi}_{S_h}$ are measured within the jet transverse plane and with respect to $x_J$, the $x$-axis that is rotated into jet coordinates
\footnote{Please see \cref{e.j_T_parameterization,e.S_hT_parameterization} for the parameterization of $\boldsymbol{j}_{\perp}$ and $\boldsymbol{S}_{h \bot}$.}.

\subsection{TMD FJFs in semi-inclusive jet productions}
\label{ss.semi_inclusive_TMD FJFs}
Again we start with the semi-inclusive TMD FJFs, which is an essential ingredient for studying the transverse momentum distribution of hadrons inside a jet in \textit{e.g.}, single inclusive jet production in proton-proton collisions ($p+p\to \mathrm{jet}(h)+X$) or in electron-proton collisions ($e+p\to \mathrm{jet}(h)+X$). Following the previous publication~\cite{Kang:2020xyq}, we define the general correlators for TMD FJFs initiated by quark or gluon as:
\begin{align}
\Delta^{h / q}\pqty{z, z_h, \boldsymbol{j}_{\perp}, S_h}
& =
\frac{z}{2 N_c} \delta \pqty{z_h - \frac{\omega_h}{\omega_J}}
\mel**{0}{\delta \pqty{\omega - \overline{n} \cdot \mathcal{P}} \delta ^2 \pqty{\mathcal{P}_{\perp} / z_h+\boldsymbol{j}_{\perp}} \chi _n (0)}{(J h) X}
\label{e.TMD_FJF_collelator_q} \\
& \quad \times
\mel**{(J h) X}{\overline{\chi}_n(0)}{0}
\nonumber , \\
\Delta^{h / g, \mu \nu} \pqty{z, z_h, \boldsymbol{j}_{\perp}, S_h}
& =
\frac{z \omega}{(d-2) \pqty{N_c^2-1}} \delta \pqty{z_h-\frac{\omega_h}{\omega_J}}
\label{e.TMD_FJF_collelator_g} \\
& \quad \times 
\mel**{0}{\delta \pqty{\omega - \overline{n} \cdot \mathcal{P}} \delta ^2 \pqty{\mathcal{P}_{\perp} / z_h + \boldsymbol{j}_{\perp}} \mathcal{B}_{n \perp}^{\mu}(0)}{(J h) X}
\mel**{(J h) X}{\mathcal{B}_{n \perp}^{\nu} (0)}{0}
\,, \nonumber
\end{align}
where we have the energy fractions $z$ and $z_h$ defined the same as in \cref{e.z_zh}. One can parameterize these correlators at the leading power~\cite{Metz:2016swz,Kang:2020xyq}:
\begin{align}
\Delta ^{h/q}(z, z_h, \boldsymbol{j}_{\perp}, S_h)
& =
\Delta ^{h/q \, [\slashed{n}]} \frac{\slashed{\overline{n}}}{2}
-
\Delta^{h/q \, [\slashed{n} \gamma _5]} \frac{\slashed{\overline{n}} \gamma _5}{2}
+
\Delta ^{h/q \, [i n_{\nu}\sigma ^{k \nu} \gamma _5]} \frac{i \overline{n}_{\mu} \sigma^{k \mu} \gamma _5}{2}
\, , \label{e.TMD_FJF_collelator_q_decomposition} \\
\Delta ^{h/g, kl} (z, z_h, \boldsymbol{j}_{\perp}, S_h)
& =
\frac{1}{2} \delta _T^{kl} \pqty{\delta _T^{mn} \Delta ^{h/g, mn}}
-
\frac{i}{2} \epsilon _T^{kl} \pqty{i \epsilon _T^{mn} \Delta ^{h/g, mn}}
+
\hat{S} \Delta ^{h/g, kl}
\, .
\label{e.TMD_FJF_collelator_g_decomposition}
\end{align}
Here we have defined $\Delta^{h/q [\Gamma]} \equiv \frac{1}{4} \Tr(\Delta ^{h/q} \Gamma)$, and we have $\hat{S}O^{kl} = \frac{1}{2} \pqty{O^{kl} + O^{lk} - \delta_T^{kl}O^{\rho\rho}}$ and $\delta_T^{kl} = -g_T^{kl}$. The three terms on the right-hand side of \cref{e.TMD_FJF_collelator_q_decomposition} correspond (in order) to the TMD FJFs with unpolarized, longitudinally polarized, and transversely polarized initial quarks. On the other hand,  the three terms on the right-hand side of \cref{e.TMD_FJF_collelator_g_decomposition} correspond to unpolarized, circularly polarized, and linearly polarized gluons. More details have been presented in the previous work \cite{Kang:2020xyq}. For completeness, here we provide the parameterization of quark TMD FJFs:
\begin{align}
\Delta ^{h/q [\slashed{n}]}
& =
\mathcal{D}_1^{h/q}(z,z_h,\boldsymbol{j}_{\perp})
-
\frac{\epsilon ^{kl}_T j_{\perp}^k S_{h \perp}^l}{z_h M_h} \mathcal{D}_{1T}^{\perp h/q}(z,z_h,\boldsymbol{j}_{\perp})
\, , \label{e.TMD_FJF_collelator_q_parameterization_u} \\
\Delta ^{h/q [\slashed{n} \gamma _5]}
& =
\lambda_h \mathcal{G}_{1L}^{h/q}(z,z_h,\boldsymbol{j}_{\perp})
-
\frac{\boldsymbol{j}_{\perp} \cdot \boldsymbol{S}_{h \perp}}{z_h M_h} \mathcal{G}_{1T}^{h/q}(z,z_h,\boldsymbol{j}_{\perp})
\, , \label{e.TMD_FJF_collelator_q_parameterization_l} \\
\Delta ^{h/q [i n_\nu \sigma^{k\nu} \gamma _5]}
& =
S_{h\perp}^k \mathcal{H}_1^{h/q}(z,z_h,\boldsymbol{j}_{\perp})
-
\frac{\epsilon^{kl}_T j_{\perp}^l}{z_h M_h} \mathcal{H}_{1}^{\perp h/q}(z,z_h,\boldsymbol{j}_{\perp})
-
\frac{j_{\perp}^k}{z_h M_h} \lambda_h {\mathcal{H}_{1L}^{\perp h/q}(z,z_h,\boldsymbol{j}_{\perp})}
\nonumber \\
& \quad +
\frac{j_{\perp}^k \boldsymbol{j}_{\perp} \cdot \boldsymbol{S}_{h \perp} - \frac{1}{2} \boldsymbol{j}_{\perp}^2 S_{h \perp}^k}{z_h^2 M_h^2} \mathcal{H}_{1T}^{\perp h/q} (z, z_h, \boldsymbol{j}_{\perp})
\, , \label{e.TMD_FJF_collelator_q_parameterization_t}
\end{align}
The functions $\mathcal{D}$, $\mathcal{G}$ and $\mathcal{H}$ on the right-hand side of \cref{e.TMD_FJF_collelator_q_parameterization_u,e.TMD_FJF_collelator_q_parameterization_l,e.TMD_FJF_collelator_q_parameterization_t} represent the TMD FJFs initiated by unpolarized, longitudinally polarized, and transversely polarized initial quark respectively.
The parameterization for gluon TMD FJFs is given by
\begin{align}
\delta _T^{kl} \Delta ^{h/g, kl}
& =
\mathcal{D}_1^{h/g} (z, z_h, \boldsymbol{j}_{\perp})
-
\frac{\epsilon ^{ij}_T j_{\perp}^i S_{h \perp}^j}{z_h M_h} \mathcal{D}_1^{\perp h/g} (z, z_h, \boldsymbol{j}_{\perp})
\label{e.TMD_FJF_collelator_g_parameterization_u}
\, , \\
i \epsilon _T^{kl} \Delta ^{h/g, kl}
& =
\lambda_h \mathcal{G}_{1L}^{h/g} (z, z_h, \boldsymbol{j}_{\perp})
-
\frac{\boldsymbol{j}_{\perp} \cdot \boldsymbol{S}_{h \perp}}{z_h M_h} \mathcal{G}_{1T}^{h/g} (z, z_h, \boldsymbol{j}_{\perp})
\label{e.TMD_FJF_collelator_g_parameterization_l}
\, , \\
\hat{S} \Delta ^{h/g, kl}
& =
\frac{j_{\perp}^k j_{\perp}^l}{2 z_h^2 M_h^2} \mathcal{H}_1^{\perp h/g} (z, z_h, \boldsymbol{j}_{\perp})
-
\frac{\epsilon_T^{j_{\perp}\{k} S_{h \perp}^{l\}} + \epsilon_T^{S_{h\perp}\{k} j_{\perp}^{l\}}}{8 z_h M_h} \mathcal{H}_{1T}^{h/g} (z, z_h, \boldsymbol{j}_{\perp})
\nonumber \\
& \quad +
\frac{\epsilon_T^{j_{\perp}\{k} j_{\perp}^{l\}}}{4 z_h^2 M_h^2}
\pqty{
\lambda_h \mathcal{H}_{1L}^{\perp, h/g} (z, z_h, \boldsymbol{j}_{\perp})
-
\frac{\boldsymbol{j}_{\perp} \cdot \boldsymbol{S}_{h \perp}}{z_h M_h} \mathcal{H}_{1T}^{\perp h/g} (z, z_h, \boldsymbol{j}_{\perp})
} \, ,\label{e.TMD_FJF_collelator_g_parameterization_t}
\end{align}
where the functions $\mathcal{D}$, $\mathcal{G}$ and $\mathcal{H}$ represent the TMD FJFs initiated by unpolarized, circularly polarized, and linearly polarized gluons, respectively. The two $\mathcal{G}$ functions with circularly polarized gluons are anti-symmetric of indices $k,\, l= 1, 2$, and the four $\mathcal{H}$ functions related to linearly polarized gluons are symmetric and traceless combinations of $k,\, l= 1, 2$. 
We have adopted the notation $v_T^{\{k}w_T^{l\}} = v_T^k w_T^l + v_T^l w_T^k$ as in~\cite{Mulders:2000sh, Boussarie:2023izj}. 

Since TMD FJFs represent the hadron fragmentation inside a fully reconstructed jet, their physical meaning is similar to that of standard TMD FFs as reviewed in \cite{Metz:2016swz}. Naturally, we adopt the calligraphic font of the letters used by the corresponding TMD FFs as the notations of TMD FJFs. For example, $\mathcal{H}_{1}^{h/q}$ in~\cref{e.TMD_FJF_collelator_q_parameterization_t} is the so-called quark transversity TMD FJFs, \textit{i.e.}, a transversely polarized quark initiates a jet, in which a transversely polarized hadron is observed.
On the other hand, $\mathcal{G}_{1L}^{h/g}$ in \cref{e.TMD_FJF_collelator_g_parameterization_l} is the helicity gluon TMD FJFs, \textit{i.e.}, a longitudinally polarized gluon initiates a jet, in which a longitudinally polarized hadron is observed.

In this study, we focus on the kinematic region $\Lambda _{\mathrm{QCD}} \lesssim j_{\perp} \ll p_T R$, necessitating the adoption of the TMD factorization~\cite{Collins:2011zzd}. In the previous work~\cite{Kang:2017glf}, the TMD factorization for transverse momentum distribution of unpolarized hadrons inside the jet was derived. Over there, the evolution equations are based on the separate rapidity renormalization group equations for the so-called ``unsubtracted'' TMD FFs and soft functions. Below, in the process of reviewing the TMD factorization, we update the formalism with the more recent language outlined in the TMD Handbook~\cite{Boussarie:2023izj}. This new formalism makes the evolution of the in-jet TMD FJFs simpler and more transparent than before. We then present the hard matching functions, where the expressions for longitudinally and transversely polarized cases are provided here for the first time.

\subsubsection{TMD Factorization}

\begin{figure}[htb]
\centering
\includegraphics[width = 0.6 \textwidth]{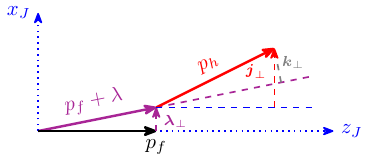}
\caption{Illustration of the soft radiation (purple) and fragmentation process (red).
The initial fragmenting quark carries a momentum $p_f$ along the jet axis $z_J$.
It undergoes a soft radiation of $\boldsymbol{\lambda}_{\bot}$, and then fragments into a hadron with momentum $p_h$.
The momentum $p_h$ has a transverse component of $\boldsymbol{k}_{\bot}$ with respect to the quark after soft radiation, or a transverse component of $\boldsymbol{j}_{\bot}$ with respect to the jet axis.}
\label{f.fragmentation}
\end{figure}

In the kinematic region under consideration, the radiation relevant at leading power is restricted to collinear radiation within the jet, characterized by the momentum that scales as $p_c = \bqty{p_c^+, p_c^-, p_{c\perp}} \sim p_c^- \bqty{\lambda ^2, 1, \lambda}$, where the power counting parameter $\lambda \sim j_{\perp} / p_T$.
Additionally, soft radiation of order $j_{\perp}$ is also relevant.
It is worth noting that harder emissions are only permitted outside the jet cone and will thus only impact the determination of the jet axis.
Consequently, the hadron transverse momentum $j_{\perp}$, which is defined with respect to the jet axis, remains intact from the radiations external to the jet.
Taking the unpolarized case as an example, a factorized formalism for the unpolarized TMD FJFs within SCET can be formulated \cite{Kang:2017glf}:
\begin{align}
\mathcal{D}_1^{h/c} \pqty{z, z_h, \boldsymbol{j}_{\perp}, p_TR, \mu, \zeta_J}
& =
\hat{H}^U_{c \to i} \pqty{z, p_T R, \mu}
\int \dd[2]{\boldsymbol{k}_{\perp}} \dd[2]{\boldsymbol{\lambda}_{\perp}} \delta ^2 \pqty{z_h \boldsymbol{\lambda}_{\perp} + \boldsymbol{k}_{\perp} - \boldsymbol{j}_{\perp}}
\nonumber \\
& \quad \times
D_1^{h/i(u)} \pqty{z_h, \boldsymbol{k}_{\perp}, \mu, \zeta/\nu^2} S_i \pqty{\boldsymbol{\lambda}_{\perp}, \mu, \nu \mathcal{R}/2}
\, ,
\end{align}
where $S_i \pqty{\boldsymbol{\lambda}_{\perp}, \mu, \nu \mathcal{R}/2}$ is the in-jet soft function, and $D_1^{h/i(u)} \pqty{z_h, \boldsymbol{k}_{\perp}, \mu, \zeta/\nu^2}$ is the ``unsubtracted'' TMD FFs with the superscript ``$(u)$'' to emphasize this fact. Here $\mu$ is the standard renormalization scale as above while $\nu$ is a rapidity scale associated with the rapidity divergence~\cite{Chiu:2011qc, Chiu:2012ir}, and $\zeta$ is the so-called Collins-Soper scale~\cite{Collins:2011zzd}. We also have a scale $\zeta_J$ to be defined below.
The $\delta$ function establishes a relationship between the hadron transverse momentum $\boldsymbol{j}_{\perp}$, relative to the jet axis, and two other momenta: the transverse momentum $\boldsymbol{\lambda}_{\perp}$ of soft radiation, and the hadron transverse momentum $\boldsymbol{k}_{\perp}$ with respect to the fragmenting parton.
Notice that $\boldsymbol{\lambda}_{\perp}$ is multiplied by $z_h$ to adjust for the dissimilarity between the fragmenting parton and the observed hadron.
The relationship among the three aforementioned transverse momenta is illustrated in \cref{f.fragmentation}.

As is common practice in TMD physics, we convert the aforementioned expression from the transverse momentum space to the coordinate $b$-space with the following transformation:
\begin{align}
\label{e.TMD_b_space}
\mathcal{D}_1^{h/c} \pqty{z, z_h, \boldsymbol{j}_{\perp}, p_TR, \mu, \zeta_J}
& =
\hat{H}^U_{c \to i} \pqty{z, p_T R, \mu}
\\
& \quad \times
\int \frac{\dd[2]{\boldsymbol{b}}}{(2 \pi)^2} e^{i \boldsymbol{j}_{\perp} \cdot \boldsymbol{b} / z_h}
\widetilde{D}_1^{h/i(u)} \pqty{z_h, \boldsymbol{b}, \mu, \zeta/\nu^2}
\widetilde{S}_i (\boldsymbol{b}, \mu, \nu \mathcal{R}/2)
\nonumber \, ,
\end{align}
where we have defined the $\widetilde{D}_1^{h/i(u)} \pqty{z_h, \boldsymbol{b}, \mu, \zeta/\nu^2}$ and $\widetilde{S}_i (\boldsymbol{b}, \mu, \nu \mathcal{R}/2)$ in Fourier $b$-space with $i=q,\, g$ as:
\begin{align}
\widetilde{D}_1^{h/i(u)} \pqty{z_h, \boldsymbol{b}, \mu, \zeta/\nu^2}
& =
\frac{1}{z_h^2} \int \dd[2]{\boldsymbol{k}_{\perp}} e^{-i \boldsymbol{k}_{\perp} \cdot \boldsymbol{b} / z_h} D_1^{h/i(u)} \pqty{z_h, k_{\perp}, \mu, \zeta/\nu^2}
\nonumber \\
& =
\frac{1}{z_h^2}
\int \dd{k_{\perp}} k_{\perp} 2 \pi J_0 \pqty{\frac{b k_{\perp}}{z_h}} D_1^{h/i(u)} \pqty{z_h, k_\perp, \mu, \zeta/\nu^2}
\, , \label{e.TMD_FF_b_space} \\
\widetilde{S}_i(\boldsymbol{b}, \mu, \nu \mathcal{R}/2)
& =
\int \dd[2]{\boldsymbol{\lambda _{\perp}}} e^{-i \boldsymbol{\lambda}_{\perp} \cdot \boldsymbol{b}} S_i \pqty{\boldsymbol{\lambda}_{\perp}, \mu, \nu \mathcal{R}/2}
\, ,
\end{align}
where $J_0 \pqty{b k_{\bot} / z_h}$ is the Bessel function of the first kind.
The $\mu$- and $\nu$-renormalization group equations for the ``unsubstracted'' TMD FFs $D_1^{h/i(u)}$ are well known:
\begin{align}
\mu \frac{\dd{}}{\dd{\mu}} \ln \widetilde{D}_1^{h/i(u)} \pqty{z_h, \boldsymbol{b}, \mu, \zeta/\nu^2}
& =
\gamma_{\mu, i}^D(\mu, \zeta/\nu^2)
\, , \\
\nu \frac{\dd{}}{\dd{\nu}} \ln \widetilde{D}_1^{h/i(u)} \pqty{z_h, \boldsymbol{b}, \mu, \zeta/\nu^2}
& =
\gamma_{\nu, i}^D(b, \mu)
\, ,
\end{align}
where the leading order $\mu$- and $\nu$-anomalous dimensions are given by:
\begin{align}
\gamma_{\mu, q}^D(\mu, \zeta/\nu^2)
& =
\frac{\alpha_s}{\pi} C_F \pqty{\ln(\frac{\nu^2}{\zeta}) + \frac{3}{2}}
\, , \\
\gamma_{\mu, g}^D(\mu, \zeta/\nu^2)
& =
\frac{\alpha_s}{\pi} C_A \pqty{\ln(\frac{\nu^2}{\zeta}) + \frac{\beta_0}{2C_A}}
\, , \\
\gamma_{\nu, i}^D(b,\mu)
& =
\frac{\alpha_s}{\pi} C_i \ln(\frac{\mu^2}{\mu_b^2})
\, , \label{e.gamma_nu_q}
\end{align}
with $C_q = C_F$, $C_g = C_A$, and $\mu _b \equiv 2 e^{- \gamma _E} / b$.

On the other hand, the in-jet soft function up to the NLO is given by~\cite{Kang:2017glf, Buffing:2018ggv, Kang:2021ffh}:
\begin{equation}
\label{e.soft}
\widetilde{S}_i \pqty{\boldsymbol{b}, \mu, \nu \mathcal{R}/2}
=
1
+
\frac{\alpha _s}{2 \pi} C_i
\bqty{
- \frac{1}{2} \ln[2](\frac{\mu ^2}{\mu _b^2})
-
\ln(\frac{\mu ^2}{\mu_b^2}) \ln(\frac{\nu ^2 \tan[2](\mathcal{R}/2)}{\mu^2})
-
\frac{\pi ^2}{12}
} \, .
\end{equation}
The corresponding $\mu$- and $\nu$-renormalization group equations are given by:
\begin{align}
\mu\frac{\dd{}}{\dd{\mu}}\ln \widetilde{S}_i \pqty{\boldsymbol{b}, \mu, \nu \mathcal{R}/2}
& =
\gamma _{\mu , i}^S \pqty{b, \mu, \nu \mathcal{R}/2}
\, , \\
\nu\frac{\dd{}}{\dd{\nu}}\ln \widetilde{S}_i \pqty{\boldsymbol{b}, \mu, \nu \mathcal{R}/2}
& =
\gamma _{\nu , i}^S \pqty{b, \mu}
\, ,
\end{align}
where the anomalous dimensions are given by:
\begin{align}
\gamma _{\mu , i}^S \pqty{b, \mu, \nu \mathcal{R}/2}
& =
- \frac{\alpha _s}{\pi} C_i
\ln(\frac{\nu ^2 \tan[2](\mathcal{R}/2)}{\mu ^2})
\, , \label{e.soft_mu_gamma} \\
\gamma _{\nu , i}^S \pqty{b, \mu}
& =
- \frac{\alpha _s}{\pi} C_i
\ln(\frac{\mu ^2}{\mu _b^2})
\, . \label{e.soft_nu_gamma}
\end{align}
It is instructive to compare this in-jet soft function $\widetilde{S}_i$ with the standard soft function $\widetilde{S}_{n\overline{n},i}\pqty{\boldsymbol{b}, \mu, \nu}$, which arises in the TMD factorization for the semi-inclusive deep inelastic scattering (SIDIS) and Drell-Yan production (involving quarks) or for the Higgs production in proton-proton collisions (involving gluons), whose expression is given by:
\begin{align}
\widetilde{S}_{n \overline{n}, i} \pqty{\boldsymbol{b}, \mu, \nu} =  1
+
\frac{\alpha _s}{\pi} C_i
\bqty{
- \frac{1}{2} \ln[2](\frac{\mu ^2}{\mu _b^2})
-
\ln(\frac{\mu ^2}{\mu _b^2}) \ln(\frac{\nu ^2}{\mu^2})
-
\frac{\pi ^2}{12}
} \, . 
\end{align}
After replacing $\nu \to \nu \tan(\mathcal{R}/2)$, the in-jet soft function $\widetilde{S}_i$ is exactly equal to $\sqrt{\widetilde{S}_{n \overline{n}, i}}$ at the NLO:
\begin{align}
\widetilde{S}_i \pqty{\boldsymbol{b}, \mu, \nu \mathcal{R}/2}
= \left.\sqrt{\widetilde{S}_{n \overline{n}, i} \pqty{\boldsymbol{b}, \mu, \nu}}\right|_{\nu \to \nu\tan(\mathcal{R}/2)}
\, .
\end{align}

Now since the standard ``subtracted'' TMD FFs are defined as a combination of ``unsubtracted'' TMD FFs $\widetilde{D}_1^{h/i(u)}$ and $\widetilde{S}_i$~\cite{Boussarie:2023izj}, we would define the ``subtracted'' in-jet TMD FFs as a product of $\widetilde{D}_1^{h/i(u)}$ and $\widetilde{S}_i$ as follows:
\begin{align}
\label{e.TMD_proper}
\widetilde{D}_1^{h/i} \pqty{z_h, \boldsymbol{b}, \mu, \zeta_J}
\equiv
\widetilde{D}_1^{h/i(u)} \pqty{z_h, \boldsymbol{b}, \mu, \zeta/\nu^2}  \widetilde{S}_i(\boldsymbol{b}, \mu, \nu \mathcal{R}/2)
\, ,
\end{align}
where the rapidity divergences cancel out between $D_1^{h/i(u)}$ and $\widetilde{S}_i$ due to the sign difference shown in \cref{e.gamma_nu_q,e.soft_nu_gamma}.
At the same time, we find that a slightly different Collins-Soper scale $\zeta_J$ arises in the in-jet ``subtracted'' TMD FFs, which is related to the Collins-Soper scale $\zeta$ for the standard TMD FFs,
\begin{align}
\sqrt{\zeta_J}
\equiv \sqrt{\zeta}\, \mathcal{R}/2
= p_T R
\, , \label{e.zeta_J}
\end{align}
where we have used $\tan(\mathcal{R}/2)\approx \mathcal{R}/2$ for narrow jets with $\mathcal{R} \ll 1$.
Note that in our regularization scheme, the natural Collins-Soper scale for TMD FFs is $\sqrt{\zeta} = \omega_J = 2 E_J = 2 p_T \cosh{\eta}$ and using $\mathcal{R} = R/\cosh{\eta}$, we obtain the natural scale $\sqrt{\zeta_J}$ for the unpolarized TMD FJFs, which is simply $p_TR$.
With this newly defined Collins-Soper scale $\zeta_J$, following~\cite{Boussarie:2023izj, Ebert:2019okf}, we can thus convert the $\nu$-RG equations above as:
\begin{align}
\frac{\dd{}}{\dd{\ln{\sqrt{\zeta_J}}}} \ln{\widetilde{D}_1^{h/i} \pqty{z_h, \boldsymbol{b}, \mu, \zeta_J}}
=
\widetilde{K}(b, \mu)
\, ,\label{e.zeta_J_D}
\end{align}
where $\widetilde{K}(b, \mu) = -\gamma_{\nu, i}^D (b,\mu)$ in \cref{e.gamma_nu_q} is the so-called Collins-Soper evolution kernel.
In the small-$b$ region where $1/b \gg \Lambda_{\mathrm{QCD}}$, it can be computed perturbatively and the four-loop expressions are available in~\cite{Moult:2022xzt, Duhr:2022yyp}. Notice that $\widetilde{K}(b, \mu)$ would become non-perturbative in the large-$b$ region, see recent numerical computations in lattice QCD~\cite{Shanahan:2021tst, LPC:2022ibr, LatticePartonLPC:2023pdv, Shu:2023cot, Avkhadiev:2023poz}.
On the other hand, the standard $\mu$-RG equation is given by:
\begin{align}
\frac{\dd{}}{\dd{\ln{\mu}}} \ln{\widetilde{D}_1^{h/i} \pqty{z_h, \boldsymbol{b}, \mu, \zeta_J}}
=
\gamma_\mu^i \bqty{\alpha_s(\mu), \zeta_J/\mu^2}
\, , \label{e.mu_D}
\end{align}
where the anomalous dimension $\gamma_\mu^i \bqty{\alpha_s(\mu), \zeta_J/\mu^2}$ are given by:
\begin{align}
\gamma_\mu^i \bqty{\alpha_s(\mu), \zeta_J/\mu^2}
= - \Gamma_{\mathrm{cusp}}^i \bqty{\alpha_s(\mu)} \,
\ln(\frac{\zeta_J}{\mu^2})
+
\gamma_\mu^i \bqty{\alpha_s(\mu)}
\, , \label{e.mu-dimension}
\end{align}
where we have written the more general form~\cite{Boussarie:2023izj} with $\Gamma_{\mathrm{cusp}}^i$ and $\gamma_\mu^i$ the cusp and non-cusp anomalous dimensions.
They are perturbatively expanded as $\Gamma_{\mathrm{cusp}}^i \bqty{\alpha_s} = \sum_{n=1} \Gamma_{n-1}^i \pqty{\frac{\alpha_s}{4\pi}}^n$ and $\gamma_\mu^i \bqty{\alpha_s} = \sum_{n=1} \gamma_{n}^i \pqty{\frac{\alpha_s}{4\pi}}^n$
and at the next-to-leading logarithmic (NLL) order, they are given by \cite{Becher:2006mr, Jain:2011xz}:
\begin{align}
\Gamma_0^q & = 4 C_F
\, , \qquad
\Gamma_1^q = 4 C_F \bqty{\pqty{\frac{67}{9}-\frac{\pi^2}{3}} C_A - \frac{20}{9} T_F n_f} \, , \qquad
\gamma_0^q = 6 C_F \, , \\
\Gamma_{0,1}^g & = \frac{C_A}{C_F}\Gamma_{0,1}^q \, , \qquad
\gamma_0^g = 2 \beta_0 \, .
\end{align}
One can solve the $\zeta_J$- and $\mu$-RG evolution equation to obtain the ``subtracted'' in-jet TMD FFs:
\begin{align}
\widetilde{D}_1^{h/i} \pqty{z_h, \boldsymbol{b}, \mu, \zeta_J}
=
\widetilde{D}_1^{h/i} \pqty{z_h, \boldsymbol{b}, \mu_0, \zeta_0} 
& \exp[\int_{\mu_0}^\mu \frac{\dd{\mu'}}{\mu'} \gamma_\mu^i \bqty{\alpha_s(\mu), \zeta_J/\mu^2}]
\nonumber\\
\times &
\exp[\widetilde{K}(b, \mu_0)\ln(\frac{\sqrt{\zeta_J}}{\sqrt{\zeta_0}})]
\, . \label{e.D_1_evolution}
\end{align}
One typically evolves the in-jet TMD FFs $\widetilde{D}_1^{h/i}$ from the initial scale $\mu_0 = \sqrt{\zeta_0} = \mu_b$ to the final scale $\mu\sim p_T$ of the jet and $\sqrt{\zeta_J} = p_T R$ as mentioned above.
On the other hand, at the initial scale $\mu_0 = \sqrt{\zeta_0} = \mu_b$, one can expand $\widetilde{D}_1^{h/i} \pqty{z_h, \boldsymbol{b}, \mu_0, \zeta_0}$ in terms of the corresponding collinear FFs when $\mu_b \gg \Lambda_{\mathrm{QCD}}$.

To generalize the above TMD factorization to the polarized TMD FJFs, let us define the following operator $\mathscr{C}$:
\begin{align}
\label{e.TMD_factorization_operator}
\mathscr{C} [\widetilde{D}^{h/i, (n)}]
\equiv
\int \frac{b^{n+1} \dd{b}}{2 \pi n !} \pqty{\frac{z_h^2 M_h^2}{j_{\perp}}}^n J_n \pqty{\frac{j_{\perp} b}{z_h}} \widetilde{D}^{h/i, (n)} \pqty{z_h, \boldsymbol{b}, \mu, \zeta_J}
\, ,
\end{align}
where $\widetilde{D}^{h/i, (n)}$ is $n$-th moment of the TMD FFs in the Fourier $b$-space:
\begin{equation}
\widetilde{D}^{h/i, (n)} \pqty{z_h, \boldsymbol{b}, \mu, \zeta_J}
=
\frac{1}{z_h^2} \frac{2 \pi n !}{\pqty{z_h^2 M_h^2}^n}
\int \dd{k_{\perp}} k_{\perp} \pqty{\frac{k_{\perp}}{b}}^n J_n \pqty{\frac{b k_{\perp}}{z_h}} D^{h / i}\pqty{z_h, \boldsymbol{k}_{\perp}, \mu, \zeta_J}
\, . \label{e.TMD_FF_b_space_nth}
\end{equation}
In addition, we suppress the superscript $(0)$ when $n = 0$. 
One can easily verify that with $n = 0$ and $D = D_1$, the combination of \cref{e.TMD_FF_b_space,e.TMD_proper} would lead to \cref{e.TMD_FF_b_space_nth}.
It is important to realize that each in-jet TMD FF $\widetilde{D}^{h/i, (n)} \pqty{z_h, \boldsymbol{b}, \mu, \zeta_J}$ would follow the same Collins-Soper evolution equation as in~\cref{e.zeta_J_D}, as well as the same $\mu$-RG equation as in~\cref{e.mu_D}.
They are identical to the standard TMD FFs with the replacement $\zeta$ by $\zeta_J$. Thus, their evolved results would be given exactly by~\cref{e.D_1_evolution}, except one replaces $\widetilde{D}_1^{h/i} \pqty{z_h, \boldsymbol{b}, \mu, \zeta_J}$ by the corresponding TMD FFs $\widetilde{D}^{h/i, (n)} \pqty{z_h, \boldsymbol{b}, \mu, \zeta_J}$ defined above and given below for specific TMD FFs.
Note that the evolved in-jet unpolarized TMD FFs $\widetilde{D}_1^{h/i} \pqty{z_h, \boldsymbol{b}, \mu , \zeta_J}$ is given by~\cref{e.D_1_evolution}, which depends on the TMD FFs $\widetilde{D}_1^{h/i} \pqty{z_h, \boldsymbol{b}, \mu_0, \zeta_0}$ at the initial scale $\mu_0$ and $\zeta_0$.
Often, one further matches the TMD FFs at the initial scales onto collinear FFs.
For example, the matching coefficients for the unpolarized TMD FFs are known up to ${\mathrm{N}^3\mathrm{LO}}$~\cite{Luo:2020epw,Ebert:2020qef}.
On the other hand, the matching coefficients for $\widetilde{H}_1^{h/q}$ are known up to NNLO~\cite{Gutierrez-Reyes:2018iod}.
We can now write down the factorization for all the TMD FJFs as:
\begin{align}
\mathcal{D}_{1}^{h/c} \pqty{z, z_h, \boldsymbol{j}_{\perp}, p_TR, \mu, \zeta_J}
& =
\hat{H}_{c \to i}^U (z, p_TR, \mu) \mathscr{C} \bqty{\widetilde{D}_{1}^{h/i}}
\, , \label{e.D_1_FJFs} \\
\mathcal{D}_{1T}^{\perp, h/c} \pqty{z, z_h, \boldsymbol{j}_{\perp}, p_TR, \mu, \zeta_J}
& =
\hat{H}_{c \to i}^U (z, p_TR, \mu) \mathscr{C} \bqty{\widetilde{D}_{1T}^{\perp, h/i, (1)}}
\, , 
\label{e.D_1T_FJFs}
\\
\mathcal{G}_{1L}^{h/c} \pqty{z, z_h, \boldsymbol{j}_{\perp}, p_TR, \mu, \zeta_J}
& =
\hat{H}_{c \to i}^L (z, p_TR, \mu) \mathscr{C} \bqty{\widetilde{G}_{1L}^{h/i}}
\, , \label{e.G_1L_FJFs} \\
\mathcal{G}_{1T}^{h/c} \pqty{z, z_h, \boldsymbol{j}_{\perp}, p_TR, \mu, \zeta_J}
& =
\hat{H}_{c \to i}^L (z, p_TR, \mu) \mathscr{C} \bqty{\widetilde{G}_{1T}^{h/i, (1)}}
\, , \label{e.G_1T_FJFs}\\
\mathcal{H}_{1}^{h/c} \pqty{z, z_h, \boldsymbol{j}_{\perp}, p_TR, \mu, \zeta_J}
& =
\hat{H}_{c \to i}^T (z, p_TR, \mu) \mathscr{C} \bqty{\widetilde{H}_{1}^{h/i}}
\, , \label{e.H_1_FJFs} \\
\mathcal{H}_{1}^{\perp, h/c} \pqty{z, z_h, \boldsymbol{j}_{\perp}, p_TR, \mu, \zeta_J}
& =
\hat{H}_{c \to i}^T (z, p_TR, \mu) \mathscr{C} \bqty{\widetilde{H}_{1}^{\perp, h/i, (1)}}
\, , \label{e.H_1perp_FJFs}\\
\mathcal{H}_{1L}^{\perp, h/c} \pqty{z, z_h, \boldsymbol{j}_{\perp}, p_TR, \mu, \zeta_J}
& =
\hat{H}_{c \to i}^T (z, p_TR, \mu) \mathscr{C} \bqty{\widetilde{H}_{1L}^{\perp, h/i, (1)}}
\, , \label{e.H_1L_FJFs} \\
\mathcal{H}_{1T}^{\perp, h/c} \pqty{z, z_h, \boldsymbol{j}_{\perp}, p_TR, \mu, \zeta_J}
& =
\hat{H}_{c \to i}^T (z, p_TR, \mu) \mathscr{C} \bqty{\widetilde{H}_{1T}^{\perp, h/i, (2)}}
\, . \label{e.H_1T_FJFs}
\end{align}
Here the superscripts $U$, $L$ and $T$ of $\hat{H}_{c \to i}$ represent unpolarized, longitudinally polarized and transversely polarized hard matching functions, respectively.
The hard matching functions will be provided in the next subsection.
The above equations also show how various TMD FJFs are matched onto their corresponding TMD FFs, with which the matching of the scenarios listed in \cref{t.FJF_and_FFs} can be performed.
\newcommand{\eqfig}[1]{\includegraphics[height = 1.5 em, valign = c]{./figures/FJFs-table/#1.pdf}}
\renewcommand{\arraystretch}{2}
\begin{table}
\begin{center}
\begin{tabular}{|c|c|c|c|c|}
\hline
\cellcolor{black} & \cellcolor{black} & \multicolumn{3}{c|}{\cellcolor{red!50} Quark polarization} \\
\hline
\cellcolor{black} & \cellcolor{black} & \cellcolor{red!50} U & \cellcolor{red!50} L & \cellcolor{red!50} T \\
\hline
\cellcolor{green!50} &
\cellcolor{green!50} U &
$\mathcal{D}_1 = \eqfig{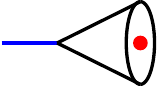}$ &
~ &
$\mathcal{H}_1^{\perp} = \eqfig{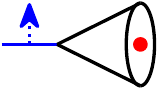} - \eqfig{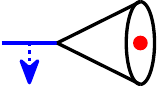}$ \\
\hhline{>{\arrayrulecolor{green!50}}|>{\arrayrulecolor{green!50}}->{\arrayrulecolor{black}}|%
>{\arrayrulecolor{black}}->{\arrayrulecolor{black}}|---}
\cellcolor{green!50} &
\cellcolor{green!50} L &
~ &
$\mathcal{G}_{1L} = \eqfig{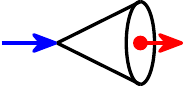} - \eqfig{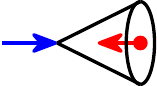}$ &
$\mathcal{H}_{1L}^{\perp} = \eqfig{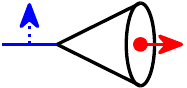} - \eqfig{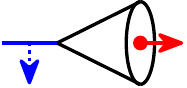}$ \\
\hhline{>{\arrayrulecolor{green!50}}|>{\arrayrulecolor{green!50}}->{\arrayrulecolor{black}}|%
>{\arrayrulecolor{black}}->{\arrayrulecolor{black}}|---}
\cellcolor{green!50} &
\cellcolor{green!50} &
&
&
$\mathcal{H}_1 = \eqfig{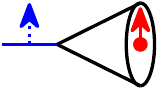} - \eqfig{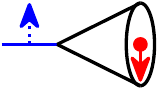}$ \\
\multirow{-4}{*}{\cellcolor{green!50} \rotatebox[origin=c]{90}{Hadron polarization}} &
\multirow{-2}{*}{\cellcolor{green!50} T} &
\multirow{-2}{*}{$\mathcal{D}_{1T}^{\perp} = \eqfig{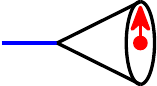} - \eqfig{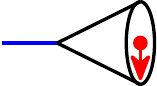}$} &
\multirow{-2}{*}{$\mathcal{G}_{1T} = \eqfig{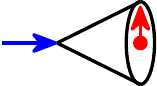} - \eqfig{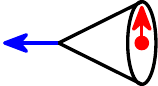}$} &
$\mathcal{H}_{1T}^{\perp} = \eqfig{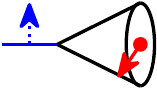} - \eqfig{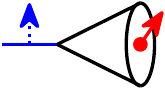}$ \\
\hline
\end{tabular}
\end{center}
\caption{
Summary of the semi-inclusive TMD FJFs. The header row represents the polarization of the quark (indicated by the blue line) that initiates the jet, while the header column indicates the corresponding polarizations of produced hadrons (indicated by the red arrow from the red dot). Shown here is for quark TMD FJFs, with $U,\, L,\, T$ representing the unpolarized case, and longitudinal and transverse polarization. For gluon TMD FJFs, one would interpret $L,\, T,$ as circular and linear polarization~\cite{Metz:2016swz}.}
\label{t.FJF_and_FFs}
\end{table}
\renewcommand{\arraystretch}{1}

\subsubsection{Hard matching functions}
The hard matching functions $\hat{H}_{c \to i}^{U,L,T}$ describe the out-of-jet radiation during which an energetic parton $c$, generated in a hard scattering event, undergoes a splitting into a parton $i$, which subsequently initiates a jet with energy $\omega _J$ and radius $R$. The hard matching functions for the unpolarized case $\hat{H}_{c \to i}^{U}$ are available previously~\cite{Kang:2016ehg, Kang:2016mcy, Kang:2017mda, Cal:2019hjc}. They describe the splitting from an unpolarized initial parton $c$ to an unpolarized parton $i$, irrespective of the polarization of the final-state hadron inside the jet. Because of that, we would have the same hard matching functions for both unpolarized TMD FJFs $\mathcal{D}_{1}^{h/c}$ and the polarized TMD FJFs $\mathcal{D}_{1T}^{\perp, h/c}$, as indicated in \cref{e.D_1_FJFs,e.D_1T_FJFs}.
We remind the reader that $\mathcal{D}_{1T}^{\perp, h/c}$ stands for the situation where an unpolarized parton $c$ splits into an unpolarized parton $i$ that fragments into a transversely polarized hadron $h$. This arises from the correlation between the hadron's transverse momentum with respect to the jet axis and the transverse spin of the hadron itself. For the same reasons, we have the same longitudinally polarized hard matching functions $\hat{H}_{c \to i}^{L}$ for the TMD FJFs $\mathcal{G}_{1L}^{h/c}$ and $\mathcal{G}_{1T}^{h/c}$, as indicated in \cref{e.G_1L_FJFs,e.G_1T_FJFs}.
Likewise, we have the same transversely polarized hard matching functions $\hat{H}_{c \to i}^{T}$ for the TMD FJFs $\mathcal{H}_{1}^{h/c}$, $\mathcal{H}_{1}^{\perp, h/c}$, $\mathcal{H}_{1L}^{\perp, h/c}$, and $\mathcal{H}_{1T}^{\perp, h/c}$, as indicated in~\cref{e.H_1_FJFs,e.H_1perp_FJFs,e.H_1L_FJFs,e.H_1T_FJFs}. Looking at~\cref{t.FJF_and_FFs}, the FJFs listed in the same column possess identical hard matching functions, due to the fact that the parton polarization is the same in the same column. 

For completeness, we list the unpolarized results $\hat{H}_{c \to i}^{U}$ here. We then provide in addition the results $\hat{H}_{c \to i}^{L,T}$ for the polarized cases. 
\begin{align}
\hat{H}_{q \to q'}^U (z, p_TR, \mu) 
& = \delta _{qq'} \delta (1-z)
+ \delta _{qq'} \frac{\alpha _s}{2 \pi} \bigg[C_F \delta (1-z) \pqty{-\frac{L^2}{2} - \frac{3}{2} L + \frac{\pi ^2}{12}}
\nonumber \\
& \quad
+ P_{qq}(z) L -2 C_F (1+z^2) \pqty{\frac{\ln(1-z)}{1-z}}_+ - C_F (1-z) \bigg]
\, , \\
\hat{H}_{q \to g}^U (z, p_TR, \mu) 
& =
\frac{\alpha _s}{2 \pi} \bigg[\Big(L - 2 \ln(1-z) \Big) P_{gq}(z) - C_F z \bigg]
\, , \\
\hat{H}_{g \to g}^U (z, p_TR, \mu)
& =
\delta(1-z)
+
\frac{\alpha _s}{2 \pi} \bigg[\delta (1-z) \pqty{-C_A \frac{L^2}{2} - \frac{\beta _0}{2} L + C_A \frac{\pi ^2}{12}}
\nonumber \\
& \quad +
P_{gg}(z) L - \frac{4 C_A \pqty{1 - z + z^2}^2}{z} \pqty{\frac{\ln(1-z)}{1-z}}_+ \bigg]
\, , \\
\hat{H}_{g \to q}^U (z, p_TR, \mu)
& =
\frac{\alpha _s}{2 \pi} \bigg[\Big(L - 2 \ln(1-z) \Big) P_{qg}(z) - 2 T_F z (1-z) \bigg]
\, ,
\end{align}
where the leading splitting kernels are given by:
\begin{align}
P_{qq}(z)
& =
C_F \bqty{\frac{1 + z^2}{(1-z)_+} + \frac{3}{2} \delta (1-z)}
\, , \\
P_{gq}(z)
& =
C_F \frac{1 + \pqty{1 - z}^2}{z}
\, , \\
P_{qg}(z)
& =
T_F \bqty{z^2 + \pqty{1-z}^2}
\, , \\
P_{gg}(z)
& =
2 C_A \bqty{\frac{z}{(1-z)_+} + \frac{1-z}{z} + z \pqty{1-z}}
+
\frac{\beta _0}{2} \delta (1-z)
\, .
\end{align}
The hard matching functions for the longitudinally polarized parton case are given here for the first time:
\begin{align}
\hat{H}^L_{q \to q'}(z, p_TR, \mu) 
& =
\delta _{qq'} \delta (1-z)
+
\delta _{qq'} \frac{\alpha _s}{2 \pi} \bigg[C_F \delta (1-z) \pqty{-\frac{L^2}{2} - \frac{3}{2} L + \frac{\pi ^2}{12}}
\nonumber \\
& \quad +
\Delta P_{qq}(z) L -2 C_F (1+z^2) \pqty{\frac{\ln(1-z)}{1-z}}_+ - C_F (1-z) \bigg]
\, , \\
\hat{H}^L_{q \to g} (z, p_TR, \mu) 
& = \frac{\alpha _s}{2 \pi} \bigg[\Big(L - 2 \ln(1-z) \Big) \Delta P_{gq}(z) + 2 C_F (1-z) \bigg]
\, , \\
\hat{H}^L_{g \to g} (z, p_TR, \mu)
& =
\delta(1-z)
+
\frac{\alpha _s}{2 \pi} \bigg[\delta (1-z) \pqty{-C_A \frac{L^2}{2} - \frac{\beta _0}{2} L + C_A \frac{\pi ^2}{12}}
\nonumber \\
& \quad +
\Delta P_{gg}(z) L + 4 C_A (1-z) - 4 C_A (2 (1-z)^2 + z) \pqty{\frac{\ln(1-z)}{1-z}}_+ \bigg]
\, , \\
\hat{H}^L_{g \to q} (z, p_TR, \mu)
& =
\frac{\alpha _s}{2 \pi} \bigg[\Big(L - 2 \ln(1-z) \Big) \Delta P_{qg}(z) - 2 T_F (1-z) \bigg]
\, , 
\end{align}
Finally, for the transversely polarized case, we only have contributions from the $q \to q$ channel, for the same reason mentioned before: gluon transversity FFs do not exist for spin-1/2 hadron.
The transversely polarized hard matching function is then given by:
\begin{align}
    \hat{H}^T_{q \to q'} (z, p_TR, \mu)
& =
\delta _{qq'} \delta(1-z)
+ \delta _{qq'} \frac{\alpha _s}{2 \pi} \bigg\{\Delta _T P_{qq}(z) L + C_F \bigg[-4 z \pqty{\frac{\ln(1-z)}{1-z}}_+
\nonumber \\
& \quad +
\pqty{-\frac{3}{2} L - \frac{L^2}{2} + \frac{\pi ^2}{12}} \delta (1-z) \bigg] \bigg\}
\, .
\end{align}
Note that the leading order splitting kernels for the polarized cases are given in \cref{e.AP_longitudinal_qq,e.AP_longitudinal_gq,e.AP_longitudinal_qg,e.AP_longitudinal_gg,e.AP_transverse_qq}.

The hard matching functions $\hat{H}_{i \to j}^{U,L,T} \pqty{z, p_TR , \mu}$ follow the RG equations below: 
\begin{align}
\mu \frac{\dd}{\dd{\mu}} \hat{H}_{i \to j}^U \pqty{z, p_TR , \mu}
& =
\sum _k \int _z^1 \frac{\dd{z'}}{z'} \gamma _{ik}^U \pqty{\frac{z}{z'}, p_TR, \mu}
\hat{H}_{k \to j}^U \pqty{z', p_TR, \mu}
\, , \\
\mu \frac{\dd}{\dd{\mu}} \hat{H}_{i \to j}^L \pqty{z, p_TR , \mu}
& =
\sum _k \int _z^1 \frac{\dd{z'}}{z'} \gamma _{ik}^L \pqty{\frac{z}{z'}, p_TR, \mu}
\hat{H}_{k \to j}^L \pqty{z', p_TR, \mu}
\, , \\
\mu \frac{\dd}{\dd{\mu}} \hat{H}_{q \to q'}^T \pqty{z, p_TR , \mu}
& =
\int _z^1 \frac{\dd{z'}}{z'} \gamma _{qq}^T \pqty{\frac{z}{z'} , p_TR , \mu}
\hat{H}_{q \to q'}^T \pqty{z', p_TR, \mu}
\, ,
\end{align}
where indices $i$, $j$ and $k$ all represent partons $q$ or $g$.
For the unpolarized hard matching functions, their anomalous dimensions $\gamma _{ij}^U \pqty{z, p_TR, \mu}$ are given by:
\begin{align}
\gamma _{qq}^U\pqty{z, p_TR, \mu}
& =
\frac{\alpha _s}{\pi} \pqty{P_{qq} \pqty{z} - C_F L \delta \pqty{1 - z} - \frac{3 C_F}{2} \delta \pqty{1 - z}}
\, , \\
\gamma _{qg}^U\pqty{z, p_TR, \mu}
& =
\frac{\alpha _s}{\pi} P_{gq} \pqty{z}
\, , \\
\gamma _{gg}^U\pqty{z, p_TR, \mu}
& =
\frac{\alpha _s}{\pi} \pqty{P_{g g} \pqty{z} - C_A L \delta \pqty{1 - z} - \frac{\beta _0}{2} \delta \pqty{1 - z}}
\, , \\
\gamma _{gq}^U\pqty{z, p_TR, \mu}
& =
\frac{\alpha _s}{\pi} P_{q g} \pqty{z}
\, .
\end{align}
On the other hand, the anomalous dimensions for the polarized cases, $\gamma _{ij}^{L \pqty{T}} \pqty{z, p_TR, \mu}$ are given by:
\begin{align}
\gamma _{qq}^L\pqty{z, p_TR, \mu}
& =
\frac{\alpha _s}{\pi} \pqty{\Delta P_{qq} \pqty{z} - C_F L \delta \pqty{1 - z} - \frac{3 C_F}{2} \delta \pqty{1 - z}}
\, , \label{e.lon_hard_function_gamma_qq} \\
\gamma _{qg}^L\pqty{z, p_TR, \mu}
& =
\frac{\alpha _s}{\pi} \Delta P_{gq} \pqty{z}
\, , \label{e.lon_hard_function_gamma_qg} \\
\gamma _{gg}^L\pqty{z, p_TR, \mu}
& =
\frac{\alpha _s}{\pi} \pqty{\Delta P_{g g} \pqty{z} - C_A L \delta \pqty{1 - z} - \frac{\beta _0}{2} \delta \pqty{1 - z}}
\, , \label{e.lon_hard_function_gamma_gg} \\
\gamma _{gq}^L\pqty{z, p_TR, \mu}
& =
\frac{\alpha _s}{\pi} \Delta P_{q g} \pqty{z}
\, , \label{e.lon_hard_function_gamma_gq} \\
\gamma _{qq}^T\pqty{z, p_TR, \mu}
& =
\frac{\alpha _s}{\pi} \pqty{\Delta _T P_{q q} \pqty{z} - C_F L \delta \pqty{1 - z} - \frac{3 C_F}{2} \delta \pqty{1 - z}}
\, . \label{e.tra_hard_function_gamma_qq}
\end{align}
Obviously, the natural scale for the hard matching functions is $\mu \sim p_T R$ as indicated in the logarithm $L$ in the perturbative results. Therefore, we can resum the large logarithms of jet radius $\ln R$ by evolving the hard matching functions from scale $\mu \sim p_T R$ to the hard scattering scale $\mu \sim p_T$ with the RG equations.
We notice that similar to the unpolarized case studied in \cite{Kang:2017glf}:
\begin{equation}
\gamma _{ij}^{U} \pqty{z, p_TR, \mu}
=
\delta _{ij} \delta (1-z) \Gamma _i \pqty{p_TR, \mu} + \frac{\alpha _s}{\pi} P_{ji}(z)
\, ,
\end{equation}
where $\Gamma _i$ are given by:
\begin{equation}
\label{e.Gamma_part_of_hard_matching_gamma}
\Gamma _q \pqty{p_TR, \mu}
=
\frac{\alpha _s}{\pi} C_F \pqty{- L - \frac{3}{2}}
\, ,
\quad
\Gamma _g \pqty{p_TR, \mu}
=
\frac{\alpha _s}{\pi} C_A \pqty{- L - \frac{\beta _0}{2 C_A}}
\, .
\end{equation}
Following the same exercise, we can express the anomalous dimensions $\gamma _{ij}^{L(T)} \pqty{z, p_TR, \mu}$ in \cref{e.lon_hard_function_gamma_qq,e.lon_hard_function_gamma_qg,e.lon_hard_function_gamma_gg,e.lon_hard_function_gamma_gq,e.tra_hard_function_gamma_qq} into a general form:
\begin{equation}
\gamma _{ij}^{L(T)} \pqty{z, p_TR, \mu}
=
\delta _{ij} \delta (1-z) \Gamma _i \pqty{p_TR, \mu} + \frac{\alpha _s}{\pi} \Delta_{(T)} P_{ji}(z)
\, ,
\end{equation}
where the functions $\Gamma _i \pqty{p_TR, \mu}$ in the first term are independent of polarization and are therefore the same as the unpolarized $\Gamma _i$ in \cref{e.Gamma_part_of_hard_matching_gamma}.

\subsection{TMD FJFs in exclusive jet production}
\label{ss.exclusive_TMD_FJFs}
For the exclusive jet production, exclusive TMD FJFs $\widetilde{\mathscr{D}}_1^{h/i} \pqty{z_h, \boldsymbol{j}_{\perp}, p_T R,  \mu, \zeta_J}$ arises in the factorization formalism. Since one cannot have out-of-jet hard radiation as explained in~\cref{ss.collinear_exclusive_FJFs}, the factorization formalism in the $j_\perp \ll p_T R$ region would be given by
\begin{align}
\widetilde{\mathscr{D}}_1^{h/i} \pqty{z_h, \boldsymbol{j}_{\perp}, p_T R,  \mu, \zeta_J}
& =
\int \dd[2]{\boldsymbol{k}_{\perp}} \dd[2]{\boldsymbol{\lambda}_{\perp}}
\delta ^2 \pqty{z_h \boldsymbol{\lambda}_{\perp} + \boldsymbol{k}_{\perp} - \boldsymbol{j}_{\perp}}
D_1^{h/i(u)} \pqty{z_h, \boldsymbol{k}_{\perp}, \mu, \frac{\zeta}{\nu^2}}
S_i \pqty{\boldsymbol{\lambda}_{\perp}, \mu, \frac{\nu \mathcal{R}}{2}}
\nonumber \\
& = 
\int \frac{\dd[2]{\boldsymbol{b}}}{(2 \pi)^2} e^{i \boldsymbol{j}_{\perp} \cdot \boldsymbol{b} / z_h} \widetilde{D}_1^{h/i} \pqty{z_h, \boldsymbol{b}, \mu, \zeta_J} 
\, ,
\end{align}
where we have used~\cref{e.TMD_proper}. Again for exclusive jet production, we no longer have the dependence on $z$ as in the semi-inclusive TMD FJFs. 
Following the same procedure as before, we can generalize this to the polarized exclusive TMD FJFs. With the operator $\mathscr{C}$ defined in \cref{e.TMD_factorization_operator}, the unpolarized exclusive TMD FJFs are given by:
\begin{equation}
\widetilde{\mathscr{D}}_1^{h/i} \pqty{z_h, p_T {R}, \boldsymbol{j}_{\perp}, \mu, \zeta_J}
=
\mathscr{C} \bqty{\widetilde{D}_1^{h/i}}
\, ,
\end{equation}
Similarly for exclusive TMD FJFs with different polarizations, we have:
\begin{align}
\widetilde{\mathscr{D}}_{1T}^{\perp, h/i} \pqty{z_h, \boldsymbol{j}_{\perp}, p_T R,  \mu, \zeta_J}
& =
\mathscr{C} \bqty{\widetilde{D}_{1T}^{\perp, h/i, (1)}}
\, , \\
\widetilde{\mathscr{G}}_{1L}^{h/i} \pqty{z_h, \boldsymbol{j}_{\perp}, p_T R,  \mu, \zeta_J}
& =
\mathscr{C} \bqty{\widetilde{G}_{1L}^{h/i}}
\, , \label{e.G_1L}\\
\widetilde{\mathscr{G}}_{1T}^{\perp, h/i} \pqty{z_h, \boldsymbol{j}_{\perp}, p_T R,  \mu, \zeta_J}
& =
\mathscr{C} \bqty{\widetilde{G}_{1T}^{\perp, h/i, (1)}}
\, , \\
\widetilde{\mathscr{H}}_1^{h/i} \pqty{z_h, \boldsymbol{j}_{\perp}, p_T R,  \mu, \zeta_J}
& =
\mathscr{C} \bqty{\widetilde{H}_{1}^{h/i}}
\, , \\
\widetilde{\mathscr{H}}_{1}^{\perp, h/i} \pqty{z_h, \boldsymbol{j}_{\perp}, p_T R,  \mu, \zeta_J}
& =
\mathscr{C} \bqty{\widetilde{H}_{1}^{\perp, h/i, (1)}}
\, , \\
\widetilde{\mathscr{H}}_{1L}^{\perp, h/i} \pqty{z_h, \boldsymbol{j}_{\perp}, p_T R,  \mu, \zeta_J}
& =
\mathscr{C} \bqty{\widetilde{H}_{1L}^{\perp, h/i, (1)}}
\, , \\
\widetilde{\mathscr{H}}_{1T}^{\perp, h/i} \pqty{z_h, \boldsymbol{j}_{\perp}, p_T R,  \mu, \zeta_J}
& =
\mathscr{C} \bqty{\widetilde{H}_{1T}^{\perp, h/i, (2)}}
\, .
\end{align}

\section{Phenomenology}
\label{s.phenomenology}
There has been growing phenomenological work for studying hadron distribution inside the jet, in particular in connection with the 3D imaging of the hadrons. For example, for single inclusive jet production in proton-proton collisions as well as electron-proton scatterings, unpolarized and polarized collinear hadron distribution inside the jet have been studied~\cite{Kaufmann:2015hma,Kang:2016ehg,Kang:2020xyq,Zhao:2023lav}. The TMD distribution of hadrons inside the jet produced in transversely polarized proton-proton collisions is sensitive to the Collins TMD FFs and has been studied in~\cite{Kang:2017btw}. For exclusive jet production, such as back-to-back $Z$+jet production in proton-proton collisions and back-to-back lepton+jet production in DIS, TMD hadron distribution inside jets has been studied~\cite{Kang:2019ahe,Kang:2021kpt,Kang:2021ffh,Arratia:2022oxd,Arratia:2020nxw}. 

Before we present the phenomenology, we first provide more details on TMD FFs. In previous sections, we have discussed the evolution of FJFs in the perturbative region, \textit{i.e.}, $1 / b \gg \Lambda _{\mathrm{QCD}}$.
However, to do calculations for phenomenology, we must take care of the non-perturbative evolution of TMDs at the large-$b$ region.
In order to do this, we adopt the $b_*$-prescription \cite{Collins:1984kg}. Alternative approaches can be found in \cite{Kulesza:2002rh, Qiu:2000hf, Catani:2015vma, Ebert:2016gcn, Monni:2016ktx}. The $b_*$ is defined as:
\begin{equation}
b_* \equiv \frac{b}{\sqrt{1 + b^2 / b_{\max}^2}}
\, ,
\end{equation}
where we choose $b_{\max} = 1.5~\mathrm{GeV}^{-1}$.
With this definition, the magnitude of $b_*$ approaches $b$ when $b \ll b_{\max}$, and approaches $b_{\max}$ in the large-$b$ region. For the perturbative contribution, we work at the NLL order, where we keep $\Gamma _{0, 1}^i$ and $\gamma_0^i$ in~\cref{e.D_1_evolution,e.mu-dimension}.
For the non-perturbative contribution, we include the following non-perturbative Sudakov factor \cite{Sun:2014dqm, Echevarria:2020hpy}:
\begin{equation}
\label{e.S_NP_q}
S_{\mathrm{NP}}^q(b, Q_0, \sqrt{\zeta_J})
=
\frac{g_2}{2} \ln(\frac{b}{b_*}) \ln(\frac{\sqrt{\zeta_J}}{Q_0}) + \frac{g_h}{z_h^2} b^2
\, ,
\end{equation}
where $Q_0^2 = 2.4~\mathrm{GeV}^2$, $g_2 = 0.84$ and $g_h = 0.042$, and $\sqrt{\zeta_J} = p_T R$ following~\cref{e.zeta_J}. For gluon TMD FFs, we use \cite{Kang:2017glf}:
\begin{equation}
\label{e.S_NP_g}
S_{\mathrm{NP}}^g(b, Q_0, \sqrt{\zeta_J})
=
\frac{C_A}{C_F} \frac{g_2}{2} \ln(\frac{b}{b_*}) \ln(\frac{\sqrt{\zeta_J}}{Q_0}) + \frac{g_h}{z_h^2} b^2
\, .
\end{equation}

Now that we have properly addressed the non-perturbative evolution, we can proceed with predicting relevant observables. In this section, we provide two additional phenomenological examples. For the case of single inclusive jet production, we compute the transverse momentum distribution for hadrons inside the jet at RHIC kinematics, where a transversely polarized proton collides with an unpolarized proton, producing a jet with a transversely polarized $\Lambda$-baryon inside. This process will be sensitive to both transversity distributions $h_1$ and transversity TMD FJFs $\mathcal{H}_1$. For the exclusive jet production, we study back-to-back lepton jet production in lepton-proton collisions, where the incoming proton is transversely polarized and the final hadron inside the jet is longitudinally polarized. This observable is sensitive to the ``worm-gear'' TMD PDFs $g_{1T}$ and longitudinally polarized TMD FJFs $\mathcal{G}_{1L}$. 

\subsection{Transverse spin transfer to $\Lambda$ in jet in polarized $pp$ collisions}

\begin{figure}[htb]
\centering
\includegraphics[width = 0.7 \textwidth]{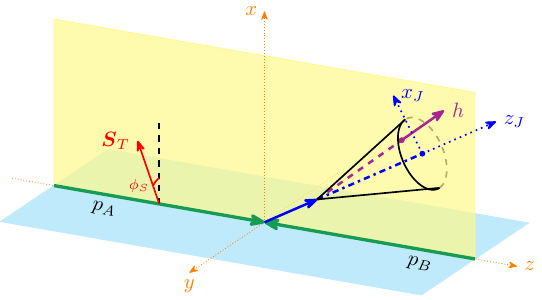}
\caption{Illustration for the distribution of hadrons inside jets in the collisions of a polarized proton and an unpolarized proton or lepton.
The reaction plane is painted yellow.
For detailed illustration of jet cone and hadron kinematics, please see \cref{f.jet_cone_with_hadron}.}
\label{f.inclusive_jet_kinematics}
\end{figure}

For semi-inclusive process, we consider the $pp$ collision in its center-of-mass frame:
\begin{equation}
p \pqty{P_A, S_A} + p \pqty{P_B}
\to
\bqty{\mathrm{jet}(p_J) h \pqty{z_h, \boldsymbol{j}_{\perp}, S_h}} + X
\, ,
\end{equation}
where $P_A$ and $P_B$ are the momenta of the incident protons, $S_A$ is the polarization of the proton $A$.
They are defined as:
\begin{align}
P_A
& =
\frac{\sqrt{s}}{2} \pqty{1, 0, 0,  1} ,
\quad
P_B
=
\frac{\sqrt{s}}{2} \pqty{1, 0, 0, -1}
\, , \\
S_A
& =
\bqty{-\lambda_A \frac{M_A}{p_A^-}, \lambda_A \frac{p_A^-}{M_A}, \boldsymbol{S}_T}
\, ,
\end{align}
where $s$ is the squared proton center of mass energy.
Notice that $S_A$ is written in light-cone coordinate defined in \cref{e.light_cone_definition}, and $\lambda_A$ is the helicity of the incoming proton with mass $M_A$, $p_A^-$ is the large light-cone momentum of the proton $A$, $\phi _{S}$ is the azimuthal angle between the transverse spin $\boldsymbol{S}_{T}$ of the initial proton and the reaction plane, with the reaction plane defined by the incoming beam direction and jet axis direction (painted yellow in \cref{f.inclusive_jet_kinematics}).
For the jet production, $p_J$ is the jet momentum, with $p_T$ being the magnitude of its transverse component:
\begin{equation}
p_J = \pqty{p_T \cosh{\eta}, p_T, 0, p_T \sinh{\eta}}
\, .
\end{equation}
Note here the $y$-component of $p_J$ is zero because we put the jet within the reaction plane ($xz$-plane).
The hadron inside the jet has a transverse momentum $\boldsymbol{j}_{\perp}$ with respect to the jet axis, and collinear momentum fraction $z_h$ of the jet momentum.
Together with the hadron's transverse spin $\boldsymbol{S}_{h \bot}$, they are given by:
\begin{align}
\boldsymbol{j}_{\perp}
& =
\abs{\boldsymbol{j}_{\perp}}
\pqty{\cos{\hat{\phi}_h} \cos{\theta _J}, \sin{\hat{\phi}_h}, -\cos{\hat{\phi}_h} \sin{\theta _J}}
\, , \label{e.j_T_parameterization} \\
\boldsymbol{S}_{h \bot}
& =
\abs{\boldsymbol{S}_{h \bot}}
\pqty{\cos{\hat{\phi}_{S_h}} \cos{\theta _J}, \sin{\hat{\phi}_{S_h}}, -\cos{\hat{\phi}_{S_h}} \sin{\theta _J}}
\, , \label{e.S_hT_parameterization}
\end{align}
where $\theta_J$ is the angle between jet axis and $z$-axis, $\hat{\phi}_h$ is the azimuthal angle between the momentum $p_h$ of the produced hadron and the reaction plane, and $\hat{\phi}_{S_h}$ is the azimuthal angle between the transverse spin $\boldsymbol{S}_{h \bot}$ and the reaction plane.
The ``hat'' that decorates $\hat{\phi}_h$ and $\hat{\phi}_{S_h}$ indicates that these angles are measured in the jet coordinates.

For single inclusive jet production, the most general azimuthal dependence
\footnote{Here we consider the situation where only one of the incoming hadrons could be polarized and the final-state hadron can be polarized.} for the hadron $z_h$ and $\boldsymbol{j}_\perp$ distributions inside the jet is given by~\cite{Kang:2020xyq}:
\begingroup
\allowdisplaybreaks
\begin{align}
\frac{\dd{\sigma ^{p \pqty{S_A} + p \to \mathrm{jet} h \pqty{S_h} + X}}}{\dd{\eta} \dd[2]{\boldsymbol{p}_T} \dd{z_h} \dd[2]{\boldsymbol{j}_{\perp}}}
& =
F_{UU, U} + \abs{\boldsymbol{S}_T} \sin(\phi _{S} - \hat{\phi}_h) F^{\sin(\phi _{S} - \hat{\phi}_h)}_{TU, U}
\nonumber \\
& \quad +
\lambda_h \bqty{\lambda_A F_{LU, L} + \abs{\boldsymbol{S}_T} \cos(\phi _{S} - \hat{\phi}_h) F^{\cos(\phi _{S} - \hat{\phi}_h)}_{TU, L}}
\nonumber \\
& \quad +
\abs{\boldsymbol{S}_{h_T}} \bigg[\sin(\hat{\phi}_h - \hat{\phi}_{S_h}) F^{\sin(\hat{\phi}_h - \hat{\phi}_{S_h})}_{UU, T}
+
\lambda_A \cos(\hat{\phi}_h - \hat{\phi}_{S_h}) F^{\cos(\hat{\phi}_h - \hat{\phi}_{S_h})}_{LU, T}
\nonumber \\
& \qquad \qquad \quad +
\abs{\boldsymbol{S}_T} \bigg(\cos(\phi _{S} - \hat{\phi}_{S_h}) F^{\cos(\phi _{S} - \hat{\phi}_{S_h})}_{TU, T}
\nonumber \\
& \qquad \qquad \qquad \quad +
\cos(2 \hat{\phi}_h - \phi _{S} - \hat{\phi}_{S_h}) F^{\cos(2 \hat{\phi}_h - \phi _{S} - \hat{\phi}_{S_h})}_{TU, T}\bigg) \bigg]
\, ,
\end{align}
\endgroup
where $F_{AB, C}$ denotes the spin-dependent structure functions, with $A$, $B$ and $C$ indicating the polarization of incoming proton $A$, incoming proton $B$ (or electron), and the fragmented hadron inside the jet.
$\lambda_A$ and $\abs{\boldsymbol{S}_{\perp}}$ are the longitudinal and transverse spin of the initial polarized proton $A$, while $\lambda_h$ and $\abs{\boldsymbol{S}_{h_T}}$ are the longitudinal and transverse polarization of the hadron inside jet measured in the fragmenting parton helicity frame.

The structure function $F_{UU, U}(z_h, j_{\perp})$ is defined by:
\begin{align}
F_{UU, U} (z_h, j_{\perp})
& =
\frac{\alpha _s^2}{s} \sum _{a, b, c}
\int _{x_1^{\min}}^1 \frac{\dd{x_1}}{x_1} f_1^{a/A}(x_1, \mu)
\int _{x_2^{\min}}^1 \frac{\dd{x_2}}{x_2} f_2^{b/B}(x_2, \mu)
\nonumber \\
& \quad \times
\int _{z^{\min}}^1 \frac{\dd{z}}{z^2}
\hat{\sigma}_{ab}^c \pqty{\hat{s}, \hat{p}_T, \hat{\eta}, \mu}
\mathcal{D}_1^{h/c} \pqty{z, z_h, \boldsymbol{j}_{\perp}, p_TR, \mu, \zeta_J}
\nonumber \\
& \equiv
\mathcal{C} \bqty{f f \mathcal{D}_1 \hat{\sigma}}
\, , \label{e.F_UUU}
\end{align}
where $f_1^{a/A} (x_1, \mu)$ and $f_2^{b/B} (x_2, \mu)$ are the collinear unpolarized PDFs in the proton with corresponding momentum fractions $x_1$ and $x_2$, and $\hat{\sigma}_{ab}^c$ is the hard function for unpolarized parton $a$, $b$ to unpolarized parton $c$. On the other hand, $\mathcal{D}_1^{h/c} \pqty{z, z_h, \boldsymbol{j}_{\perp}, p_TR, \mu, \zeta_J}$ are the unpolarized TMD FJFs in~\cref{e.TMD_b_space} and we have studied extensively in the previous section. The lower integration limits $x_1^{\min}$, $x_2^{\min}$ and $z^{\min}$ can be found in \cite{Kaufmann:2015hma, Kang:2016ehg}.
The variables $\hat{s}$, $\hat{p}_T$ and $\hat{\eta}$ are the squared parton center of mass energy, transverse momentum and rapidity of parton $c$, respectively, and are related to their hadron analogues as:
\begin{equation}
\hat{s} = x_1 x_2 s
\, ,
\quad
\hat{p}_T = p_T / z
\, ,
\quad
\hat{\eta} = \eta - \frac{1}{2} \ln(\frac{x_1}{x_2})
\, .
\end{equation}
In \cref{e.F_UUU}, we also defined the notation $\mathcal{C} \bqty{f f \mathcal{D}_1 \hat{\sigma}}$, where parton flavors are summed for PDFs and FJFs along with their corresponding unpolarized hard functions. Using this notation, we can write down the expressions for all the polarized structure functions:
\begin{align}
F^{\sin(\phi _S - \hat{\phi}_h)}_{TU, U} (z_h, j_{\perp})
& =
\mathcal{C} \bqty{\frac{j_{\perp}}{z_h M_h} h_{1} f_1 \mathcal{H}_{1}^{\perp} \Delta_T \hat{\sigma}}
\, , \\
F_{LU, L} (z_h, j_{\perp})
& =
\mathcal{C} \Big[g_{1 L} f_1 \mathcal{G}_{1 L} \Delta \hat{\sigma} \Big]
\, , \\
F^{\cos(\phi _S - \hat{\phi}_h)}_{TU, L} (z_h, j_{\perp})
& =
- \mathcal{C} \bqty{\frac{j_{\perp}}{z_h M_h} h_1 f_1 \mathcal{H}_{1 L}^{\perp} \Delta _T \hat{\sigma}}
\, , \\
F^{\sin(\hat{\phi}_h - \hat{\phi}_{S_h})}_{UU, T} (z_h, j_{\perp})
& =
- \mathcal{C} \bqty{\frac{j_{\perp}}{z_h M_h} f_1 f_1 \mathcal{D}_{1 T}^{\perp} \hat{\sigma}}
\, , \\
F^{\cos(\hat{\phi}_h - \hat{\phi}_{S_h})}_{LU, T} (z_h, j_{\perp})
& =
- \mathcal{C} \bqty{\frac{j_{\perp}}{z_h M_h} g_{1 L} f_1 \mathcal{G}_{1 T} \Delta \hat{\sigma}}
\, , \\
F^{\cos(\phi _S - \hat{\phi}_{S_h})}_{TU, T} (z_h, j_{\perp})
& =
\mathcal{C} \Big[h_1 f_1 \mathcal{H}_{1} \Delta _T \hat{\sigma} \Big]
\, , \label{e.F_TUT} \\
F^{\cos(2 \hat{\phi}_h - \phi _S - \hat{\phi}_{S_h})}_{TU, T} (z_h, j_{\perp})
& =
- \mathcal{C} \bqty{\frac{j^2_{\perp}}{2 z_h^2 M_h^2} h_{1} f_1 \mathcal{H}_{1 T}^{\perp} \Delta_T \hat{\sigma}}
\, .
\end{align}
The corresponding polarizations for hard functions $\hat{\sigma}_{ab}^c$, $\Delta\hat{\sigma}_{ab}^c$ and $\Delta _T\hat{\sigma}_{ab}^c$ are given in \cref{t.hard_functions}. The expressions for $F_{UU, U}$ and $F^{\sin(\hat{\phi}_h - \hat{\phi}_{S_h})}_{UU, T}$ were given in the previous publication~\cite{Kang:2020xyq} while~\cite{Kang:2017btw} performed a phenomenological study for $F^{\sin(\phi _S - \hat{\phi}_h)}_{TU, U}$, that is related to the Collins TMD FFs. The detailed expressions for all the other structure functions are written down here for the first time. 
\begin{table}
\centering
\begin{tabular}{|c|c|c|c|}
\hline
\diagbox{$c$}{$a$} & U & L & T \\
\hline
U & $\hat{\sigma}_{ab}^c$ & & \\
\hline
L &  & $\Delta \hat{\sigma}_{ab}^c$ & \\
\hline
T & & & $\Delta _T \hat{\sigma}_{ab}^c$ \\
\hline
\end{tabular}
\caption{Hard functions for parton $a$, $b$ to parton $c$.
The header row represents the polarization of the parton $a$ while the header column indicates the polarizations of the parton $c$.
Parton $b$ is always unpolarized.
}
\label{t.hard_functions}
\end{table}

Relevant measurements have been performed for the $\pi^{\pm}$-in-jet production in the case of Collins FFs \cite{STAR:2022hqg}.
More relevantly, a measurement of transverse spin transfer to $\Lambda/\overline{\Lambda}$ hyperons in polarized $pp$ collisions (no jet is involved) was performed by the STAR collaboration at RHIC~\cite{STAR:2018fqv}. This suggests that measuring the transverse polarization of $\Lambda/\overline{\Lambda}$ hyperons inside the jet in transversely polarized $pp$ collisions could be achieved at RHIC. This process would correspond to the structure function $F^{\cos(\phi _S - \hat{\phi}_{S_h})}_{TU, T}$, which is sensitive to both transversity PDFs $h_1$ and transversity TMD FJFs $\mathcal{H}_{1}^{h/q}$. Below, we focus on this structure function and define the following spin asymmetry for $\Lambda$-in-jet productions:
\begin{align}
A_{TU, T}^{\cos(\phi _S - \hat{\phi}_{S_h})}
\equiv
F_{TU, T}^{\cos(\phi _S - \hat{\phi}_{S_h})} \bigg/ F_{UU, U} \,.
\end{align}
 In order to compute $F_{TU, T}^{\cos(\hat{\phi}_h - \hat{\phi}_{S_h})}$, we will need to parameterize the collinear transversity PDF $h_1(x, \mu)$, which appears in the initial state of $F_{TU, T}^{\cos(\phi _S - \hat{\phi}_{S_h})}$ in \cref{e.F_TUT}.
For this, we follow the parameterization in \cite{Kang:2015msa} and write the quark transversity as:
\begin{equation}
h_1^q \pqty{x, Q_0}
=
N_q^h x^{a_q} \pqty{1 - x}^{b_q}
\frac{\pqty{a_q + b_q}^{a_q + b_q}}{a_q^{a_q} b_q^{b_q}}
\frac{1}{2} \pqty{f_1^q \pqty{x, Q_0} + g_1^q \pqty{x, Q_0}}
\, ,
\end{equation}
at initial scale $Q_0 = 1.27~\mathrm{GeV}$, for up and down quarks $q = u$ and $d$ only.
The $f_1^q$ and $g_1^q$ are the collinear unpolarized and helicity parton distributions respectively, for which we use the parameters from \cite{Cocuzza:2022jye}. Lastly, in the generation of TMD FFs grids, we also need the transversity FFs.
In this work we use the parameterization from \cite{Kang:2021kpt}, which assumes a simple normalization of the collinear unpolarized FFs:
\begin{equation}
\label{e.transversity_FFs_parameterization}
H_{\Lambda / q} \pqty{z, Q}
=
N_q^H D_{\Lambda / q} \pqty{z, Q}
\, .
\end{equation}
Here $H_{\Lambda / q}$ and $D_{\Lambda / q}$ are the transversity and unpolarized FFs of the $\Lambda$ baryon, and $N_q^H$ is the fitted parameter. 
We also follow the assumptions made in \cite{Callos:2020qtu, Kang:2021kpt} and demand:
\begin{equation}
D_{\Lambda / q}
=
D_{\overline{\Lambda} / q}
=
\frac{1}{2} D_{\Lambda / \overline{\Lambda} \leftarrow q}
\, ,
\end{equation}
as for the unpolarized $\Lambda$ baryon FFs that appear in the denominator of $A_{TU,T}^{\cos(\phi _S - \hat{\phi}_{S_h})}$, we use the AKK parametrization~\cite{Albino:2005mv}. For both the transversity PDFs and transversity FFs, we also vary their fitted parameters according to the extracted uncertainties, weighted by Gaussian random number with a mean of 0 and variance of 1.

Using the above setup, we now present the first prediction for the spin asymmetry $A_{TU,T}^{\cos(\phi _S - \hat{\phi}_{S_h})}$.
We choose the jet kinematics consistent with the available data from STAR \cite{STAR:2022hqg}, with the center of mass energy at $\sqrt{s} = 200~\mathrm{GeV}$, rapidity ranges $\eta \in \pqty{0, 0.9}$ and $\eta \in \pqty{-0.9, 0}$.
The jets are reconstructed using the anti-$k_T$ algorithm with radius $R = 0.6$.
The numerical implementation is very similar to that used in the longitudinal momentum distribution of hadrons inside jets \cite{Kang:2016ehg}, with the numerical DGLAP evolution tool developed in \textsc{Pegasus} \cite{Vogt:2004ns}.
The RG evolution of the various parts of the cross sections is performed as outlined in \cite{Kang:2017glf}.

\begin{figure}
\centering
\includegraphics[width = 0.95 \textwidth]{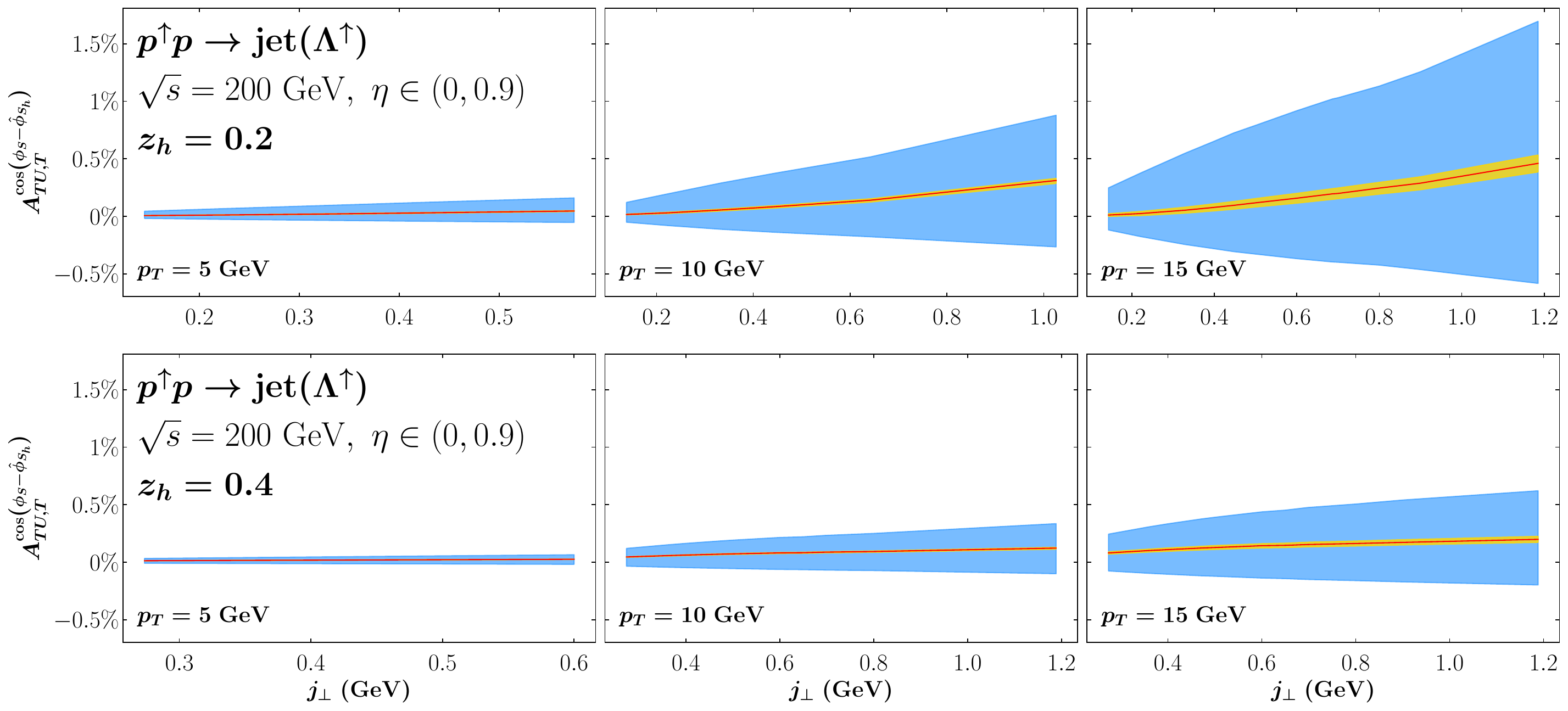}
\caption{
Hadron $j_{\perp}$-distribution within jets in $pp$ collisions at $\sqrt{s} = 200~\mathrm{GeV}$.
Jets are detected in the rapidity interval $\eta \in \pqty{0, 0.9}$ (\textit{i.e.}, the jet is scattered forward relative to the polarized beam), and are reconstructed using the anti-$k_T$ algorithm with $R = 0.6$.
We vary the jet transverse momentum from 5 to 15 GeV, and choose average $z_h$ at 0.2 (top row) and 0.4 (bottom row), which is within the typical kinematic region according to \cite{STAR:2022hqg}.
The blue band is obtained by varying the parameters for transversity PDFs and transversity FFs within their fitted uncertainties.
While the yellow band is obtained by keeping the fitted parameters of transversity FFs fixed at their central values and varying only the parameters of transversity PDFs.
The red lines show the theory obtained from the central of the fitted parameters.
}
\label{f.A_UT_positive_eta}
\end{figure}

In \cref{f.A_UT_positive_eta}, the prediction to $A_{TU,T}^{\cos(\phi _S - \hat{\phi}_{S_h})}$ for the jet$(\Lambda)$ production at RHIC kinematics is presented.
The blue band is obtained by varying the fitted parameters of both transversity PDFs and transversity FFs, while the yellow band is obtained by varying the parameters of transversity PDFs only (with parameters for transversity FFs fixed at central values).
We can therefore deduce that the uncertainty of theoretical prediction for the observable $A_{TU,T}^{\cos(\phi _S - \hat{\phi}_{S_h})}$ comes largely from the transversity FFs, \textit{i.e.}, variation of $N_q^H$ in \cref{e.transversity_FFs_parameterization}.
Given the large uncertainty resulting from the currently available extraction of the transversity FFs for $\Lambda$ baryon, this observable, when measured in the future, can be used to constrain the uncertainty on the $\Lambda$ transversity FFs.
\begin{figure}
\centering
\includegraphics[width = 0.95 \textwidth]{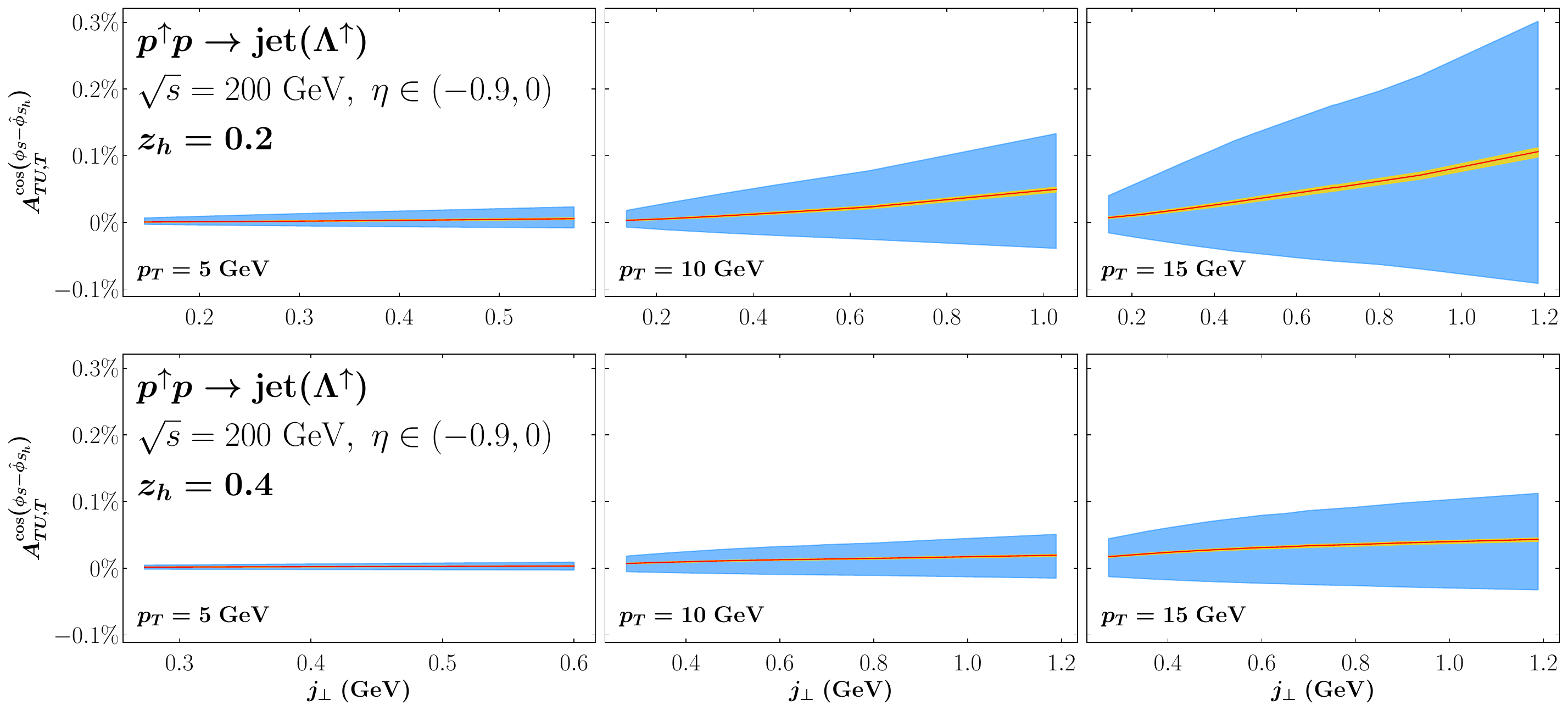}
\caption{Same plot as \cref{f.A_UT_positive_eta}, but with jet scattered backward relative to the polarized beam.}
\label{f.A_UT_negative_eta}
\end{figure}
In \cref{f.A_UT_negative_eta}, the same predictions are presented but with a negative rapidity range.
Similar patterns occur as in \cref{f.A_UT_positive_eta}, but with a much smaller magnitude, indicating a much smaller probability of producing polarized final particles in the opposite direction of the polarized beam.

Overall, in proton-proton collisions for $\Lambda$-in-jet production, where one of the proton beams is transversely polarized, and the resulting Lambda particles in the jet also exhibit transverse polarization, a distinct observation emerges.
Specifically, in the forward region (positive rapidity), a moderate spin asymmetry $A_{TU,T}^{\cos(\phi _S - \hat{\phi}_{S_h})}\sim 1.5\%$ can be observed at RHIC kinematics, while in the backward region (negative rapidity), the spin asymmetry $A_{TU,T}^{\cos(\phi _S - \hat{\phi}_{S_h})}$ appears to be very small. This is largely due to the probing $x$ range for the transversity PDF $h_1(x, \mu)$: in the forward rapidity region $x\sim 0.2$ at $p_T\sim 15$ GeV where $h_1(x, \mu)$ is sizable~\cite{Kang:2015msa}; in the backward rapidity region $x\sim 0.05$ for $p_T\sim 15$ GeV where $h_1(x, \mu)$ is very small. 

\subsection{Back-to-back electron-jet production with longitudinally polarized $\Lambda$ in jet}
\label{ss.exclusive_phenomenology}
\begin{figure}[htb]
\centering
\includegraphics[width = 0.67 \textwidth]{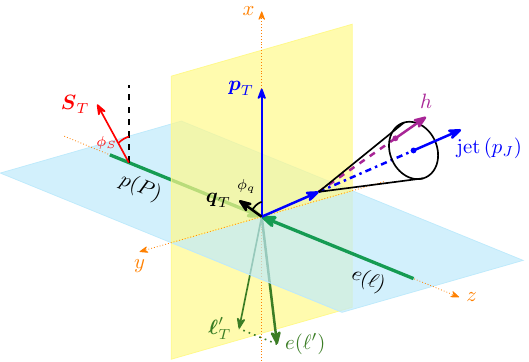}
\caption{Illustration for the distribution of hadrons inside jets in the collisions of a polarized proton and an unpolarized lepton.
For a detailed illustration of jet and hadron kinematics, please see \cref{f.jet_cone_with_hadron}.}
\label{f.exclusive_jet_kinematics}
\end{figure}

For exclusive jet production, we choose to study back-to-back electron-jet production in deep inelastic $ep$ scattering:
\begin{equation}
\label{e.ep_to_ejet}
e \pqty{\ell , \lambda _e} + p \pqty{P, S}
\to
e \pqty{\ell '} + \bqty{\mathrm{jet}(p_J) h\pqty{z_h, \boldsymbol{j}_{\perp}, S_h}} + X
\, ,
\end{equation}
where $\ell$ and $\ell '$ are the momenta of the incident and scattered lepton respectively.
The incident electron could carry a helicity $\lambda _e$ and the scattered lepton is assumed to be unpolarized but with a transverse momentum $\boldsymbol{\ell}'_{T}$.
Here we use $P$ and $S$ to represent the momentum and polarization of the incoming proton, and the jet has a momentum $p_J$ with transverse component $\boldsymbol{p}_T$.
The hadron inside the jet has a transverse momentum $\boldsymbol{j}_{\perp}$ with respect to the jet axis and collinear momentum fraction $z_h$ of the jet, as well as a polarization $\boldsymbol{S}_h$ with respect to the jet axis.
The kinematics is shown in~\cref{f.exclusive_jet_kinematics}. In the back-to-back region where $\abs{\boldsymbol{q}_T} \ll \abs{\boldsymbol{p}_T}$ with the transverse momentum imbalance $\boldsymbol{q}_T \equiv \boldsymbol{\ell}'_{T} + \boldsymbol{p}_T$, TMD factorization formalism for the full differential cross section has been given in the previous publication~\cite{Kang:2021ffh}, where various structure functions were defined and studied. For example, the structure function $F_{UU,U}^{\cos(\phi_q - \hat{\phi}_h)}$ for $\pi^{\pm}$-in-jet production is sensitive to Boer-Mulder function and Collins function, while $F_{UU,T}^{\cos(\hat{\phi}_h - \hat{\phi}_{S_h})}$ from $\Lambda$-in-jet production has the sensitivity to the polarizing TMD FFs $D_{1T}^{\bot}$. At the same time, in \cite{Arratia:2020nxw}, the structure function $F_{TU,U}^{\,\sin(\phi_{S} - \hat{\phi}_h)}$ is computed for $\pi^{+/-}$-in-jet production, and is found to be able to further constrain the Collins FFs.

In this section, we studied yet another observable - the spin asymmetry $A_{TU,L}^{\cos(\phi _q - \phi _{S})}$ defined as
\begin{align} \label{e.A_LT}
A_{TU,L}^{\cos(\phi _q - \phi _{S})}
\equiv
\frac{F_{TU, L}^{\cos(\phi _q - \phi _{S})}}{F_{UU, U}}
\, ,
\end{align}
where the structure $F_{TU, L}^{\cos(\phi _q - \phi _{S})}$ in the numerator denotes the spin-dependent structure functions, with $T$, $U$ and $L$ being the polarization of incoming proton, incoming electron, and the final-state hadron inside the jet, respectively. Here, $\phi _{S}$ is the angle between the transverse spin of the incoming proton $\boldsymbol{S}_T$ and the $x$-axis, while $\phi _q$ is the azimuthal angle between the transverse momentum imbalance $\boldsymbol{q}_T$ and the $x$-axis, see~\cref{f.exclusive_jet_kinematics}.
Following~\cite{Kang:2021ffh}, this structure function can be factorized as follows:
\begin{align}
F_{TU, L}^{\cos(\phi _q - \phi _{S})} (z_h, j_{\perp})
& =
\hat{\sigma}_0 H \pqty{Q, \mu}
\sum _{q} e_q^2 \mathscr{G}_{1L}^{h/q} \pqty{z_h, \boldsymbol{j}_{\perp}, p_T R,  \mu, \zeta_J}
\nonumber \\
&\times
\int \frac{b^2 \dd{b}}{2 \pi}
J_1 \pqty{q_T b} x M\,\widetilde{g}_{1T}^{(1)q} \pqty{x, \boldsymbol{b}, \mu, \zeta}
\overline{S}_{\mathrm{global}} \pqty{b^2 , \mu}
\overline{S}_{\mathrm{cs}} \pqty{b^2 , R , \mu}
\, , \label{e.F_TUL}
\end{align}
where the global soft function $\overline{S}_{\mathrm{global}}$ and the collinear-soft function $\overline{S}_{\mathrm{cs}}$ are given in~\cite{Kang:2021ffh}. The bar in $\overline{S}_{\mathrm{global}/\mathrm{cs}}$ indicates that their dependence on azimuthal angle is averaged. $\hat{\sigma}_0$ is the partonic cross section of electron-quark scattering:
\begin{equation}
\hat{\sigma}_0
=
\frac{\alpha _{\mathrm{em}} \alpha _s}{\hat{s} Q^2}
\frac{2 \pqty{\hat{u}^2 + \hat{s}^2}}{\hat{t}^2}
\, ,
\end{equation}
with the Mandelstam variables $\hat{s} \equiv \pqty{xP + \ell}^2$, $\hat{t} \equiv \pqty{\ell - \ell '}^2$ and $\hat{u} \equiv \pqty{xP - \ell '}^2$, where $x$ is the fraction of collinear momentum carried by the parton inside the incoming proton.
The electromagnetic and strong coupling constants are denoted as $\alpha _{\mathrm{em}}$ and $\alpha _s$, and the virtuality of the exchanged photon is defined as $Q^2 \equiv - \pqty{\ell - \ell '}^2 = - \hat{t}$.

It is evident from~\cref{e.F_TUL} that this spin asymmetry $A_{TU,L}^{\cos(\phi _q - \phi _{S})}$ has sensitivity on both ``worm-gear'' TMD PDFs $\widetilde{g}_{1T}^{q} \pqty{x, \boldsymbol{b}, \mu, \zeta}$, as well as the longitudinally polarized exclusive TMD FJFs $\mathscr{G}_{1L} \pqty{z_h, \boldsymbol{j}_{\perp}, p_T R,  \mu, \zeta_J}$ in~\cref{e.G_1L}, which is connected to the helicity TMD FFs $G_{1L}^{h/q}$. The observable describes the following scattering process: a longitudinally polarized quark inside the transversely polarized proton (as denoted by the TMD PDFs $g_{1T}^q$) scatters with an unpolarized electron and then initiates a jet in which a longitudinally polarized hadron is observed.
To generate a prediction, for the first moment of ``worm-gear'' distribution $\widetilde{g}_{1T}^{(1)q}$, we adopt its parameterization in $\boldsymbol{k}_{\perp}$-space from \cite{Bhattacharya:2021twu}, and Fourier transform it to the conjugate $\boldsymbol{b}$-space:
\begin{equation}
\widetilde{g}_{1T}^{(1)q} \pqty{x, \boldsymbol{b}, \mu, \zeta}
=
\frac{2\pi}{M^2} \int \dd{k}_{\perp}\frac{{k}_{\perp}^2}{b}J_1({k}_{\perp}b) g_{1T}^{q} (x, \boldsymbol{k}_{\perp}^2)
= g_{1T}^{\pqty{1} q} \pqty{x}
\exp(-\frac{1}{4} b^2 \langle k_{\perp}^2\rangle \big|_{g_{1T}^q})
\, ,
\end{equation}
where $M \approx 938.272~\mathrm{MeV}$ is the proton mass and $g_{1T}^{\pqty{1}q} \pqty{x}$ is the collinear function that is parameterized at $Q_0 = 2~\mathrm{GeV}$ \cite{Bhattacharya:2021twu}. Note that this is a simple parton model extraction without TMD evolution, we thus also do not take into account TMD evolution for $\widetilde{g}_{1T}^{q}$. Since CT10 PDFs \cite{Lai:2010vv} are used in the extraction of $\widetilde{g}_{1T}^{q}$, we will use them in the computation of $F_{UU, U}$ as well. For the final-state hadron, we observe the longitudinally polarized $\Lambda/\overline{\Lambda}$ production. The corresponding helicity TMD FFs is given by:
\begin{align}
\widetilde{G}_{1L}^{\Lambda/q} \pqty{z_h, \boldsymbol{b}, \mu, \zeta_J}
=
{G}_{1L}^{\Lambda/q} \pqty{z_h, \mu_{b_{*}}} 
\exp[\int_{\mu_{b_*}}^\mu \frac{\dd{\mu'}}{\mu'} \gamma_\mu^i \bqty{\alpha_s(\mu), \zeta_J/\mu^2}]
S_{NP}^q \pqty{b, Q_0, \sqrt{\zeta_J}}
\, ,
\end{align}
where we use the TMD evolution in \cref{e.D_1_evolution} and we assume the non-perturbative Sudakov is the same as the unpolarized case in~\cref{e.S_NP_q}~\cite{Koike:2006fn}. For the collinear helicity FFs for $\Lambda/\overline{\Lambda}$, we take the parameterizations from \cite{deFlorian:1997zj}, in which three scenarios are provided, each with a different assumption for the source of contribution to $\Lambda$ longitudinal polarization. Specifically,

\begin{itemize}
\item Scenario 1: only polarized $s$ quark contributes;
\item Scenario 2: $u$ and $d$ quarks have equal contributions, while the distribution of $s$ quark has an opposite sign relative to $u$ and $d$ quarks; $u$ and $d$ helicity FFs are negative
\item Scenario 3: $u$, $d$, and $s$ quarks have the same distribution functions, which are all positive.
\end{itemize}

\begin{figure}[htb]
\centering
\hfil
\includegraphics[width = 0.47 \textwidth]{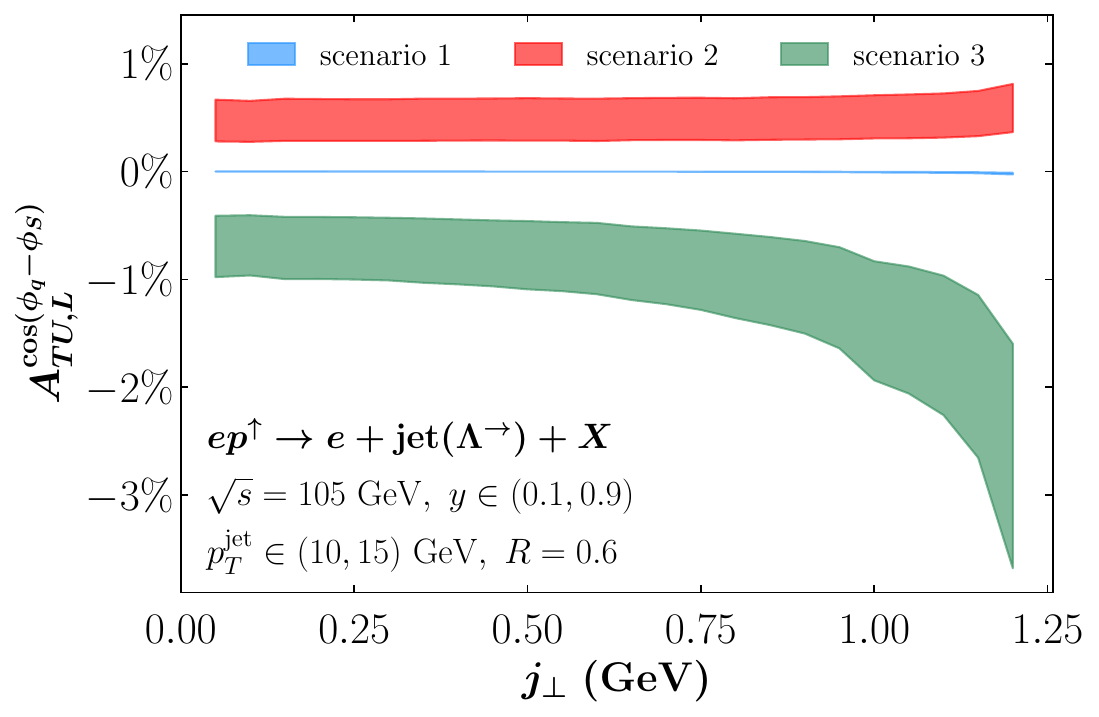}
\hfil
\includegraphics[width = 0.47 \textwidth]{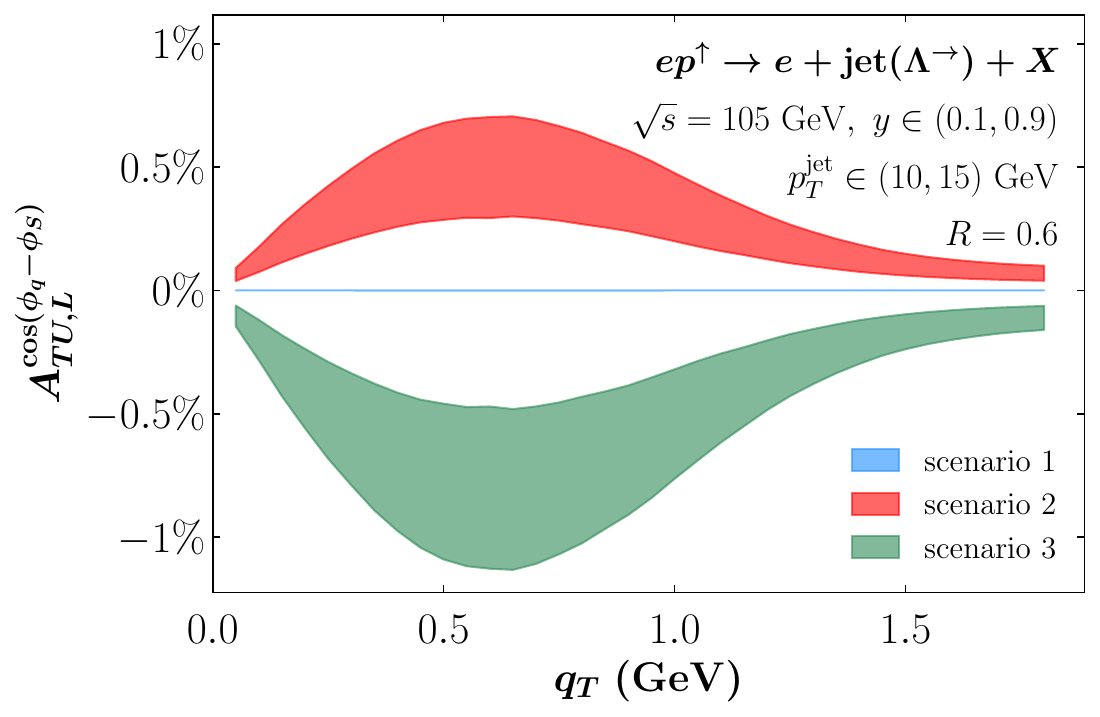}
\hfil
\caption{Longitudinally polarized hadron $j_{\perp}$-distribution (left) and $q_T$-distribution (right) within jets in $ep$ collisions at $\sqrt{s} = 105~\mathrm{GeV}$.
The event inelasticity $y$ is constrained within $\pqty{0.1, 0.9}$.
Jets are detected with transverse momentum $p_T \in \pqty{10, 15}~\mathrm{GeV}$, and are reconstructed using the anti-$k_T$ algorithm with $R = 0.6$.
For the $j_{\bot}$-distribution, $q_T$ is chosen to be 0.5 GeV, while for the $q_T$-distribution, $j_{\bot}$ is chosen to be 0.5 GeV.
The uncertainty band is obtained by computing the 1-$\sigma$ band of the fitted parameters for the ``worm-gear'' PDFs $g_{1T}$, and the three scenarios for $\Lambda/\overline{\Lambda}$ are shown by blue (scenario 1), red (scenario 2) and green (scenario 3).
We collectively denote $\Lambda/\overline{\Lambda}$ as $\Lambda$.
Scenario 1 is consistent with zero because the current extraction of the strange quark worm-gear function is zero.}
\label{f.A_TL}
\end{figure}

With the above setup, we are able to make predictions for the exclusive process $ep \to e + \mathrm{jet}\pqty{\Lambda/\overline{\Lambda}}+X$, in which we measure the longitudinal polarization of the final state $\Lambda/\overline{\Lambda}$ inside the jet. In \cref{f.A_TL}, we plot the spin asymmetry $A_{TU, L}^{\cos(\phi _q - \phi _{S})}$ defined in \cref{e.A_LT} with EIC kinematics, where $\sqrt{s} = 105~\mathrm{GeV}$, jet transverse momentum $p_T \in \pqty{10, 15}~\mathrm{GeV}$, and jet radius parameter $R = 0.6$.
We also introduce the quantity event inelasticity $y = 1 - \frac{\ell_T'}{\sqrt{s}} e^{- y_e} \in \pqty{0.1, 0.9}$, where $\ell_T'$ and $y_e$ are the measured transverse momentum and rapidity of the scattered electron. Constraining the event inelasticity ensures that our calculation stays in the region with reasonable resolution on $x$ and $Q^2$, as well as avoiding the phase where QED radiative corrections are important \cite{Arratia:2020nxw}. On the left panel, we plot $A_{TU, L}^{\cos(\phi _q - \phi _{S})}$ as a function of hadron transverse momentum $j_{\bot}$, while on the right panel, we plot the same spin asymmetry as a function of the electron-jet transverse momentum imbalance $q_T$. For the $j_{\bot}$-distribution (left), $q_T$ is chosen to be 0.5 GeV, while for the $q_T$-distribution (right), $j_{\bot}$ is chosen to be 0.5 GeV. The uncertainty band is obtained by computing the 1-$\sigma$ band of the fitted parameters for the ``worm-gear'' PDFs $g_{1T}$, and the three scenarios for the helicity fragmentation functions of $\Lambda/\overline{\Lambda}$ are shown by blue (scenario 1), red (scenario 2) and green (scenario 3).
We collectively denote $\Lambda/\overline{\Lambda}$ as $\Lambda$.

As can be inferred from \cref{f.A_TL}, the different scenarios for helicity FFs provided in \cite{deFlorian:1997zj} can result in drastically different values in $A_{TU, L}^{\cos(\phi _q - \phi _{S})}$.
Notably, Scenario 1 is consistent with zero because only the $s$ quark helicity FFs contribute in this scenario. However, the current extraction of the ``worm-gear'' function ${g}_{1T}^{(1)q} (x)$ of the incident proton receives contributions solely from $u$ and $d$ quarks, and the $s$ quark worm-gear function vanish~\cite{Bhattacharya:2021twu}. As for Scenarios 2 and 3, distinct measurable asymmetries emerge with different signs, simply because in Scenario 2, both $u$ and $d$ helicity FFs are both negative, while they are positive in Scenario 3. This observable thus enables the discrimination of previously indistinguishable helicity fragmentation functions for $\Lambda/\overline{\Lambda}$ from different quark flavors.

\section{Conclusion}
\label{s.conclusion}

The study of fragmenting jet functions (FJFs) has become a topic of paramount importance in high-energy physics, particularly at the LHC and RHIC, where measurements have been made for a broad range of identified particles within jets.
These studies not only provide crucial information at the LHC and RHIC, but also hold the potential to yield novel insights at the upcoming Electron-Ion Collider (EIC). 

It is thus important to study both longitudinal and transverse momentum distribution of hadrons inside the jet, which are characterized by the so-called semi-inclusive fragmenting jet functions and exclusive fragmenting jet functions.
They are critical ingredients in the factorization theorems.
The semi-inclusive fragmenting jet functions arise in the situation \textit{e.g.}, single inclusive jet production in proton-proton or electron-proton collisions, while the exclusive fragmenting jet functions appear in the situation \textit{e.g.}, back-to-back electron-jet production in electron-proton collisions.
There have been partial results available in previous publications, but never the complete full results, which we set up to do in this paper.
In this work, we have studied the hadron longitudinal momentum fraction $z_h$ distribution, as well as the hadron transverse momentum $j_{\perp}$ distribution within jets, for both semi-inclusive and exclusive fragmenting jet functions.
We set up factorization formalism within SCET framework that allows for the systematic determination of these distributions.
We first calculated all the components of the factorization theorem up to NLO, then we resummed all the associated large logarithms $\ln{R}$ and $\ln(p_T R / j_{\perp})$.

In the phenomenology for semi-inclusive process, we study the transverse polarization of Lambda hyperons inside the jet produced in transversely polarized proton-proton collisions, which is sensitive to the elusive TMD transversity fragmentation functions.
Our numerical estimate shows that this observable is promising at the RHIC.
For the exclusive jet production process, we study the longitudinal polarization of Lambda hyperon inside the jet, in the back-to-back electron-jet production in electron-proton collisions where the proton is transversely polarized.
We find that this observable is sensitive to the ``worm-gear'' TMD PDF $g_{1T}$ and the TMD helicity fragmentation functions and its measurement is promising at the future EIC.
With the various applications for probing the 3D QCD structures, the future for fragmenting jet observables is promising with all the planned measurements at the RHIC and the EIC.

\acknowledgments

We thank Y. He, H. Wang and S. Yang for useful discussions.
H.X. and Y.Z. are supported by the Guangdong Major Project of Basic and Applied Basic Research No. 2020B0301030008, No. 2022A1515010683, the National Natural Science Foundation of China under Grants No. 12022512, No. 12035007.
Z.K. is supported by the National Science Foundation under Grant No. PHY-1945471.
F.Z. is supported by U.S. Department of Energy, Office of Science, Office of Nuclear Physics under grant Contract Number DESC0011090 and U.S. Department of Energy, Office of Science, National Quantum Information Science Research Centers, Co-design Center for Quantum Advantage (C2QA) under Contract No. DESC0012704.

\appendix

\section{Perturbative NLO semi-inclusive FJFs}
\label{appendix.siFJFs}

As defined in \cref{sss.collinear_siFJFs_calculation}, we have the leading order and one-loop bare semi-inclusive FJFs for longitudinally and transversely polarized quarks and gluons.
To get the final expression as given in \cref{e.bare_collinear_G_qq,e.bare_collinear_G_qg,e.bare_collinear_G_gq,e.bare_collinear_G_gg,e.bare_collinear_G_qq_T}, we first perform the $q_{\perp}$ integral:
\begin{align}
\int \frac{\dd{q_{\perp}^2}}{\pqty{q_{\perp}^2}^{1 + \epsilon}}
\Theta^{\text{anti-}k_T}_{\textrm{both}} \pqty{q_{\perp}}
& =
\frac{-1}{\epsilon} \pqty{\omega _J^2 \tan[2](\frac{\mathcal{R}}{2})}^{- \epsilon}
z_h^{-2 \epsilon} \pqty{1 - z_h}^{-2 \epsilon} \, , \\
\int \frac{\dd{q_{\perp}^2}}{\pqty{q_{\perp}^2}^{1 + \epsilon}}
\Theta^{\text{anti-}k_T}_{j} \pqty{q_{\perp}}
& =
\frac{1}{\epsilon} \pqty{\omega _J^2 \tan[2](\frac{\mathcal{R}}{2})}^{- \epsilon}
\pqty{1 - z}^{-2 \epsilon}
\, , 
\end{align}
then apply the following expansions as $\epsilon \to 0$:
\begin{align}
\frac{1}{\pqty{1 - z}^{1 + 2 \epsilon}}
& =
\frac{-1}{2 \epsilon} \delta \pqty{1 - z} + \frac{1}{\pqty{1 - z}_+} - 2 \epsilon \pqty{\frac{\ln(1 - z)}{\pqty{1 - z}}}_+ + \order{\epsilon ^2}
\, , \\
\frac{1}{z^{2 \epsilon}}
& =
1 - 2 \epsilon \ln(z) + \order{\epsilon ^2}
\, , \\
\pqty{\frac{e^{\gamma _E} \mu ^2}{\omega _J^2 \tan[2](\frac{\mathcal{R}}{2})}}^{\epsilon}
\frac{1}{\Gamma \pqty{1 - \epsilon}}
& =
1 + L \epsilon + \frac{1}{12} \pqty{6 L^2 - \pi ^2} \epsilon ^2 + \order{\epsilon ^3}
\, , 
\end{align}
where $\displaystyle L \equiv \ln(\frac{\mu ^2}{\omega _J^2 \tan[2](\frac{\mathcal{R}}{2})}) = \ln(\frac{\mu ^2}{(p_TR)^2})$.
 
\section{Calculation of collinear exclusive FJFs}
\label{s.exclusive_FJF_calculation}

In this section, we provide the derivation for the matching coefficients $\Delta_{(T)}\mathscr{J}_{i j}$ of the exclusive fragmenting jet functions $\Delta_{(T)}\mathscr{G}_i^h(z_h, p_TR, \mu)$ to the collinear fragmentation function $\Delta_{(T)}D_i^h(z_h, \mu)$ for anti-${k}_{T}$ jets.
The unpolarized results were first written down in the appendix of \cite{Waalewijn:2012sv}.
We start by specifying the phase space constraint from the jet algorithm, which was nicely outlined in \cite{Ellis:2010rwa}.
Consider a parton splitting process, $i(\ell) \rightarrow j(q) + k(\ell-q)$, where an incoming parton $i$ with momentum $\ell$ splits into a parton $j$ with momentum $q$ and a parton $k$ with momentum $\ell-q$.
The four-vector $\ell^{\mu}$ can be decomposed in light-cone coordinates as $\ell^{\mu} = \pqty{\ell^+, \ell^- = \omega, 0_{\perp}}$ where $\ell^{\pm} = \ell^0 \mp \ell^z$.
The constraint for anti-$k_T$ algorithm with radius $R$ is given by:
\begin{align}
\Theta^{\text{anti-}k_T}
& =
\theta \pqty{\tan[2](\frac{\mathcal{R}}{2}) - \frac{q^+ \omega^2}{q^- \pqty{\omega - \ell^-}^2}}
\, .
\end{align}
For fragmenting jet functions, the above constraint leads to a constraint on the jet invariant mass $m_J^2 = \omega \ell^{+}$ \cite{Procura:2011aq}, which is derived and listed as follows:
\begin{align}
\delta_{\text{anti-}k_T}
& =
\theta \pqty{z_h (1-z_h) \omega ^2 \tan[2](\frac{\mathcal{R}}{2}) - m_J^2}
\theta \pqty{m_J^2} \,,
\end{align}
where $z_h = \omega _h / \omega _J = \omega _h / \omega$, $\omega _h$ is the large light-cone momentum of the final hadron, and $\omega = \omega _J$ since both partons are in the jet (see \cref{f.jet_NLO_Feynman} (A)).
\renewcommand{\arraystretch}{2}
\begin{table}
\begin{center}
\begin{tabular}{|c|c|c|c|c|}
\hline
\cellcolor{black} & \cellcolor{black} & \multicolumn{3}{c|}{\cellcolor{red!50} Quark polarization} \\
\hline
\cellcolor{black} & \cellcolor{black} & \cellcolor{red!50} U & \cellcolor{red!50} L & \cellcolor{red!50} T \\
\hline
\cellcolor{green!50} &
\cellcolor{green!50} U &
$\mathscr{G}_q^h = \eqfig{1-1}$ &
~ &
\\
\hhline{>{\arrayrulecolor{green!50}}|>{\arrayrulecolor{green!50}}->{\arrayrulecolor{black}}|%
>{\arrayrulecolor{black}}->{\arrayrulecolor{black}}|---}
\cellcolor{green!50} &
\cellcolor{green!50} L &
~ &
$\Delta \mathscr{G}_q^h = \eqfig{2-2-1} - \eqfig{2-2-2}$ &
\\
\hhline{>{\arrayrulecolor{green!50}}|>{\arrayrulecolor{green!50}}->{\arrayrulecolor{black}}|%
>{\arrayrulecolor{black}}->{\arrayrulecolor{black}}|---}
\multirow{-3}{*}{\cellcolor{green!50} \rotatebox[origin=c]{90}{Hadron pol.}} &
\cellcolor{green!50} T &
&
&
$\Delta _T \mathscr{G}_q^h = \eqfig{3-3-1} - \eqfig{3-3-2}$ \\
\hline
\end{tabular}
\end{center}
\caption{
Summary of the exclusive collinear FJFs.
The header row represents the polarization of the quark (indicated by the blue line) that initiates the jet, while the header column indicates the corresponding polarizations of produced hadrons (indicated by the red arrow from the red dot).
Shown here is for quark FJFs, with $U,\, L,\, T$ representing the unpolarized case, and longitudinal and transverse polarization.
For gluon FJFs, one would interpret $L,\, T,$ as circular and linear polarization.}
\label{t.collinear_FJFs}
\end{table}
\renewcommand{\arraystretch}{1}

The exclusive FJFs $\Delta_{(T)}\mathscr{G}_i^h(z_h, p_TR, \mu)$ can be matched onto corresponding fragmentation functions $\Delta_{(T)}D_i^h(z_h, \mu)$ as listed in \cref{t.collinear_FJFs}:
\begin{align} \label{e.collinear_exclusive_FJF_matching}
\Delta _{(T)} \mathscr{G}_i^h (z_h, p_TR, \mu)
& =
\sum_j \int _{z_h}^1 \frac{\dd{z'_h}}{z'_h}
\Delta _{(T)} \mathscr{J}_{ij} (z'_h, p_TR, \mu)
\Delta _{(T)} D_{h/j} \pqty{\frac{z_h}{z'_h}, \mu}
\nonumber \\
& \quad +
\order{\frac{\Lambda _{\mathrm{QCD}}^2}{(p_TR)^2}}
\, ,
\end{align}
where $\Delta _{(T)} \mathscr{J}_{ij}$ are the matching coefficients.
The exclusive FJFs $\Delta _{(T)} \mathscr{G}_i^j \pqty{z_h, m_J^2, \mu}$ with $i, j \in \Bqty{q, g}$ has been extensively studied in \cite{Jain:2011xz, Ritzmann:2014mka}.
Using pure dimensional regularization with $4 - 2 \epsilon$ dimensions in the $\overline{\mathrm{MS}}$ scheme, the bare results at $\order{\alpha _s}$ can be written in the following compact form \cite{ Giele:1991vf,Ritzmann:2014mka,Chien:2015ctp}:
\begin{equation}
\Delta _{(T)} \mathscr{G}_{i, \mathrm{bare}}^j \pqty{z_h, m_J^2}
=
\frac{\alpha _s}{2 \pi} \frac{\pqty{e^{\gamma _E} \mu^2}^\epsilon}{\Gamma (1 - \epsilon)}
\Delta _{(T)} \hat{P}_{ji}(z_h, \epsilon)
z_h^{-\epsilon} (1-z_h)^{-\epsilon} \pqty{m_J^2}^{-1-\epsilon}
\, .
\end{equation}
Generalized from \cite{Chien:2015ctp}, $\Delta _{(T)} \mathscr{G}_{i, \mathrm{bare}}^j \pqty{z_h, m_J^2, \mu}$ is related to $\Delta _{(T)}\mathscr{G}_{i,  \mathrm{bare}}^h (z_h, p_TR, \mu)$ by:
\begin{equation}
\Delta _{(T)} \mathscr{G}_{i,  \mathrm{bare}}^h (z_h, p_T R)
=
\int \dd{m_J^2} \Delta _{(T)} \mathscr{G}_{i, \mathrm{bare}}^h \pqty{z_h, m_J^2} \delta_{\text{anti-}k_T}
\, ,
\end{equation}
notice that we reinstated hadronic FJFs.
The splitting functions $\Delta _{(T)} \hat{P}_{ji} (z_h, \epsilon)$ are given in \cref{e.Delta_P_qq,e.Delta_P_gq,e.Delta_P_qg,e.Delta_P_gg,e.Delta_T_P_qq}.
By performing the integration over $m_J^2$ with the constraints imposed by the jet algorithm $\delta _{\text{anti-}k_T}$, one obtains the bare exclusive FJFs $\Delta_{(T)}\mathscr{G}_{i, \mathrm{bare }}^j(z_h, p_T R)$.
We present the results for anti-$k_T$ jets here, as their explicit expressions are not available in the literature:
\begingroup
\allowdisplaybreaks
\begin{align}
\Delta \mathscr{G}_{q, \text{bare}}^q (z_h, p_T R)
& =
\frac{\alpha _s}{2 \pi} \bigg\{C_F \pqty{\frac{1}{\epsilon ^2} + \frac{3}{2 \epsilon} + \frac{L}{\epsilon}} \delta (1-z_h) - \frac{1}{\epsilon} \Delta P_{qq} (z_h) + C_F (1-z_h)
\nonumber \\
& \quad + C_F \delta (1-z_h) \pqty{\frac{L^2}{2}-\frac{\pi ^2}{12}} + \Delta P_{qq} (z_h) (2 \ln(z_h) - L) \nnu
& \quad + \frac{3 C_F L}{2} \delta \pqty{1-z_h} + 2 C_F \pqty{1+z_h^2} \pqty{\frac{\ln (1-z_h)}{1-z_h}}_+ \bigg\}
\, , \label{e.bare_exclusive_FJFs_qq} \\
\Delta \mathscr{G}_{q, \text{bare}}^g (z_h, p_T R)
& =
\frac{\alpha _s}{2 \pi} \bigg\{\frac{-1}{\epsilon} \Delta P_{gq}(z_h) + \Delta P_{gq}(z_h) \Big[2 \ln(z_h (1-z_h)) - L\Big] - 2 C_F (1-z_h) \bigg\}
\, , \label{e.bare_exclusive_FJFs_qg} \\
\Delta \mathscr{G}_{g, \text{bare}}^q (z_h, p_T R)
& =
\frac{\alpha _s}{2 \pi} \bigg\{\frac{-1}{\epsilon} \Delta P_{qg}(z_h)+\Delta P_{qg}(z_h) \Big[2\ln(z_h (1-z_h)) - L \Big] + 2 T_F (1-z_h) \bigg\}
\, , \label{e.bare_exclusive_FJFs_gq} \\
\Delta \mathscr{G}_{g, \text{bare}}^g (z_h, p_T R)
& =
\frac{\alpha _s}{2 \pi} \bigg\{C_A \pqty{\frac{1}{\epsilon ^2}+\frac{1}{\epsilon} \frac{\beta _0}{2 C_A}+\frac{L}{\epsilon}} \delta(1-z_h) - \frac{1}{\epsilon} \Delta P_{gg}(z_h)
\nonumber \\
& \quad + C_A \delta (1-z_h) \pqty{\frac{L^2}{2}-\frac{\pi^2}{12}} + \Delta P_{gg}(z_h)(2\ln(z_h) - L) + \frac{\beta _0 L}{2} \delta (1-z_h)
\nonumber \\
& \quad - 4 C_A (1-z_h) + 4 C_A \Big[2(1-z_h)^2+z_h\Big] \pqty{\frac{\ln(1-z_h)}{1-z_h}}_+ \bigg\}
\, , \label{e.bare_exclusive_FJFs_gg} \\
\Delta _T \mathscr{G}_{q, \text{bare}}^q (z_h, p_T R)
& =
\frac{\alpha _s}{2 \pi} \bigg\{C_F \pqty{\frac{1}{\epsilon ^2}+\frac{3}{2 \epsilon }+\frac{L}{\epsilon}} \delta (1-z_h) - \frac{1}{\epsilon} \Delta _T P_{qq}(z_h) \nonumber \\
& \quad + C_F \delta (1-z_h) \pqty{\frac{L^2}{2}-\frac{\pi ^2}{12}} + \Delta _T P_{qq}(z_h)(2\ln(z_h) - L) \nonumber \\
& \quad + \frac{3 C_F L}{2} \delta \pqty{1-z_h} + 4 C_F z_h \pqty{\frac{\ln (1-z_h)}{1-z_h}}_+ \bigg\}
\, , \label{e.bare_exclusive_FJFs_qq_T}
\end{align}
\endgroup
where, as given in the main text, $\beta_0$ is defined as:
\begin{equation}
\beta _0 \equiv \frac{11}{3} C_A - \frac{4}{3} T_F n_f\,
\end{equation}
and $\Delta_{(T)}{P}_{j i}(z_h)$ are given in \cref{e.AP_longitudinal_qq,e.AP_longitudinal_gq,e.AP_longitudinal_qg,e.AP_longitudinal_gg,e.AP_transverse_qq}.
It is instructive to point out that the $\epsilon$ poles in the first term of \cref{e.bare_exclusive_FJFs_qq,e.bare_exclusive_FJFs_gg,e.bare_exclusive_FJFs_qq_T} correspond to ultraviolet (UV) divergences, and they are related to the renormalization of the FJFs $\Delta_{(T)}\mathscr{G}_{i, \text{bare}}^j (z_h, p_T R)$.
All the remaining $\epsilon$ poles in \cref{e.bare_exclusive_FJFs_qq,e.bare_exclusive_FJFs_qg,e.bare_exclusive_FJFs_gq,e.bare_exclusive_FJFs_gg,e.bare_exclusive_FJFs_qq_T} are infrared (IR) poles, and they match exactly with those in the fragmentation functions $\Delta_{(T)}D_{j/i}(z_h, \mu)$, which we will show below.
$\Delta_{(T)}\mathscr{G}_{i, \text{bare}}^h (z_h, p_T R)$ is renormalized by:
\begin{equation}
\Delta _{(T)} \mathscr{G}_{i, \text{bare}}^h (z_h, p_T R)
=
\mathcal{Z}_{\mathscr{G}}^i(\mu)
\Delta _{(T)} \mathscr{G}_i^h (z_h, p_TR, \mu)\,,
\end{equation}
where $i$ is not summed over in the above equation.
The corresponding renormalization group (RG) equation is:
\begin{equation}
\mu \frac{\dd{}}{\dd{\mu}} \Delta _{(T)} \mathscr{G}_i^h (z_h, p_TR, \mu)
=
\gamma _{\mathscr{G}}^i(\mu) \Delta _{(T)} \mathscr{G}_i^h (z_h, p_TR, \mu)\,,
\end{equation}
where the anomalous dimension $\gamma_{\mathscr{G}}^i(\mu)$ is:
\begin{equation} \label{e.exclusive_FJFs_gamma}
\gamma _{\mathscr{G}}^i(\mu)
=
- \pqty{\mathcal{Z}_{\mathscr{G}}^i(\mu)}^{-1}
\mu \frac{\dd{}}{\dd{\mu}} \mathcal{Z}_{\mathscr{G}}^i(\mu)\, .
\end{equation}
The solution to \cref{e.exclusive_FJFs_gamma} is then:
\begin{equation}
\Delta _{(T)}\mathscr{G}_i^h (z_h, p_TR, \mu)
=
\Delta _{(T)}\mathscr{G}_i^h \pqty{z_h, p_TR, \mu_{\mathscr{G}}}
\exp(\int _{\mu_{\mathscr{G}}}^{\mu} \frac{\dd{\mu '}}{\mu '} \gamma _{\mathscr{G}}^i \pqty{\mu '}) \,,
\end{equation}
where the scale $\mu_{\mathscr{G}}$ should be the characteristic scale chosen such that large logarithms in the fixed-order calculation vanish.
The counter terms $\mathcal{Z}_{\mathscr{G}}^i(\mu)$ are given by
\footnote{Note here the counter terms for polarized quark and gluon FJFs are the same as those of the unpolarized ones as shown in \cite{Chien:2015ctp}.
For explanation, see the main text in \cref{ss.collinear_exclusive_FJFs}.}:
\begin{align}
\mathcal{Z}_{\mathscr{G}}^q (\mu)
& =
1 + \frac{\alpha _s}{2 \pi} C_F \bqty{\frac{1}{\epsilon^2}+\frac{3}{2 \epsilon}+\frac{L}{\epsilon}} , \label{e.exclusive_FJFs_counter_q} \\
\mathcal{Z}_{\mathscr{G}}^g (\mu)
& =
1 + \frac{\alpha _s}{2 \pi} C_A \bqty{\frac{1}{\epsilon^2} + \frac{1}{\epsilon} \frac{\beta _0}{2 C_A} + \frac{L}{\epsilon}}\, . \label{e.exclusive_FJFs_counter_g}
\end{align}
From these results we obtain the anomalous dimension $\gamma _{\mathscr{G}}^i (\mu)$ with the following form:
\begin{align}
\gamma _{\mathscr{G}}^i(\mu)
=
\Gamma _{\text{cusp}}^i \pqty{\alpha _s} \ln(\frac{\mu^2}{(p_TR)^2})
+
\gamma ^i \pqty{\alpha _s}\,,
\end{align}
where $\Gamma _{\text{cusp}}^i = \sum _n \Gamma _{n-1}^i \pqty{\frac{\alpha _s}{4 \pi}}^n$ and $\gamma ^i = \sum _n \gamma _{n-1}^i \pqty{\frac{\alpha _s}{4 \pi}}^n$.
The lowest-order coefficients can be extracted from the above calculations:
\begin{align}
\Gamma _0^q = 4 C_F, & \quad \gamma _0^q = 6 C_F
\, , \\
\Gamma _0^g = 4 C_A, & \quad \gamma _0^g = 2 \beta _0
\, ,
\end{align}
and higher-order results can be found in \cite{Jain:2011xz, Becher:2006mr, Becher:2009th, Echevarria:2012pw, Moch:2004pa}.
After the subtraction of the UV counter terms in \cref{e.exclusive_FJFs_counter_q,e.exclusive_FJFs_counter_g}, the renormalized FJFs $\Delta _{(T)}\mathscr{G}_i^j (\omega, R, z_h, \mu)$ are given by:
\begingroup
\allowdisplaybreaks
\begin{align}
\Delta \mathscr{G}_q^q (z_h, p_TR, \mu)
& =
\frac{\alpha _s}{2 \pi} \bigg\{\frac{-1}{\epsilon} \Delta P_{qq}(z_h) + C_F \delta (1-z_h) \pqty{\frac{L^2}{2}-\frac{\pi ^2}{12}} + C_F (1-z_h) \nnu
& \quad + \Delta P_{qq}(z_h)(2\ln(z_h) - L) + \frac{3 C_F L}{2} \delta \pqty{1 - z_h} \nonumber \\
& \quad + 2 C_F \pqty{1 + z_h^2} \pqty{\frac{\ln(1-z_h)}{1-z_h}}_+ \bigg\} \,, \\
\Delta \mathscr{G}_{q}^g (z_h, p_TR, \mu)
& =
\frac{\alpha _s}{2 \pi} \bigg\{\frac{-1}{\epsilon} \Delta P_{gq}(z_h)+\Delta P_{gq}(z_h) \Big[2\ln(z_h(1-z_h)) - L\Big]- 2 C_F (1-z_h) \bigg\} \,, \\
\Delta \mathscr{G}_g^q (z_h, p_TR, \mu)
& =
\frac{\alpha _s}{2 \pi} \bigg\{\frac{-1}{\epsilon} \Delta{P}_{q g}(z_h)+\Delta P_{qg}(z_h)\Big[2\ln(z_h(1-z_h)) - L\Big] + 2 T_F (1-z_h) \bigg\} \,, \\
\Delta \mathscr{G}_g^g (z_h, p_TR, \mu)
& =
\frac{\alpha _s}{2 \pi} \bigg\{\frac{-1}{\epsilon} \Delta P_{gg}(z_h) + C_A \delta (1-z_h) \pqty{\frac{L^2}{2}-\frac{\pi^2}{12}} \nnu
& \quad + \Delta P_{gg}(z_h)(2\ln(z_h) - L) + \frac{\beta _0 L}{2} \delta \pqty{1 - z_h} - 4 C_A (1-z_h) \nnu
& \quad + 4 C_A \Big[2(1-z_h)^2+z_h\Big] \pqty{\frac{\ln (1-z_h)}{1-z_h}}_+ \bigg\} \,, \\
\Delta _T \mathscr{G}_q^q (z_h, p_TR, \mu)
& =
\frac{\alpha _s}{2 \pi} \bigg\{\frac{-1}{\epsilon} \Delta _T P_{qq}(z_h) + C_F \delta (1-z_h) \pqty{\frac{L^2}{2}-\frac{\pi^2}{12}} + \frac{3 C_F L}{2} \delta \pqty{1 - z_h} \nnu
& \quad + \Delta _T P_{qq}(z_h)(2\ln(z_h) - L) + 4 C_F z_h \pqty{\frac{\ln (1-z_h)}{1-z_h}}_+ \bigg\}\, ,
\end{align}
\endgroup
where we can eliminate all large logarithms $L$ by choosing $\mu \sim p_T R$.
At the intermediate scale $\mu_{\mathscr{G}} \gg \Lambda_{\mathrm{QCD}}$, one can match the FJFs $\Delta_{(T)}\mathscr{G}_i^h(z_h, p_TR, \mu)$ onto the longitudinally (transversely) polarized fragmentation functions $\Delta_{(T)}D_{h/j}(z_h, \mu)$ as in \cref{e.collinear_exclusive_FJF_matching}.
In order to perform the matching calculation and determine the coefficients $\mathscr{J}_{i j}$, we simply need the perturbative results of the fragmentation functions $\Delta_{(T)} D_{j/i}(z_h, \mu)$ for a parton $i$ fragmenting into a parton $j$.
The renormalized $\Delta_{(T)}D_{j/i}(z_h, \mu)$ at $\order{\alpha _s}$ using pure dimensional regularization are given by:
\begin{align}
\Delta D_q^q (z_h, \mu)
& =
\delta (1-z_h) + \frac{\alpha _s C_F}{2 \pi} \pqty{-\frac{1}{\epsilon}} \Delta P_{qq}(z_h)\, , \\
\Delta D_q^g (z_h, \mu)
& =
\frac{\alpha _s C_F}{2 \pi} \pqty{-\frac{1}{\epsilon}} \Delta P_{gq}(z_h)\, , \\
\Delta D_g^q (z_h, \mu)
& =
\frac{\alpha _s T_F}{2 \pi} \pqty{-\frac{1}{\epsilon}} \Delta P_{qg}(z_h) \,, \\
\Delta D_g^g (z_h, \mu)
& =
\delta (1-z_h) + \frac{\alpha _s C_A}{2 \pi} \pqty{-\frac{1}{\epsilon}} \Delta P_{gg}(z_h) \,, \\
\Delta _T D_q^q (z_h, \mu)
& =
\delta (1-z_h) + \frac{\alpha _s C_F}{2 \pi} \pqty{-\frac{1}{\epsilon}} \Delta _T P_{qq}(z_h) \,.
\end{align}
Using the results for $\Delta _{(T)} \mathscr{G}_i^j (z_h, p_TR, \mu)$ and $\Delta _{(T)} D_{j/i} (z_h, \mu)$, we obtain the following matching coefficients:
\begin{align}
\Delta \mathscr{J}_{q q} (z_h, p_TR, \mu)
& =
\delta (1-z_h) + \frac{\alpha _s}{2 \pi} \bigg[C_F \delta (1-z_h) \pqty{\frac{L^2}{2}-\frac{\pi^2}{12}} + C_F (1-z_h) - \Delta P_{qq}(z_h) L
\nonumber \\
& \qquad \qquad \qquad \qquad +
\frac{3 C_F L}{2} \delta \pqty{1 - z_h} + \Delta \hat{\mathscr{I}}_{qq}^{\text{anti-}k_T} \pqty{z_h} \bigg]
\, , \\
\Delta \mathscr{J}_{qg} (z_h, p_TR, \mu)
& =
\frac{\alpha _s}{2\pi} \bigg[-\Delta P_{gq}(z_h) L- 2 C_F (1-z_h) + \Delta \hat{\mathscr{I}}_{qg}^{\text{anti-}k_T}(z_h) \bigg]
\, , \\
\Delta \mathscr{J}_{gq} (z_h, p_TR, \mu)
& =
\frac{\alpha _s}{2 \pi} \bigg[-\Delta P_{qg}(z_h) L + 2 T_F (1-z_h) + \Delta \hat{\mathscr{I}}_{gq}^{\text{anti-}k_T}(z_h)\bigg]
\, , \\
\Delta \mathscr{J}_{gg} (z_h, p_TR, \mu)
& =
\delta (1-z_h) + \frac{\alpha _s}{2 \pi} \bigg[C_A \delta (1-z_h) \pqty{\frac{L^2}{2}-\frac{\pi^2}{12}} - 4 C_A (1-z_h) - \Delta P_{gg}(z_h) L
\nonumber \\
& \qquad \qquad \qquad \qquad +
\frac{\beta _0 L}{2} \delta \pqty{1 - z_h} + \Delta\hat{\mathscr{I}}_{g g}^{\text{anti-}k_T}(z_h)\bigg]
\, , \\
\Delta _T \mathscr{J}_{qq} (z_h, p_TR, \mu)
& =
\delta (1-z_h) + \frac{\alpha _s}{2 \pi} \bigg[C_F \delta (1-z_h) \pqty{\frac{L^2}{2}-\frac{\pi^2}{12}} - \Delta _T P_{qq}(z_h) L
\nonumber \\
& \qquad \qquad \qquad \qquad +
\frac{3 C_F L}{2} \delta \pqty{1 - z_h} + \Delta _T \hat{\mathscr{I}}_q^{\text{anti-}k_T} \bigg]
\, ,
\end{align}
where $\Delta _{(T)} \hat{\mathscr{I}}_{i j}^{{\text{anti-}k_T}}(z_h)$ are jet-algorithm dependent.
For anti-$k_T$ jets, we have:
\begin{align}
\Delta \hat{\mathscr{I}}_{qq}^{\text{anti}-k_T} \pqty{z_h}
& =
2 \Delta P_{qq}(z_h) \ln(z_h) + 2 C_F \pqty{1+z_h^2} \pqty{\frac{\ln(1-z_h)}{1-z_h}}_+
\, , \\
\Delta \hat{\mathscr{I}}_{qg}^{\text{anti}-k_T} \pqty{z_h}
& =
2 \Delta P_{gq}(z_h) \Big[\ln(z_h(1-z_h)) \Big]
\, , \\
\Delta \hat{\mathscr{I}}_{gq}^{\text{anti}-k_T} \pqty{z_h}
& =
2 \Delta P_{qg}(z_h) \Big[\ln(z_h(1-z_h)) \Big]
\, , \\
\Delta \hat{\mathscr{I}}_{gg}^{\text{anti}-k_T} \pqty{z_h}
& =
2 \Delta P_{gg}(z_h) \ln(z_h) + 4 C_A \Big[2(1-z_h)^2 + z_h\Big] \pqty{\frac{\ln(1-z_h)}{1-z_h}}_+
\, , \\
\Delta _T \hat{\mathscr{I}}_{qq}^{\text{anti}-k_T} \pqty{z_h}
& =
2 \Delta _T P_{qq}(z_h) \ln(z_h) + 4 C_F z_h \pqty{\frac{\ln (1-z_h)}{1-z_h}}_+
\, .
\end{align}
Substituting the matching coefficients $\Delta _{(T)} \mathscr{J}_{ij}$ into \cref{e.collinear_exclusive_FJF_matching}, and writing out explicitly the plus functions, one obtains:
\begin{align}
\Delta _{(T)} \mathscr{G}_q^h (z_h, p_TR, \mu)
& =
\Bqty{1 + \frac{\alpha _s}{2 \pi} C_F \bqty{2 \ln[2](\frac{p_TR(1-z_h)}{\mu}) - \frac{\pi^2}{12}}} \Delta _{(T)} D_q^h (z_h, \mu) + \cdots
\, , \\
\Delta \mathscr{G}_g^h (z_h, p_TR, \mu)
& =
\Bqty{1 + \frac{\alpha _s}{2 \pi} C_A \bqty{2 \ln[2](\frac{p_T R (1-z_h)}{\mu}) - \frac{\pi^2}{12}}} \Delta D_g^h (z_h, \mu) + \cdots
\, . 
\end{align}
Here the ellipses represent terms that are regular as $z_h \rightarrow 1$.
Under the large $z_h$ limit, one can find additional logarithms $\sim \ln(1-z_h)$. By choosing the scale $\mu = p_T R (1-z_h)$, we can simultaneously resum both logarithms of $R$ and $(1-z_h)$ \cite{Procura:2011aq}. A more rigorous resummation at the threshold limit is given by~\cite{Kaufmann:2019ksh}.

\section{Separation of hard matching functions}

In this work we follow the methodology outlined in \cite{Kang:2017glf} and separately evolve the hard matching functions.
Doing this have the benefit of simplifying the final evolution of the TMD FJFs, for more details, please refer to \cite{Kang:2017glf}.
In this section, we will only provide the result of such separation.
We start by noticing that the anomalous dimensions $\gamma_{i j}^{L(T)} \pqty{z, p_TR, \mu}$ in \cref{e.lon_hard_function_gamma_qq,e.lon_hard_function_gamma_qg,e.lon_hard_function_gamma_gg,e.lon_hard_function_gamma_gq,e.tra_hard_function_gamma_qq} include a purely diagonal piece $\delta _{ij} \delta (1-z) \Gamma _i \pqty{p_TR, \mu}$ and an off-diagonal Altarelli-Parisi splitting function $\Delta _{(T)} P_{ji}(z)$ that has polarization dependence.
Therefore following the approach in \cite{Kang:2017glf}, we can separate these two contributions by rewriting the functions $\hat{H}^{U (L, T)}_{i \rightarrow j}$ into two parts that follow different evolution equations:
\begin{equation}
\hat{H}^{U (L, T)}_{i \rightarrow j} \pqty{z, p_TR, \mu}
=
\mathcal{E}_i \pqty{p_TR, \mu}
\hat{\mathcal{C}}^{U (L, T)}_{i \rightarrow j} \pqty{z, p_TR, \mu}
\, . \label{e.hard_function_separation}
\end{equation}
The coefficients $\hat{\mathcal{C}}^{U (L, T)}_{i \rightarrow j} \pqty{z, p_TR, \mu}$ follow the evolution equations governed by the Altarelli-Parisi splitting functions:
\begin{align}
\mu \frac{\dd{}}{\dd{\mu}} \hat{\mathcal{C}}^{U}_{i \rightarrow j} \pqty{z, p_TR, \mu}
& =
\frac{\alpha _s}{2 \pi} \sum _k \int _z^1 \frac{\dd{z'}}{z'} 
P_{k i} \pqty{\frac{z}{z'}}\hat{\mathcal{C}}^{U}_{k \rightarrow j} \pqty{z', p_TR, \mu}
\, , \\
\mu \frac{\dd{}}{\dd{\mu}} \hat{\mathcal{C}}^{L(T)}_{i \rightarrow j} \pqty{z, p_TR, \mu}
& =
\frac{\alpha _s}{2 \pi} \sum _k \int _z^1 \frac{\dd{z'}}{z'} 
\Delta _{(T)} P_{k i} \pqty{\frac{z}{z'}}
\hat{\mathcal{C}}^{L(T)}_{k \rightarrow j} \pqty{z', p_TR, \mu}
\, , \label{e.hard_function_evolution}
\end{align}
and as pointed out in \cite{Kang:2017glf}, although the above evolution equations look like DGLAP equations, it is only the combined TMD FJFs that will satisfy the standard timelike DGLAP evolution equations.
As for the functions $\mathcal{E}_i \pqty{p_TR, \mu}$, they follow the multiplicative RG equations:
\begin{equation} \label{e.hard_function_E_evolution}
\mu \frac{\dd{}}{\dd{\mu}} \ln{\mathcal{E}_i \pqty{p_TR, \mu}}
=
\Gamma _i \pqty{p_TR, \mu}
\, ,
\end{equation}
with $\Gamma_i$ given in~\cref{e.Gamma_part_of_hard_matching_gamma} and their solutions are:
\begin{equation}
\mathcal{E}_i \pqty{p_TR, \mu}
=
\mathcal{E}_i \pqty{p_TR, \mu _J}
\exp(\int _{\mu _J}^{\mu} \frac{\dd{\mu'}}{\mu'} 
\Gamma _i \pqty{p_TR, \mu'})
\, ,
\end{equation}
with the fixed-order results provided in \cite{Kang:2017glf}.
By choosing $\mu _J = p_T R$, we obtain the initial condition $\mathcal{E}_i \pqty{p_TR, \mu_J} = 1$ for the evolution given in \cref{e.hard_function_E_evolution}.

Collecting the results,  for unpolarized hard matching function $\hat{H}_{i \to j}^U \pqty{z', p_TR , \mu}$, one find the coefficients $\hat{\mathcal{C}}^{U}_{i \rightarrow j} \pqty{z, p_TR, \mu}$ in \cref{e.hard_function_separation} to be the same as in \cite{Kang:2017glf}, here we list the results for completeness:
\begin{align}
\hat{\mathcal{C}}^U_{q \rightarrow q^{\prime}} \pqty{z, p_TR, \mu}
& =
\delta_{q q'} \delta(1-z)
+
\delta_{q q'} \frac{\alpha_s}{2 \pi} \left[C_F \delta(1-z) \frac{\pi^2}{12}+P_{q q}(z) L \right.
\nonumber \\
& \qquad \qquad \qquad \qquad \qquad
\left. -2 C_F \pqty{1+z^2} \pqty{\frac{\ln (1-z)}{1-z}}_+ - C_F(1-z) \right]
\, , \\
\hat{\mathcal{C}}^U_{q \rightarrow g} \pqty{z, p_TR, \mu}
& =
\frac{\alpha_s}{2 \pi} \bqty{(L-2 \ln(1-z)) P_{g q}(z)-C_F z}
\, , \\
\hat{\mathcal{C}}^U_{g \rightarrow g} \pqty{z, p_TR, \mu}
& =
\delta(1-z)
+
\frac{\alpha_s}{2 \pi} \bqty{\delta(1-z) \frac{\pi^2}{12} + P_{g g}(z) L - \frac{4 C_A \pqty{1-z+z^2}^2}{z} \pqty{\frac{\ln (1-z)}{1-z}}_+}
\, , \\
\hat{\mathcal{C}}^U_{g \rightarrow q} \pqty{z, p_TR, \mu}
& =
\frac{\alpha_s}{2 \pi} \bqty{(L-2 \ln(1-z)) P_{q g}(z) - 2 T_F z (1-z)}
\, .
\end{align}
As for the coefficients of polarized functions $\hat{H}^{L(T)}_{i \rightarrow j} \pqty{z, p_TR, \mu}$, given that $\mathcal{E}_i \pqty{p_TR, \mu}$ follow the same evolution as in \cref{e.hard_function_E_evolution}, we can write \cref{e.hard_function_separation} as:
\begin{equation} \label{e.hard_function_evolution_final}
\hat{H}^{L(T)}_{i \rightarrow j} \pqty{z, p_TR, \mu}
=
\exp(\int _{\mu _J}^{\mu} \frac{\dd{\mu'}}{\mu'} 
\Gamma _i \pqty{p_TR, \mu'})
\hat{\mathcal{C}}^{L(T)}_{i \rightarrow j} \pqty{z, p_TR, \mu}
\, .
\end{equation}
And we therefore only need to evolve the functions $\hat{\mathcal{C}}^{L(T)}_{i \rightarrow j} \pqty{z, p_TR, \mu}$ from scale $\mu \sim \mu _J = p_T R$ to $\mu \sim p_T$ following \cref{e.hard_function_evolution}.
In particular, the fixed-order results are:
\begin{align}
\hat{\mathcal{C}}^L_{q \rightarrow q'} \pqty{z, p_TR, \mu}
& =
\delta _{qq'} \delta (1-z)
+
\delta _{qq'} \frac{\alpha _s}{2 \pi}
\bigg[C_F \delta (1-z) \frac{\pi ^2}{12}
+ \Delta P_{qq}(z) L
\nonumber \\
& \qquad \qquad \qquad \qquad \qquad
-2 C_F (1+z^2) \pqty{\frac{\ln(1-z)}{1-z}}_+ - C_F (1-z) \bigg]
\, , \\
\hat{\mathcal{C}}^L_{q \rightarrow g} \pqty{z, p_TR, \mu}
& =
\frac{\alpha _s}{2 \pi} \Big[\pqty{L - 2 \ln(1-z)} \Delta P_{gq}(z) + 2 C_F (1-z) \Big]
\, , \\
\hat{\mathcal{C}}^L_{g \rightarrow q} \pqty{z, p_TR, \mu}
& =
\frac{\alpha _s}{2 \pi} \Big[\pqty{L - 2 \ln(1-z)} \Delta P_{qg}(z) - 2 T_F (1-z) \Big]
\, , \\
\hat{\mathcal{C}}^L_{g \rightarrow g} \pqty{z, p_TR, \mu}
& =
\delta (1-z)
+
\frac{\alpha _s}{2 \pi} \bigg[C_A \delta (1-z) \frac{\pi ^2}{12} + \Delta P_{gg}(z) L
\nonumber \\
& \qquad \qquad \qquad \quad +
4 C_A (1-z) - 4 C_A (2(1-z)^2 + z) \pqty{\frac{\ln(1-z)}{1-z}}_+ \bigg]
\, , \\
\hat{\mathcal{C}}^T_{q \rightarrow q'} \pqty{z, p_TR, \mu}
& =
\delta _{qq'} \delta(1-z)
\nonumber \\
& \quad +
\delta _{qq'} \frac{\alpha _s}{2 \pi} \Bqty{\Delta_T P_{qq}(z) L + C_F \bqty{-4z \pqty{\frac{\ln(1-z)}{1-z}}_+ + \frac{\pi ^2}{12} \delta (1-z)}}
\, .
\end{align}
So far, we have presented the perturbative expressions for polarized hard matching functions in \cref{e.hard_function_evolution_final} up to NLO.

\bibliographystyle{JHEP}
\bibliography{main}

\providecommand{\href}[2]{#2}\begingroup\raggedright\begin{thebibliography}{100}

\bibitem{Butterworth:2008iy}
J.~M. Butterworth, A.~R. Davison, M.~Rubin, and G.~P. Salam, {\it {Jet
  substructure as a new Higgs search channel at the LHC}},  {\em Phys. Rev.
  Lett.} {\bf 100} (2008) 242001, [\href{http://arxiv.org/abs/0802.2470}{{\tt
  arXiv:0802.2470}}].

\bibitem{Newman:2013ada}
P.~Newman and M.~Wing, {\it {The Hadronic Final State at HERA}},  {\em Rev.
  Mod. Phys.} {\bf 86} (2014), no.~3 1037,
  [\href{http://arxiv.org/abs/1308.3368}{{\tt arXiv:1308.3368}}].

\bibitem{Kogler:2018hem}
R.~Kogler et~al., {\it {Jet Substructure at the Large Hadron Collider:
  Experimental Review}},  {\em Rev. Mod. Phys.} {\bf 91} (2019), no.~4 045003,
  [\href{http://arxiv.org/abs/1803.06991}{{\tt arXiv:1803.06991}}].

\bibitem{Larkoski:2017jix}
A.~J. Larkoski, I.~Moult, and B.~Nachman, {\it {Jet Substructure at the Large
  Hadron Collider: A Review of Recent Advances in Theory and Machine
  Learning}},  {\em Phys. Rept.} {\bf 841} (2020) 1--63,
  [\href{http://arxiv.org/abs/1709.04464}{{\tt arXiv:1709.04464}}].

\bibitem{Connors:2017ptx}
M.~Connors, C.~Nattrass, R.~Reed, and S.~Salur, {\it {Jet measurements in heavy
  ion physics}},  {\em Rev. Mod. Phys.} {\bf 90} (2018) 025005,
  [\href{http://arxiv.org/abs/1705.01974}{{\tt arXiv:1705.01974}}].

\bibitem{Cunqueiro:2021wls}
L.~Cunqueiro and A.~M. Sickles, {\it {Studying the QGP with Jets at the LHC and
  RHIC}},  {\em Prog. Part. Nucl. Phys.} {\bf 124} (2022) 103940,
  [\href{http://arxiv.org/abs/2110.14490}{{\tt arXiv:2110.14490}}].

\bibitem{Belmont:2023fau}
R.~Belmont et~al., {\it {Predictions for the sPHENIX physics program}},  in
  {\em {RBRC Workshop:~Predictions for sPHENIX}}, 5, 2023.
\newblock \href{http://arxiv.org/abs/2305.15491}{{\tt arXiv:2305.15491}}.

\bibitem{AbdulKhalek:2021gbh}
R.~Abdul~Khalek et~al., {\it {Science Requirements and Detector Concepts for
  the Electron-Ion Collider}: {EIC Yellow Report}},  {\em Nucl. Phys. A} {\bf
  1026} (2022) 122447, [\href{http://arxiv.org/abs/2103.05419}{{\tt
  arXiv:2103.05419}}].

\bibitem{AbdulKhalek:2022hcn}
R.~Abdul~Khalek et~al., {\it {Snowmass 2021 White Paper: Electron Ion Collider
  for High Energy Physics}},  \href{http://arxiv.org/abs/2203.13199}{{\tt
  arXiv:2203.13199}}.

\bibitem{Aschenauer:2019kzf}
E.~C. Aschenauer, I.~Borsa, R.~Sassot, and C.~Van~Hulse, {\it {Semi-inclusive
  Deep-Inelastic Scattering, Parton Distributions and Fragmentation Functions
  at a Future Electron-Ion Collider}},  {\em Phys. Rev. D} {\bf 99} (2019),
  no.~9 094004, [\href{http://arxiv.org/abs/1902.10663}{{\tt
  arXiv:1902.10663}}].

\bibitem{Wang:2019bvb}
B.~Wang, J.~O. Gonzalez-Hernandez, T.~C. Rogers, and N.~Sato, {\it {Large
  Transverse Momentum in Semi-Inclusive Deeply Inelastic Scattering Beyond
  Lowest Order}},  {\em Phys. Rev. D} {\bf 99} (2019), no.~9 094029,
  [\href{http://arxiv.org/abs/1903.01529}{{\tt arXiv:1903.01529}}].

\bibitem{Whitehill:2022mpq}
{\bf Jefferson Lab Angular Momentum (JAM)} Collaboration, R.~M. Whitehill,
  Y.~Zhou, N.~Sato, and W.~Melnitchouk, {\it {Accessing gluon polarization with
  high-PT hadrons in SIDIS}},  {\em Phys. Rev. D} {\bf 107} (2023), no.~3
  034033, [\href{http://arxiv.org/abs/2210.12295}{{\tt arXiv:2210.12295}}].

\bibitem{Boussarie:2023izj}
R.~Boussarie et~al., {\it {TMD Handbook}},
  \href{http://arxiv.org/abs/2304.03302}{{\tt arXiv:2304.03302}}.

\bibitem{LHCb:2022rky}
{\bf LHCb} Collaboration, {\it {Multidifferential study of identified charged
  hadron distributions in $Z$-tagged jets in proton-proton collisions at
  $\sqrt{s}=$13 TeV}},  \href{http://arxiv.org/abs/2208.11691}{{\tt
  arXiv:2208.11691}}.

\bibitem{ATLAS:2017pgl}
{\bf ATLAS} Collaboration, M.~Aaboud et~al., {\it {Measurement of jet
  fragmentation in 5.02 TeV proton-lead and proton-proton collisions with the
  ATLAS detector}},  {\em Nucl. Phys. A} {\bf 978} (2018) 65,
  [\href{http://arxiv.org/abs/1706.02859}{{\tt arXiv:1706.02859}}].

\bibitem{ATLAS:2019dsv}
{\bf ATLAS} Collaboration, M.~Aaboud et~al., {\it {Comparison of Fragmentation
  Functions for Jets Dominated by Light Quarks and Gluons from $pp$ and Pb+Pb
  Collisions in ATLAS}},  {\em Phys. Rev. Lett.} {\bf 123} (2019), no.~4
  042001, [\href{http://arxiv.org/abs/1902.10007}{{\tt arXiv:1902.10007}}].

\bibitem{ATLAS:2011myc}
{\bf ATLAS} Collaboration, G.~Aad et~al., {\it {Measurement of the jet
  fragmentation function and transverse profile in proton-proton collisions at
  a center-of-mass energy of 7 TeV with the ATLAS detector}},  {\em Eur. Phys.
  J. C} {\bf 71} (2011) 1795, [\href{http://arxiv.org/abs/1109.5816}{{\tt
  arXiv:1109.5816}}].

\bibitem{CMS:2014jjt}
{\bf CMS} Collaboration, S.~Chatrchyan et~al., {\it {Measurement of Jet
  Fragmentation in PbPb and pp Collisions at $\sqrt{s_{NN}}= 2.76$ TeV}},  {\em
  Phys. Rev. C} {\bf 90} (2014), no.~2 024908,
  [\href{http://arxiv.org/abs/1406.0932}{{\tt arXiv:1406.0932}}].

\bibitem{ALICE:2023jgm}
{\bf ALICE} Collaboration, {\it {Exploring the non-universality of charm
  hadronisation through the measurement of the fraction of jet longitudinal
  momentum carried by $\Lambda_{\rm c}^+$ baryons in pp collisions}},
  \href{http://arxiv.org/abs/2301.13798}{{\tt arXiv:2301.13798}}.

\bibitem{Collins:2011zzd}
J.~Collins, {\em {Foundations of perturbative QCD}}, vol.~32.
\newblock Cambridge University Press, 11, 2013.

\bibitem{Kang:2016ehg}
Z.-B. Kang, F.~Ringer, and I.~Vitev, {\it {Jet substructure using
  semi-inclusive jet functions in SCET}},  {\em JHEP} {\bf 11} (2016) 155,
  [\href{http://arxiv.org/abs/1606.07063}{{\tt arXiv:1606.07063}}].

\bibitem{Anderle:2017cgl}
D.~P. Anderle, T.~Kaufmann, M.~Stratmann, F.~Ringer, and I.~Vitev, {\it {Using
  hadron-in-jet data in a global analysis of $D^{*}$ fragmentation functions}},
   {\em Phys. Rev. D} {\bf 96} (2017), no.~3 034028,
  [\href{http://arxiv.org/abs/1706.09857}{{\tt arXiv:1706.09857}}].

\bibitem{Kang:2017glf}
Z.-B. Kang, X.~Liu, F.~Ringer, and H.~Xing, {\it {The transverse momentum
  distribution of hadrons within jets}},  {\em JHEP} {\bf 11} (2017) 068,
  [\href{http://arxiv.org/abs/1705.08443}{{\tt arXiv:1705.08443}}].

\bibitem{Arleo:2013tya}
F.~Arleo, M.~Fontannaz, J.-P. Guillet, and C.~L. Nguyen, {\it {Probing
  fragmentation functions from same-side hadron-jet momentum correlations in
  p-p collisions}},  {\em JHEP} {\bf 04} (2014) 147,
  [\href{http://arxiv.org/abs/1311.7356}{{\tt arXiv:1311.7356}}].

\bibitem{Kaufmann:2015hma}
T.~Kaufmann, A.~Mukherjee, and W.~Vogelsang, {\it {Hadron Fragmentation Inside
  Jets in Hadronic Collisions}},  {\em Phys. Rev. D} {\bf 92} (2015), no.~5
  054015, [\href{http://arxiv.org/abs/1506.01415}{{\tt arXiv:1506.01415}}].
  [Erratum: Phys.Rev.D 101, 079901 (2020)].

\bibitem{Dai:2016hzf}
L.~Dai, C.~Kim, and A.~K. Leibovich, {\it {Fragmentation of a Jet with Small
  Radius}},  {\em Phys. Rev. D} {\bf 94} (2016), no.~11 114023,
  [\href{http://arxiv.org/abs/1606.07411}{{\tt arXiv:1606.07411}}].

\bibitem{Kang:2019ahe}
Z.-B. Kang, K.~Lee, J.~Terry, and H.~Xing, {\it {Jet fragmentation functions
  for $Z$-tagged jets}},  {\em Phys. Lett. B} {\bf 798} (2019) 134978,
  [\href{http://arxiv.org/abs/1906.07187}{{\tt arXiv:1906.07187}}].

\bibitem{Zhang:2022bhq}
S.-L. Zhang, H.~Xing, and B.-W. Zhang, {\it {Hadron productions and jet
  substructures associated with $Z^0/\gamma$ in Pb+Pb collisions at the LHC}},
  \href{http://arxiv.org/abs/2209.15336}{{\tt arXiv:2209.15336}}.

\bibitem{Liu:2023fsq}
C.~Liu, X.~Shen, B.~Zhou, and J.~Gao, {\it {Automated calculation of jet
  fragmentation at NLO in QCD}},  {\em JHEP} {\bf 09} (2023) 108,
  [\href{http://arxiv.org/abs/2305.14620}{{\tt arXiv:2305.14620}}].

\bibitem{Procura:2009vm}
M.~Procura and I.~W. Stewart, {\it {Quark Fragmentation within an Identified
  Jet}},  {\em Phys. Rev. D} {\bf 81} (2010) 074009,
  [\href{http://arxiv.org/abs/0911.4980}{{\tt arXiv:0911.4980}}]. [Erratum:
  Phys.Rev.D 83, 039902 (2011)].

\bibitem{Jain:2011xz}
A.~Jain, M.~Procura, and W.~J. Waalewijn, {\it {Parton Fragmentation within an
  Identified Jet at NNLL}},  {\em JHEP} {\bf 05} (2011) 035,
  [\href{http://arxiv.org/abs/1101.4953}{{\tt arXiv:1101.4953}}].

\bibitem{Jain:2011iu}
A.~Jain, M.~Procura, and W.~J. Waalewijn, {\it {Fully-Unintegrated Parton
  Distribution and Fragmentation Functions at Perturbative $k_T$}},  {\em JHEP}
  {\bf 04} (2012) 132, [\href{http://arxiv.org/abs/1110.0839}{{\tt
  arXiv:1110.0839}}].

\bibitem{Chien:2015ctp}
Y.-T. Chien, Z.-B. Kang, F.~Ringer, I.~Vitev, and H.~Xing, {\it {Jet
  fragmentation functions in proton-proton collisions using soft-collinear
  effective theory}},  {\em JHEP} {\bf 05} (2016) 125,
  [\href{http://arxiv.org/abs/1512.06851}{{\tt arXiv:1512.06851}}].

\bibitem{Kang:2020xyq}
Z.-B. Kang, K.~Lee, and F.~Zhao, {\it {Polarized jet fragmentation functions}},
   {\em Phys. Lett. B} {\bf 809} (2020) 135756,
  [\href{http://arxiv.org/abs/2005.02398}{{\tt arXiv:2005.02398}}].

\bibitem{Bain:2016rrv}
R.~Bain, Y.~Makris, and T.~Mehen, {\it {Transverse Momentum Dependent
  Fragmenting Jet Functions with Applications to Quarkonium Production}},  {\em
  JHEP} {\bf 11} (2016) 144, [\href{http://arxiv.org/abs/1610.06508}{{\tt
  arXiv:1610.06508}}].

\bibitem{Makris:2017arq}
Y.~Makris, D.~Neill, and V.~Vaidya, {\it {Probing Transverse-Momentum Dependent
  Evolution With Groomed Jets}},  {\em JHEP} {\bf 07} (2018) 167,
  [\href{http://arxiv.org/abs/1712.07653}{{\tt arXiv:1712.07653}}].

\bibitem{Neill:2018wtk}
D.~Neill, A.~Papaefstathiou, W.~J. Waalewijn, and L.~Zoppi, {\it {Phenomenology
  with a recoil-free jet axis: TMD fragmentation and the jet shape}},  {\em
  JHEP} {\bf 01} (2019) 067, [\href{http://arxiv.org/abs/1810.12915}{{\tt
  arXiv:1810.12915}}].

\bibitem{Bacchetta:2000jk}
A.~Bacchetta and P.~J. Mulders, {\it {Deep inelastic leptoproduction of
  spin-one hadrons}},  {\em Phys. Rev. D} {\bf 62} (2000) 114004,
  [\href{http://arxiv.org/abs/hep-ph/0007120}{{\tt hep-ph/0007120}}].

\bibitem{Mulders:2000sh}
P.~J. Mulders and J.~Rodrigues, {\it {Transverse momentum dependence in gluon
  distribution and fragmentation functions}},  {\em Phys. Rev. D} {\bf 63}
  (2001) 094021, [\href{http://arxiv.org/abs/hep-ph/0009343}{{\tt
  hep-ph/0009343}}].

\bibitem{Metz:2016swz}
A.~Metz and A.~Vossen, {\it {Parton Fragmentation Functions}},  {\em Prog.
  Part. Nucl. Phys.} {\bf 91} (2016) 136--202,
  [\href{http://arxiv.org/abs/1607.02521}{{\tt arXiv:1607.02521}}].

\bibitem{Kang:2021ffh}
Z.-B. Kang, K.~Lee, D.~Y. Shao, and F.~Zhao, {\it {Spin asymmetries in
  electron-jet production at the future electron ion collider}},  {\em JHEP}
  {\bf 11} (2021) 005, [\href{http://arxiv.org/abs/2106.15624}{{\tt
  arXiv:2106.15624}}].

\bibitem{Kang:2021kpt}
Z.-B. Kang, J.~Terry, A.~Vossen, Q.~Xu, and J.~Zhang, {\it {Transverse Lambda
  production at the future Electron-Ion Collider}},  {\em Phys. Rev. D} {\bf
  105} (2022), no.~9 094033, [\href{http://arxiv.org/abs/2108.05383}{{\tt
  arXiv:2108.05383}}].

\bibitem{Zhao:2023lav}
F.~Zhao, {\it {3D Imaging via Polarized Jet Fragmentation Functions and Quantum
  Simulation of the QCD Phase Diagram}},  other thesis, 9, 2023.

\bibitem{Bacchetta:2023njc}
A.~Bacchetta, M.~Radici, and L.~Rossi, {\it {Analogies between hadron-in-jet
  and dihadron fragmentation}},  {\em Phys. Rev. D} {\bf 108} (2023), no.~1
  014005, [\href{http://arxiv.org/abs/2303.04314}{{\tt arXiv:2303.04314}}].

\bibitem{Chien:2021yol}
Y.-T. Chien, A.~Deshpande, M.~M. Mondal, and G.~Sterman, {\it {Probing
  hadronization with flavor correlations of leading particles in jets}},  {\em
  Phys. Rev. D} {\bf 105} (2022), no.~5 L051502,
  [\href{http://arxiv.org/abs/2109.15318}{{\tt arXiv:2109.15318}}].

\bibitem{Neill:2016vbi}
D.~Neill, I.~Scimemi, and W.~J. Waalewijn, {\it {Jet axes and universal
  transverse-momentum-dependent fragmentation}},  {\em JHEP} {\bf 04} (2017)
  020, [\href{http://arxiv.org/abs/1612.04817}{{\tt arXiv:1612.04817}}].

\bibitem{Gutierrez-Reyes:2019vbx}
D.~Gutierrez-Reyes, I.~Scimemi, W.~J. Waalewijn, and L.~Zoppi, {\it {Transverse
  momentum dependent distributions in $e^+e^-$ and semi-inclusive
  deep-inelastic scattering using jets}},  {\em JHEP} {\bf 10} (2019) 031,
  [\href{http://arxiv.org/abs/1904.04259}{{\tt arXiv:1904.04259}}].

\bibitem{Makris:2021drz}
Y.~Makris, {\it {Revisiting the role of grooming in DIS}},  {\em Phys. Rev. D}
  {\bf 103} (2021), no.~5 054005, [\href{http://arxiv.org/abs/2101.02708}{{\tt
  arXiv:2101.02708}}].

\bibitem{Makris:2018npl}
Y.~Makris and V.~Vaidya, {\it {Transverse Momentum Spectra at Threshold for
  Groomed Heavy Quark Jets}},  {\em JHEP} {\bf 10} (2018) 019,
  [\href{http://arxiv.org/abs/1807.09805}{{\tt arXiv:1807.09805}}].

\bibitem{Gutierrez-Reyes:2019msa}
D.~Gutierrez-Reyes, Y.~Makris, V.~Vaidya, I.~Scimemi, and L.~Zoppi, {\it
  {Probing Transverse-Momentum Distributions With Groomed Jets}},  {\em JHEP}
  {\bf 08} (2019) 161, [\href{http://arxiv.org/abs/1907.05896}{{\tt
  arXiv:1907.05896}}].

\bibitem{Gao:2019ojf}
A.~Gao, H.~T. Li, I.~Moult, and H.~X. Zhu, {\it {Precision QCD Event Shapes at
  Hadron Colliders: The Transverse Energy-Energy Correlator in the Back-to-Back
  Limit}},  {\em Phys. Rev. Lett.} {\bf 123} (2019), no.~6 062001,
  [\href{http://arxiv.org/abs/1901.04497}{{\tt arXiv:1901.04497}}].

\bibitem{Li:2020bub}
H.~T. Li, I.~Vitev, and Y.~J. Zhu, {\it {Transverse-Energy-Energy Correlations
  in Deep Inelastic Scattering}},  {\em JHEP} {\bf 11} (2020) 051,
  [\href{http://arxiv.org/abs/2006.02437}{{\tt arXiv:2006.02437}}].

\bibitem{Kang:2023big}
Z.-B. Kang, K.~Lee, D.~Y. Shao, and F.~Zhao, {\it {Probing Transverse Momentum
  Dependent Structures with Azimuthal Dependence of Energy Correlators}},
  \href{http://arxiv.org/abs/2310.15159}{{\tt arXiv:2310.15159}}.

\bibitem{Bauer:2000ew}
C.~W. Bauer, S.~Fleming, and M.~E. Luke, {\it {Summing Sudakov logarithms in $B
  \to X_s \gamma$ in effective field theory}},  {\em Phys. Rev. D} {\bf 63}
  (2000) 014006, [\href{http://arxiv.org/abs/hep-ph/0005275}{{\tt
  hep-ph/0005275}}].

\bibitem{Bauer:2000yr}
C.~W. Bauer, S.~Fleming, D.~Pirjol, and I.~W. Stewart, {\it {An Effective field
  theory for collinear and soft gluons: Heavy to light decays}},  {\em Phys.
  Rev. D} {\bf 63} (2001) 114020,
  [\href{http://arxiv.org/abs/hep-ph/0011336}{{\tt hep-ph/0011336}}].

\bibitem{Bauer:2001ct}
C.~W. Bauer and I.~W. Stewart, {\it {Invariant operators in collinear effective
  theory}},  {\em Phys. Lett. B} {\bf 516} (2001) 134--142,
  [\href{http://arxiv.org/abs/hep-ph/0107001}{{\tt hep-ph/0107001}}].

\bibitem{Bauer:2001yt}
C.~W. Bauer, D.~Pirjol, and I.~W. Stewart, {\it {Soft collinear factorization
  in effective field theory}},  {\em Phys. Rev. D} {\bf 65} (2002) 054022,
  [\href{http://arxiv.org/abs/hep-ph/0109045}{{\tt hep-ph/0109045}}].

\bibitem{Bauer:2002nz}
C.~W. Bauer, S.~Fleming, D.~Pirjol, I.~Z. Rothstein, and I.~W. Stewart, {\it
  {Hard scattering factorization from effective field theory}},  {\em Phys.
  Rev. D} {\bf 66} (2002) 014017,
  [\href{http://arxiv.org/abs/hep-ph/0202088}{{\tt hep-ph/0202088}}].

\bibitem{Kang:2016mcy}
Z.-B. Kang, F.~Ringer, and I.~Vitev, {\it {The semi-inclusive jet function in
  SCET and small radius resummation for inclusive jet production}},  {\em JHEP}
  {\bf 10} (2016) 125, [\href{http://arxiv.org/abs/1606.06732}{{\tt
  arXiv:1606.06732}}].

\bibitem{vonKuk:2023jfd}
R.~von Kuk, J.~K.~L. Michel, and Z.~Sun, {\it {Transverse momentum
  distributions of heavy hadrons and polarized heavy quarks}},  {\em JHEP} {\bf
  09} (2023) 205, [\href{http://arxiv.org/abs/2305.15461}{{\tt
  arXiv:2305.15461}}].

\bibitem{Dai:2018ywt}
L.~Dai, C.~Kim, and A.~K. Leibovich, {\it {Heavy Quark Jet Fragmentation}},
  {\em JHEP} {\bf 09} (2018) 109, [\href{http://arxiv.org/abs/1805.06014}{{\tt
  arXiv:1805.06014}}].

\bibitem{Cacciari:2008gp}
M.~Cacciari, G.~P. Salam, and G.~Soyez, {\it {The anti-$k_t$ jet clustering
  algorithm}},  {\em JHEP} {\bf 04} (2008) 063,
  [\href{http://arxiv.org/abs/0802.1189}{{\tt arXiv:0802.1189}}].

\bibitem{Barone:2003fy}
V.~Barone and P.~G. Ratcliffe, {\em {Transverse spin physics}}.
\newblock 2003.

\bibitem{Ellis:2010rwa}
S.~D. Ellis, C.~K. Vermilion, J.~R. Walsh, A.~Hornig, and C.~Lee, {\it {Jet
  Shapes and Jet Algorithms in SCET}},  {\em JHEP} {\bf 11} (2010) 101,
  [\href{http://arxiv.org/abs/1001.0014}{{\tt arXiv:1001.0014}}].

\bibitem{Kang:2017mda}
Z.-B. Kang, F.~Ringer, and W.~J. Waalewijn, {\it {The Energy Distribution of
  Subjets and the Jet Shape}},  {\em JHEP} {\bf 07} (2017) 064,
  [\href{http://arxiv.org/abs/1705.05375}{{\tt arXiv:1705.05375}}].

\bibitem{Vogelsang:1996im}
W.~Vogelsang, {\it {The Spin dependent two loop splitting functions}},  {\em
  Nucl. Phys. B} {\bf 475} (1996) 47--72,
  [\href{http://arxiv.org/abs/hep-ph/9603366}{{\tt hep-ph/9603366}}].

\bibitem{Vogelsang:1997ak}
W.~Vogelsang, {\it {Next-to-leading order evolution of transversity
  distributions and Soffer's inequality}},  {\em Phys. Rev. D} {\bf 57} (1998)
  1886--1894, [\href{http://arxiv.org/abs/hep-ph/9706511}{{\tt
  hep-ph/9706511}}].

\bibitem{Altarelli:1977zs}
G.~Altarelli and G.~Parisi, {\it {Asymptotic Freedom in Parton Language}},
  {\em Nucl. Phys. B} {\bf 126} (1977) 298--318.

\bibitem{Kang:2020xez}
Z.-B. Kang, K.~Lee, D.~Y. Shao, and J.~Terry, {\it {The Sivers Asymmetry in
  Hadronic Dijet Production}},  {\em JHEP} {\bf 02} (2021) 066,
  [\href{http://arxiv.org/abs/2008.05470}{{\tt arXiv:2008.05470}}].

\bibitem{Gao:2023ulg}
M.-S. Gao, Z.-B. Kang, D.~Y. Shao, J.~Terry, and C.~Zhang, {\it {QCD
  resummation of dijet azimuthal decorrelations in pp and pA collisions}},
  {\em JHEP} {\bf 10} (2023) 013, [\href{http://arxiv.org/abs/2306.09317}{{\tt
  arXiv:2306.09317}}].

\bibitem{Liu:2018trl}
X.~Liu, F.~Ringer, W.~Vogelsang, and F.~Yuan, {\it {Lepton-jet Correlations in
  Deep Inelastic Scattering at the Electron-Ion Collider}},  {\em Phys. Rev.
  Lett.} {\bf 122} (2019), no.~19 192003,
  [\href{http://arxiv.org/abs/1812.08077}{{\tt arXiv:1812.08077}}].

\bibitem{Waalewijn:2012sv}
W.~J. Waalewijn, {\it {Calculating the Charge of a Jet}},  {\em Phys. Rev. D}
  {\bf 86} (2012) 094030, [\href{http://arxiv.org/abs/1209.3019}{{\tt
  arXiv:1209.3019}}].

\bibitem{Baumgart:2014upa}
M.~Baumgart, A.~K. Leibovich, T.~Mehen, and I.~Z. Rothstein, {\it {Probing
  Quarkonium Production Mechanisms with Jet Substructure}},  {\em JHEP} {\bf
  11} (2014) 003, [\href{http://arxiv.org/abs/1406.2295}{{\tt
  arXiv:1406.2295}}].

\bibitem{Procura:2011aq}
M.~Procura and W.~J. Waalewijn, {\it {Fragmentation in Jets: Cone and Threshold
  Effects}},  {\em Phys. Rev. D} {\bf 85} (2012) 114041,
  [\href{http://arxiv.org/abs/1111.6605}{{\tt arXiv:1111.6605}}].

\bibitem{Chiu:2011qc}
J.-y. Chiu, A.~Jain, D.~Neill, and I.~Z. Rothstein, {\it {The Rapidity
  Renormalization Group}},  {\em Phys. Rev. Lett.} {\bf 108} (2012) 151601,
  [\href{http://arxiv.org/abs/1104.0881}{{\tt arXiv:1104.0881}}].

\bibitem{Chiu:2012ir}
J.-Y. Chiu, A.~Jain, D.~Neill, and I.~Z. Rothstein, {\it {A Formalism for the
  Systematic Treatment of Rapidity Logarithms in Quantum Field Theory}},  {\em
  JHEP} {\bf 05} (2012) 084, [\href{http://arxiv.org/abs/1202.0814}{{\tt
  arXiv:1202.0814}}].

\bibitem{Buffing:2018ggv}
M.~G.~A. Buffing, Z.-B. Kang, K.~Lee, and X.~Liu, {\it {A transverse momentum
  dependent framework for back-to-back photon+jet production}},
  \href{http://arxiv.org/abs/1812.07549}{{\tt arXiv:1812.07549}}.

\bibitem{Ebert:2019okf}
M.~A. Ebert, I.~W. Stewart, and Y.~Zhao, {\it {Towards Quasi-Transverse
  Momentum Dependent PDFs Computable on the Lattice}},  {\em JHEP} {\bf 09}
  (2019) 037, [\href{http://arxiv.org/abs/1901.03685}{{\tt arXiv:1901.03685}}].

\bibitem{Moult:2022xzt}
I.~Moult, H.~X. Zhu, and Y.~J. Zhu, {\it {The four loop QCD rapidity anomalous
  dimension}},  {\em JHEP} {\bf 08} (2022) 280,
  [\href{http://arxiv.org/abs/2205.02249}{{\tt arXiv:2205.02249}}].

\bibitem{Duhr:2022yyp}
C.~Duhr, B.~Mistlberger, and G.~Vita, {\it {Four-Loop Rapidity Anomalous
  Dimension and Event Shapes to Fourth Logarithmic Order}},  {\em Phys. Rev.
  Lett.} {\bf 129} (2022), no.~16 162001,
  [\href{http://arxiv.org/abs/2205.02242}{{\tt arXiv:2205.02242}}].

\bibitem{Shanahan:2021tst}
P.~Shanahan, M.~Wagman, and Y.~Zhao, {\it {Lattice QCD calculation of the
  Collins-Soper kernel from quasi-TMDPDFs}},  {\em Phys. Rev. D} {\bf 104}
  (2021), no.~11 114502, [\href{http://arxiv.org/abs/2107.11930}{{\tt
  arXiv:2107.11930}}].

\bibitem{LPC:2022ibr}
{\bf LPC} Collaboration, M.-H. Chu et~al., {\it {Nonperturbative determination
  of the Collins-Soper kernel from quasitransverse-momentum-dependent wave
  functions}},  {\em Phys. Rev. D} {\bf 106} (2022), no.~3 034509,
  [\href{http://arxiv.org/abs/2204.00200}{{\tt arXiv:2204.00200}}].

\bibitem{LatticePartonLPC:2023pdv}
{\bf Lattice Parton (LPC)} Collaboration, M.-H. Chu et~al., {\it {Lattice
  calculation of the intrinsic soft function and the Collins-Soper kernel}},
  {\em JHEP} {\bf 08} (2023) 172, [\href{http://arxiv.org/abs/2306.06488}{{\tt
  arXiv:2306.06488}}].

\bibitem{Shu:2023cot}
H.-T. Shu, M.~Schlemmer, T.~Sizmann, A.~Vladimirov, L.~Walter, M.~Engelhardt,
  A.~Sch\"afer, and Y.-B. Yang, {\it {Universality of the Collins-Soper kernel
  in lattice calculations}},  {\em Phys. Rev. D} {\bf 108} (2023), no.~7
  074519, [\href{http://arxiv.org/abs/2302.06502}{{\tt arXiv:2302.06502}}].

\bibitem{Avkhadiev:2023poz}
A.~Avkhadiev, P.~Shanahan, M.~Wagman, and Y.~Zhao, {\it {Collins-Soper kernel
  from lattice QCD at the physical pion mass}},
  \href{http://arxiv.org/abs/2307.12359}{{\tt arXiv:2307.12359}}.

\bibitem{Becher:2006mr}
T.~Becher, M.~Neubert, and B.~D. Pecjak, {\it {Factorization and Momentum-Space
  Resummation in Deep-Inelastic Scattering}},  {\em JHEP} {\bf 01} (2007) 076,
  [\href{http://arxiv.org/abs/hep-ph/0607228}{{\tt hep-ph/0607228}}].

\bibitem{Luo:2020epw}
M.-x. Luo, T.-Z. Yang, H.~X. Zhu, and Y.~J. Zhu, {\it {Unpolarized quark and
  gluon TMD PDFs and FFs at N$^{3}$LO}},  {\em JHEP} {\bf 06} (2021) 115,
  [\href{http://arxiv.org/abs/2012.03256}{{\tt arXiv:2012.03256}}].

\bibitem{Ebert:2020qef}
M.~A. Ebert, B.~Mistlberger, and G.~Vita, {\it {TMD fragmentation functions at
  N$^{3}$LO}},  {\em JHEP} {\bf 07} (2021) 121,
  [\href{http://arxiv.org/abs/2012.07853}{{\tt arXiv:2012.07853}}].

\bibitem{Gutierrez-Reyes:2018iod}
D.~Gutierrez-Reyes, I.~Scimemi, and A.~Vladimirov, {\it {Transverse momentum
  dependent transversely polarized distributions at
  next-to-next-to-leading-order}},  {\em JHEP} {\bf 07} (2018) 172,
  [\href{http://arxiv.org/abs/1805.07243}{{\tt arXiv:1805.07243}}].

\bibitem{Cal:2019hjc}
P.~Cal, F.~Ringer, and W.~J. Waalewijn, {\it {The jet shape at NLL'}},  {\em
  JHEP} {\bf 05} (2019) 143, [\href{http://arxiv.org/abs/1901.06389}{{\tt
  arXiv:1901.06389}}].

\bibitem{Kang:2017btw}
Z.-B. Kang, A.~Prokudin, F.~Ringer, and F.~Yuan, {\it {Collins azimuthal
  asymmetries of hadron production inside jets}},  {\em Phys. Lett. B} {\bf
  774} (2017) 635--642, [\href{http://arxiv.org/abs/1707.00913}{{\tt
  arXiv:1707.00913}}].

\bibitem{Arratia:2022oxd}
M.~Arratia, Z.-B. Kang, S.~J. Paul, A.~Prokudin, F.~Ringer, and F.~Zhao, {\it
  {Neutrino-tagged jets at the Electron-Ion Collider}},  {\em Phys. Rev. D}
  {\bf 107} (2023), no.~9 094036, [\href{http://arxiv.org/abs/2212.02432}{{\tt
  arXiv:2212.02432}}].

\bibitem{Arratia:2020nxw}
M.~Arratia, Z.-B. Kang, A.~Prokudin, and F.~Ringer, {\it {Jet-based
  measurements of Sivers and Collins asymmetries at the future electron-ion
  collider}},  {\em Phys. Rev. D} {\bf 102} (2020), no.~7 074015,
  [\href{http://arxiv.org/abs/2007.07281}{{\tt arXiv:2007.07281}}].

\bibitem{Collins:1984kg}
J.~C. Collins, D.~E. Soper, and G.~F. Sterman, {\it {Transverse Momentum
  Distribution in Drell-Yan Pair and W and Z Boson Production}},  {\em Nucl.
  Phys. B} {\bf 250} (1985) 199--224.

\bibitem{Kulesza:2002rh}
A.~Kulesza, G.~F. Sterman, and W.~Vogelsang, {\it {Joint resummation in
  electroweak boson production}},  {\em Phys. Rev. D} {\bf 66} (2002) 014011,
  [\href{http://arxiv.org/abs/hep-ph/0202251}{{\tt hep-ph/0202251}}].

\bibitem{Qiu:2000hf}
J.-w. Qiu and X.-f. Zhang, {\it {Role of the nonperturbative input in QCD
  resummed Drell-Yan $Q_T$ distributions}},  {\em Phys. Rev. D} {\bf 63} (2001)
  114011, [\href{http://arxiv.org/abs/hep-ph/0012348}{{\tt hep-ph/0012348}}].

\bibitem{Catani:2015vma}
S.~Catani, D.~de~Florian, G.~Ferrera, and M.~Grazzini, {\it {Vector boson
  production at hadron colliders: transverse-momentum resummation and leptonic
  decay}},  {\em JHEP} {\bf 12} (2015) 047,
  [\href{http://arxiv.org/abs/1507.06937}{{\tt arXiv:1507.06937}}].

\bibitem{Ebert:2016gcn}
M.~A. Ebert and F.~J. Tackmann, {\it {Resummation of Transverse Momentum
  Distributions in Distribution Space}},  {\em JHEP} {\bf 02} (2017) 110,
  [\href{http://arxiv.org/abs/1611.08610}{{\tt arXiv:1611.08610}}].

\bibitem{Monni:2016ktx}
P.~F. Monni, E.~Re, and P.~Torrielli, {\it {Higgs Transverse-Momentum
  Resummation in Direct Space}},  {\em Phys. Rev. Lett.} {\bf 116} (2016),
  no.~24 242001, [\href{http://arxiv.org/abs/1604.02191}{{\tt
  arXiv:1604.02191}}].

\bibitem{Sun:2014dqm}
P.~Sun, J.~Isaacson, C.~P. Yuan, and F.~Yuan, {\it {Nonperturbative functions
  for SIDIS and Drell-Yan processes}},  {\em Int. J. Mod. Phys. A} {\bf 33}
  (2018), no.~11 1841006, [\href{http://arxiv.org/abs/1406.3073}{{\tt
  arXiv:1406.3073}}].

\bibitem{Echevarria:2020hpy}
M.~G. Echevarria, Z.-B. Kang, and J.~Terry, {\it {Global analysis of the Sivers
  functions at NLO+NNLL in QCD}},  {\em JHEP} {\bf 01} (2021) 126,
  [\href{http://arxiv.org/abs/2009.10710}{{\tt arXiv:2009.10710}}].

\bibitem{STAR:2022hqg}
{\bf STAR} Collaboration, M.~Abdallah et~al., {\it {Azimuthal transverse
  single-spin asymmetries of inclusive jets and identified hadrons within jets
  from polarized $pp$ collisions at $\sqrt{s} = 200~\mathrm{GeV}$}},  {\em
  Phys. Rev. D} {\bf 106} (2022), no.~7 072010,
  [\href{http://arxiv.org/abs/2205.11800}{{\tt arXiv:2205.11800}}].

\bibitem{STAR:2018fqv}
{\bf STAR} Collaboration, J.~Adam et~al., {\it {Transverse spin transfer to
  $\Lambda$ and $\bar{\Lambda}$ hyperons in polarized proton-proton collisions
  at $\sqrt{s}=200\,\mathrm{GeV}$}},  {\em Phys. Rev. D} {\bf 98} (2018), no.~9
  091103, [\href{http://arxiv.org/abs/1808.08000}{{\tt arXiv:1808.08000}}].

\bibitem{Kang:2015msa}
Z.-B. Kang, A.~Prokudin, P.~Sun, and F.~Yuan, {\it {Extraction of Quark
  Transversity Distribution and Collins Fragmentation Functions with QCD
  Evolution}},  {\em Phys. Rev. D} {\bf 93} (2016), no.~1 014009,
  [\href{http://arxiv.org/abs/1505.05589}{{\tt arXiv:1505.05589}}].

\bibitem{Cocuzza:2022jye}
{\bf Jefferson Lab Angular Momentum (JAM)} Collaboration, C.~Cocuzza,
  W.~Melnitchouk, A.~Metz, and N.~Sato, {\it {Polarized antimatter in the
  proton from a global QCD analysis}},  {\em Phys. Rev. D} {\bf 106} (2022),
  no.~3 L031502, [\href{http://arxiv.org/abs/2202.03372}{{\tt
  arXiv:2202.03372}}].

\bibitem{Callos:2020qtu}
D.~Callos, Z.-B. Kang, and J.~Terry, {\it {Extracting the transverse momentum
  dependent polarizing fragmentation functions}},  {\em Phys. Rev. D} {\bf 102}
  (2020), no.~9 096007, [\href{http://arxiv.org/abs/2003.04828}{{\tt
  arXiv:2003.04828}}].

\bibitem{Albino:2005mv}
S.~Albino, B.~A. Kniehl, and G.~Kramer, {\it {Fragmentation functions for K0(S)
  and Lambda with complete quark flavor separation}},  {\em Nucl. Phys. B} {\bf
  734} (2006) 50--61, [\href{http://arxiv.org/abs/hep-ph/0510173}{{\tt
  hep-ph/0510173}}].

\bibitem{Vogt:2004ns}
A.~Vogt, {\it {Efficient evolution of unpolarized and polarized parton
  distributions with QCD-\textsc{Pegasus}}},  {\em Comput. Phys. Commun.} {\bf
  170} (2005) 65--92, [\href{http://arxiv.org/abs/hep-ph/0408244}{{\tt
  hep-ph/0408244}}].

\bibitem{Bhattacharya:2021twu}
S.~Bhattacharya, Z.-B. Kang, A.~Metz, G.~Penn, and D.~Pitonyak, {\it {First
  global QCD analysis of the TMD g1T from semi-inclusive DIS data}},  {\em
  Phys. Rev. D} {\bf 105} (2022), no.~3 034007,
  [\href{http://arxiv.org/abs/2110.10253}{{\tt arXiv:2110.10253}}].

\bibitem{Lai:2010vv}
H.-L. Lai, M.~Guzzi, J.~Huston, Z.~Li, P.~M. Nadolsky, J.~Pumplin, and C.~P.
  Yuan, {\it {New parton distributions for collider physics}},  {\em Phys. Rev.
  D} {\bf 82} (2010) 074024, [\href{http://arxiv.org/abs/1007.2241}{{\tt
  arXiv:1007.2241}}].

\bibitem{Koike:2006fn}
Y.~Koike, J.~Nagashima, and W.~Vogelsang, {\it {Resummation for polarized
  semi-inclusive deep-inelastic scattering at small transverse momentum}},
  {\em Nucl. Phys. B} {\bf 744} (2006) 59--79,
  [\href{http://arxiv.org/abs/hep-ph/0602188}{{\tt hep-ph/0602188}}].

\bibitem{deFlorian:1997zj}
D.~de~Florian, M.~Stratmann, and W.~Vogelsang, {\it {QCD analysis of
  unpolarized and polarized Lambda baryon production in leading and
  next-to-leading order}},  {\em Phys. Rev. D} {\bf 57} (1998) 5811--5824,
  [\href{http://arxiv.org/abs/hep-ph/9711387}{{\tt hep-ph/9711387}}].

\bibitem{Ritzmann:2014mka}
M.~Ritzmann and W.~J. Waalewijn, {\it {Fragmentation in Jets at NNLO}},  {\em
  Phys. Rev. D} {\bf 90} (2014), no.~5 054029,
  [\href{http://arxiv.org/abs/1407.3272}{{\tt arXiv:1407.3272}}].

\bibitem{Giele:1991vf}
W.~T. Giele and E.~W.~N. Glover, {\it {Higher order corrections to jet
  cross-sections in $e^+ e^-$ annihilation}},  {\em Phys. Rev. D} {\bf 46}
  (1992) 1980--2010.

\bibitem{Becher:2009th}
T.~Becher and M.~D. Schwartz, {\it {Direct photon production with effective
  field theory}},  {\em JHEP} {\bf 02} (2010) 040,
  [\href{http://arxiv.org/abs/0911.0681}{{\tt arXiv:0911.0681}}].

\bibitem{Echevarria:2012pw}
M.~G. Echevarria, A.~Idilbi, A.~Sch\"afer, and I.~Scimemi, {\it
  {Model-Independent Evolution of Transverse Momentum Dependent Distribution
  Functions (TMDs) at NNLL}},  {\em Eur. Phys. J. C} {\bf 73} (2013), no.~12
  2636, [\href{http://arxiv.org/abs/1208.1281}{{\tt arXiv:1208.1281}}].

\bibitem{Moch:2004pa}
S.~Moch, J.~A.~M. Vermaseren, and A.~Vogt, {\it {The Three loop splitting
  functions in QCD: The Nonsinglet case}},  {\em Nucl. Phys. B} {\bf 688}
  (2004) 101--134, [\href{http://arxiv.org/abs/hep-ph/0403192}{{\tt
  hep-ph/0403192}}].

\bibitem{Kaufmann:2019ksh}
T.~Kaufmann, X.~Liu, A.~Mukherjee, F.~Ringer, and W.~Vogelsang, {\it
  {Hadron-in-jet production at partonic threshold}},  {\em JHEP} {\bf 02}
  (2020) 040, [\href{http://arxiv.org/abs/1910.11746}{{\tt arXiv:1910.11746}}].

\end{thebibliography}\endgroup
\end{document}